\documentclass[prd,aps,reprint,groupedaddress,floatfix,preprintnumbers,flushbottom,nobibnotes,nofootinbib,showpacs,amsmath,floatfix]{revtex4-1}
\usepackage{amssymb}
\usepackage{graphicx}
\usepackage{longtable}
\usepackage{hyperref}
\usepackage{braket}
\usepackage{mathbbol}
\usepackage{color}
\usepackage{xcolor}
\usepackage{cancel}
\usepackage[utf8]{inputenc}
\usepackage{soul}
\usepackage{subfigure}
\usepackage{comment}
\usepackage{bm}

\begin{abstract}
We present a lattice QCD study of charmonium resonances and bound states with $J^{PC}=1^{--}$ and $3^{--}$ near 
the open-charm threshold, taking into account their strong transitions to $\bar DD$. Vector charmonia are the most 
abundant in the experimentally established charmonium spectrum, while recently LHCb reported also the first 
discovery of a charmonium with likely spin three. The $\bar DD$ scattering amplitudes for partial waves $l=1$ 
and $l=3$ are extracted on the lattice by means of the L\"uscher formalism, using multiple volumes and inertial 
frames. Parameterizations of the scattering amplitudes provide masses and widths of the resonances, as well as 
the masses of bound states. CLS ensembles with $2+1$ dynamical flavors of non-perturbatively $O(a)$ improved 
Wilson quarks are employed with $m_\pi\simeq 280~$MeV, a single lattice spacing of $a=0.09$~fm and two lattice 
spatial extents of $L=24$ and $32$. Two values of the charm quark mass are considered to examine the influence of 
the position of the $\bar{D}D$ threshold on the hadron masses. For the lighter charm quark mass we find the 
vector resonance $\psi(3770)$ with mass $m=3780(7)$ MeV and coupling $g=16.0(^{+2.1}_{-0.2})$~(related to the 
width by $\Gamma=g^2p^3/6\pi m^2$). Both quantities are consistent with their experimental values, 
$m^{\textrm{exp}}=3773.13(35)$ MeV and $g^{\textrm{exp}}=18.7(9)$. The vector $\psi(2S)$ appears as a bound state 
with $m=3666(10)~$MeV. The charmonium resonance with $J^{PC}=3^{--}$ is found at $m=3831(^{+10}_{-16})$~MeV, 
consistent with the $X(3842)$ recently discovered by LHCb. At our heavier charm-quark mass the $\psi(2S)$ as well 
as the $\psi(3770)$ are bound states and the $X(3842)$ remains a resonance. We stress that all quoted uncertainties 
are only statistical, while lattice spacing effects and the approach to the physical point~(for the light 
and strange quarks) still need to be explored. This study of conventional charmonia sets the stage for more 
challenging future studies of unconventional charmonium-like states.
\end{abstract}

\begin{document}

\def\Regensburg{Institute for Theoretical Physics, University of Regensburg, 93040 Regensburg, Germany}
\def\Mainz{Helmholtz-Institut Mainz, 55099 Mainz, Germany}
\title{Charmonium resonances with $J^{PC}=1^{--}$ and $3^{--}$ from  $\bar DD$ scattering on the lattice}
\author{Stefano Piemonte}
\email{stefano.piemonte@ur.de}
\affiliation{\Regensburg}
\author{Sara~Collins}
\email{sara.collins@ur.de}
\affiliation{\Regensburg}
\author{Daniel~Mohler}
\email{damohler@uni-mainz.de}
\affiliation{\Mainz}
\affiliation{Johannes Gutenberg-Universit\"at Mainz, 55099 Mainz, Germany}
\author{M.~Padmanath}
\email{Padmanath.M@physik.uni-regensburg.de}
\affiliation{\Regensburg}
\author{Sasa~Prelovsek}
\email{sasa.prelovsek@ijs.si}
\affiliation{\Regensburg}
\affiliation{Faculty of Mathematics and Physics, University of Ljubljana, 1000 Ljubljana, Slovenia}
\affiliation{Jozef Stefan Institute, 1000 Ljubljana, Slovenia}

\date{\today}

\maketitle
\preprint{MITP/19-029}

\section{Introduction}

The charmonium spectrum determined from experiment challenges our theoretical understanding of the internal dynamics of mesons containing charm quarks. While charmonia below the open-charm decay threshold are found to fit within a quark-antiquark picture, the properties of several resonances above the threshold, generally referred to as $XYZ$ states, call for exotic interpretations. Various suggestions have been made based on phenomenological approaches, effective field theories, potential models, etc. These include compact diquark-antidiquark states, mesonic molecules, hybrid mesons with gluonic excitations, or resonances generated through coupled channel scattering. The theoretical approaches are motivated
by phenomenology and approximate certain regimes of the strong interaction. However, a complete description of the observed 
charmonium spectrum from Quantum ChromoDynamics (QCD) also requires 
complementary first principles non-perturbative calculations, which can be performed using lattice QCD. 
 {\it Ab-initio} determinations of ground state charmonia, such as in Refs. \cite{Burch:2009az,Donald:2012ga,Becirevic:2012dc,Briceno:2012wt,Yang:2014sea,Galloway:2014tta,Mathur:2018epb,DeTar:2018uko}, illustrate the power of these non-perturbative
approaches in precisely determining masses and splittings.
Lattice calculations of other conventional 
charmonium bound states and resonances aim to satisfactorily describe well understood excitations and demonstrate the efficacy of the methods employed. This gives confidence in investigations of unconventional charmonium-like states and of those for which the experimental situation is less clear.

To date, several lattice QCD calculations of the excited charmonium
spectrum with isospin zero have been performed within the single
hadron approach, where one employs only mesonic interpolators of $\bar
cc$ type and assumes that this qualitatively captures the single meson
spectrum in a finite volume
\cite{Dudek:2007wv,Bali:2011rd,Liu:2012ze,Mohler:2012na,
  Cheung:2016bym}. So far, only one lattice study has gone beyond the
single-hadron approach, taking into account the strong decays of
charmonium resonances above the open-charm threshold.\footnote{The strong decay of the $Z_c(3900)$ resonance with isospin 1 has been studied in the HALQCD approach \cite{Ikeda:2016zwx,Ikeda:2017mee}.} In that study an exploratory determination of the
$\bar DD$ scattering amplitude in the $l=1$ and $l=0$ partial waves
was performed \cite{Lang:2015sba}.  The low-lying vector and scalar
charmonium spectra were calculated in a single lattice volume in the
center of momentum frame (CMF). The authors extracted the resonance
and bound state parameters in the vector channel, while the
conclusions in the scalar channel indicated different possible
interpretations and left several unresolved puzzles for both theory
and experiment. 

Note that all previous lattice studies of charmonium were limited to
the CMF
\cite{Dudek:2007wv,Bali:2011rd,Liu:2012ze,Mohler:2012na,Prelovsek:2013cra,Padmanath:2015era,Lang:2015sba,
  Cheung:2016bym}. However, lattice calculations in the moving frames ({\it
  i.e.} with non-zero total momentum) can provide additional
information on the relevant two-meson scattering amplitudes (for a
review, see Ref. \cite{Briceno:2017max}).  The challenge in
spectroscopy using lattice techniques is that, due to the reduced
symmetry on the lattice, several $J$ can contribute (in the CMF) to any
given lattice irrep. In the moving frames, the situation is much worse
as the finite volume spectrum in any lattice irrep gets a contribution
from states with different $J^P$ (in their rest frame), as they are
not good quantum numbers already in the infinite volume continuum and
also on the lattice. As the first step, we extracted the excited
charmonium spectrum within the single hadron approach on the lattice
in multiple inertial frames and identified the continuum quantum
numbers \cite{Padmanath:2018tuc}. Such a spin identified single-hadron
spectrum tells us where to expect effects from resonances in different
partial waves, which in turn provides valuable information for the
parameterization of scattering amplitudes. Thus this study allows one
to fully exploit the data in moving frames in the present and related
future studies.

The current article presents a lattice QCD investigation of charmonium resonances taking into account their most important
strong decay, which goes beyond the single hadron approach. We focus on the scattering of a pair of $\bar D$ and $D$ mesons 
in the $l=1$ partial wave, which features charmonium resonances and bound states with $J^{PC}=1^{--}$, to determine the 
respective masses as well as the couplings with the $\bar DD$ scattering channel.  We also consider the $\bar DD$ scattering 
in partial wave $l=3$ yielding information on the lowest charmonium resonance with $J^{PC}=3^{--}$, for which LHCb 
discovered a candidate named $X(3842)$ \cite{Aaij:2019evc}.

The $\bar DD$ scattering channel in partial wave $l=1$ is the dominant hadronic decay mode of the $\psi(3770)$~(with branching ratio $Br~=~93\pm9\%$). It is 
a well-established vector resonance generally accepted to be predominantly a conventional $\bar cc$ state. We therefore assume 
elastic $\bar DD$ scattering, neglecting the influence of all other allowed decay modes of the $\psi(3770)$ and of higher thresholds. The determination of the relevant scattering amplitudes and resonance parameters, such as the mass and 
decay width of the $\psi(3770)$, will serve as a demonstration of our realization of L\"uscher's finite volume treatment. In 
contrast to the only previous study \cite{Lang:2015sba}, the $\bar DD$ scattering amplitude in $p$-wave is determined utilizing 
the finite volume spectrum from two lattice ensembles with different physical volumes, each in three different inertial frames. 
The study is performed for two charm quark masses, one below and one above the physical $m_c$. In this way we explore the 
dependence of the spectra on the position of the $\bar DD$ threshold.

The recent LHCb discovery of the charmonium $X(3842)$ with likely quantum numbers $J^{PC}=3^{--}$ \cite{Aaij:2019evc} in part
motivates our investigation of this narrow state with $\Gamma\simeq  3~$MeV. This resonance is also considered since it inevitably 
affects the spectrum of vector excitations in a finite hyper-cubic volume. While we are able to determine the mass of this 
charmonium resonance, the finite volume spectrum in our lattice setup is not dense enough to reliably estimate the very narrow 
width of this state.

The layout of this paper is as follows. In Section~\ref{luescher_formalism}, the relevant quantization condition relating the infinite 
volume scattering amplitudes to the finite volume energy spectrum is discussed. The details of the lattice setup and related technical 
information are briefly outlined in Section~\ref{sec:setup}. In Section~\ref{sec:spec}, our procedure for the determination of the excited 
charmonium spectrum and our approach for dealing with discretization errors in hadron observables involving charm quarks is discussed. 
We summarize our findings in Section~\ref{sec:res} and conclude in Section~\ref{sec:concl}.

\section{Scattering in a finite box: L\"uscher formalism}\label{luescher_formalism}

In a Euclidean box of finite size, the QCD spectrum is discrete. Energy levels corresponding to bound states are affected by exponentially suppressed finite (spatial) volume effects, while a power law dependence on the size of the box is expected for energy levels related to two-particle states above the strong decay threshold. L\"uscher's finite volume formalism and its extensions \cite{Luscher:1986pf,Luscher:1990ux,Luscher:1991cf,Briceno:2014oea} (see the review Ref. \cite{Briceno:2017max} to find more references) show that the energy dependence of infinite volume scattering amplitudes determines the discrete energy spectrum in a finite-sized box with periodic boundary conditions.

We are interested in $\bar DD$ scattering, {\it i.e.} single-channel scattering of spin-zero particles with equal masses, in inertial frames with zero and non-zero total momenta $\vec P$. In this case the scattering matrix $S_l(p) = e^{2i \delta_l(p)}$ can be expressed in terms of a single phase shift $\delta_l(p)$ in the partial-wave $l$, where $p$ is the momentum of each hadron in the center of momentum frame. The spin of the system is zero, since the scattering particles are spin-less and the total angular momentum equals the orbital angular momentum ($J=l$).

The quantization condition connecting the infinite volume scattering amplitudes with the finite volume discrete energy $E$ of the eigenstates can be expressed in our case as
\begin{equation}\label{eq:det}
 \det[\tilde{K}_{l}^{-1}(E_{cm}) ~\delta_{l^{\prime}l}-
B^{\vec P;\Lambda}_{l^{\prime}l}(E_{cm})] = 0.
\end{equation}
Here the determinant is computed over all partial waves $l$ and $E_{cm}$ is the eigenenergy in the center of momentum frame. 
The real function $\tilde{K}_l$ parametrizes the infinite volume scattering amplitude $S_l$ and the transition
amplitude $t_l$ in the partial wave $l$ as 
\begin{equation}\label{eq:TKSmatrices}
(\rho ~t_l)^{-1} = 2i~(S_{l} - 1)^{-1} = p^{-2l-1}\tilde{K}_{l}^{-1} - i\,,
\end{equation}
with $\rho = 2p/E_{cm}$, or equivalently 
\begin{equation}
 \tilde K^{-1}_l(E_{cm}) = p^{2l+1} \cot\delta_l(p)\,.
\end{equation}
The kinematic variable $p$ is the momentum of each $D$ meson in its center-of-momentum frame
 \begin{equation}
 p^2 = \frac{s}{4} - m_D^2\,, 
\end{equation}
with $s=E_{cm}^2$. We follow the definitions of $\tilde{K}$ in Ref.~\cite{Brett:2018jqw} and the definition of $t_l$ used by the Hadron Spectrum Collaboration (see \emph{e.g.} Ref. \cite{Dudek:2014qha}).\footnote{However, unlike in Ref.~\cite{Brett:2018jqw}, we do not divide our energy levels or physical quantities by the mass of the scattering particles.} The factor $p^{-2l-1}$ in Eq.~(\ref{eq:TKSmatrices}) ensures a smooth kinematic behaviour of $\tilde{K}$ at the threshold.

The box-matrix $B^{\vec P;\Lambda}_{l^{\prime}l}(E_{cm})$, as introduced in Ref.~\cite{Morningstar:2017spu}, depends on $E_{cm}$, the lattice spatial size $L$ and the lattice irreducible representation~(irrep) $\Lambda$ under investigation. Its entries are customarily written in terms of the known $\mathcal{Z}$-functions defined in Ref.~\cite{Luscher:1986pf}. Note that $B$ also has off-diagonal entries in the partial wave indices $l^{\prime}$ and $l$, due to the reduced rotational symmetries in a hyper-cubic box. The latter leads to the finite volume spectrum for any lattice irrep containing information from scattering in multiple partial waves.

The main focus of the present work are the resonances and bound states of charmonium in the vector channel that are related to $\bar DD$ mesons scattering in $p$-wave. In the simplest approximation, in which the influence of higher partial waves $l > 1$ are neglected, the quantization condition in Eq.~(\ref{eq:det}) allows one to compute $\delta_1$ directly from a given measured energy level in a finite box. In this case, the determinant condition reduces to the well-known L\"uscher relation \cite{Luscher:1986pf}
\begin{equation}
 p \cot\delta_1(p)-\frac{2\mathcal{Z}_{00}(p^2)}{\sqrt{\pi}L}=0\,,
\end{equation}
for zero total momentum.

In the case when the effects of higher partial waves are non-negligible, the properties of the scattering amplitudes $S_l$ are accessible only after finding and fitting a suitable parametrization of \emph{all} the relevant $\tilde{K}_l$, so that the determinant condition in Eq.~(\ref{eq:det}) is solved for each of the energy levels in the finite volume spectrum. The finite volume mixing of even and odd partial waves is not allowed if the scattering particles have equal masses \cite{Rummukainen:1995vs,Dudek:2012xn}, as in our case. Hence the contamination of higher partial waves in the $l=1$ excitations due to broken rotational symmetry can occur only from odd partial waves $l=3,5,\dots$. The contributions from the $l>5$ partial waves are expected to be strongly suppressed in the near-threshold region explored here and it can be neglected. We will therefore investigate partial waves $l=1$ and $l=3$~(as well as their mutual influence) through the finite volume quantization condition of Eq.~(\ref{eq:det}). %In this study we determine $l=1$ and $l=3$ partial waves, which are related to $J^{PC}=1^{--}$ and $3^{--}$ charmonium resonances and bound states. {\color{blue} }  In particular, we also investigate the influence of the low lying $l=3$ partial wave excitations on the $l=1$ partial wave, where the influence arises due to   the quantization condition (\ref{eq:det}). 
We use the package TwoHadronsInBox to compute the determinant in Eq.~(\ref{eq:det}) when fitting the parametrization of the $\tilde K$-matrix to our energy levels, following the ``determinant residual'' method \cite{Morningstar:2017spu}.

\section{Lattice setup}\label{sec:setup}

The results presented in this work have been determined from two $N_f = 2+1$ ensembles labeled U101 and H105 generated by the Coordinated Lattice Simulation (CLS) consortium \cite{Bruno:2014jqa,Bali:2016umi}. The inverse gauge coupling is $\beta=6/g^2=3.4$ and the lattice spacing is $a=0.08636(98)(40)$~fm \cite{Bruno:2016plf}. The lattice volume is $T\times L^3=96\times 32^3$ for H105 and $128\times 24^3$ for U101. The Wilson-clover action ~\cite{Sheikholeslami:1985ij} is non-perturbatively improved to remove the $O(a)$ lattice artefacts, with the clover term set equal to $c_{\textrm{sw}} = 1.986246$. The strange and light hopping parameters are $\kappa_l = 0.13697$ and $\kappa_s = 0.13634079$, respectively, leading to pion and the kaon masses of $m_\pi = 280(3)$~MeV and $m_K = 467(2)$~MeV. Note that the light quark mass~($m_l$) is heavier than its physical value while the strange quark mass~($m_s$) is lighter. This is because the U101 and H105 ensembles lie on a trajectory where the physical point is approached keeping the average quark mass $2 m_l + m_s$ fixed. The gauge and the fermion fields fulfill open boundary conditions in the time direction and periodic boundary conditions in the spatial directions. The total number of configurations analyzed is 255 for the U101 ensemble and 492 for H105, with one configuration taken every 20 and 16 MDUs, respectively, from two different replica in each case.

\subsection{Choosing the charm quark mass}

The bare charm quark mass, set by the hopping parameter $\kappa_c = 1/(2m_c+8)$ for Wilson-type actions, is a parameter to be determined with care. As we aim to study the strong decay of near-threshold charmonium resonances, the particles relevant for tuning the charm quark mass are the lowest $\bar{c}c$-states, the $J/\psi$ and the $\eta_c$, and the decay products of their excited states, such as the $D$ and $D_s$ mesons. Away from the physical point, the masses of all these particles cannot be tuned simultaneously to their experimental values. However, the most critical quantity for our current investigation is the position of the strong decay threshold relative to the $J/\psi$ and the $\eta_c$, rather than the mass of an individual charmonium state. We aim to explore how charmonia near and above the open-charm threshold depend on its position. We also want to investigate the systematics coming from the tuning of the charm quark mass. To this end we employ two different values of $\kappa_c$ given by 0.12522 and 0.12315, corresponding to the $D$ meson mass being below and above its physical value, see Table \ref{tab:states}. The main results for the position of the resonances in physical units will be presented relative to the spin average mass of the $J/\psi$ and the $\eta_c$,
\begin{equation}\label{mav}
 M_{\textrm{av}} = \frac{1}{4}(m_{\eta_c} + 3 m_{J/\psi})\,.
\end{equation}
This is motivated by the fact that some of the discretization effects (which can be significant at a lattice spacing of $a \approx 0.09$ fm) are canceled when computing energy differences. The spin-average mass is equal to 3103(3) MeV for $\kappa_c=0.12315$ and 2820(3) MeV for $\kappa_c=0.12522$. The energy gap between $M_{\textrm{av}}$ and 2$m_D$ is 
\begin{eqnarray*}
 M_{\textrm{av}} - 2m_D & = & -750(3) \textrm{~MeV} ~~~~ \kappa_c=0.12315\, \\
% M_{\textrm{av}} - 2m_{D_s} & = & -856(2) \textrm{~MeV} ~~~~ \kappa_c=0.12315\, \\
 M_{\textrm{av}} - 2m_D & = & -703(3) \textrm{~MeV} ~~~~ \kappa_c=0.12522\, \\
% M_{\textrm{av}} - 2m_{D_s} & = & -813(3) \textrm{~MeV} ~~~~ \kappa_c=0.12522\, \\
 M_{\textrm{av}} - 2m_{D} & = & -665 \textrm{~MeV} ~~~~ \textrm{~~~~experimental}\,
% M_{\textrm{av}} - 2m_{D_s} & = & -868.2 \textrm{~MeV} ~~~~ \textrm{experimental}\,.
\end{eqnarray*}
Note that the relative position of the $\bar{D} D$ threshold with respect to $M_{\textrm{av}}$ is closer to the experimental value for $\kappa_c=0.12522$ than for $\kappa_c=0.12315$.

\subsection{Setup of the distillation algorithm}

\begin{figure}
\centering
 \subfigure{\includegraphics[width=0.235\textwidth]{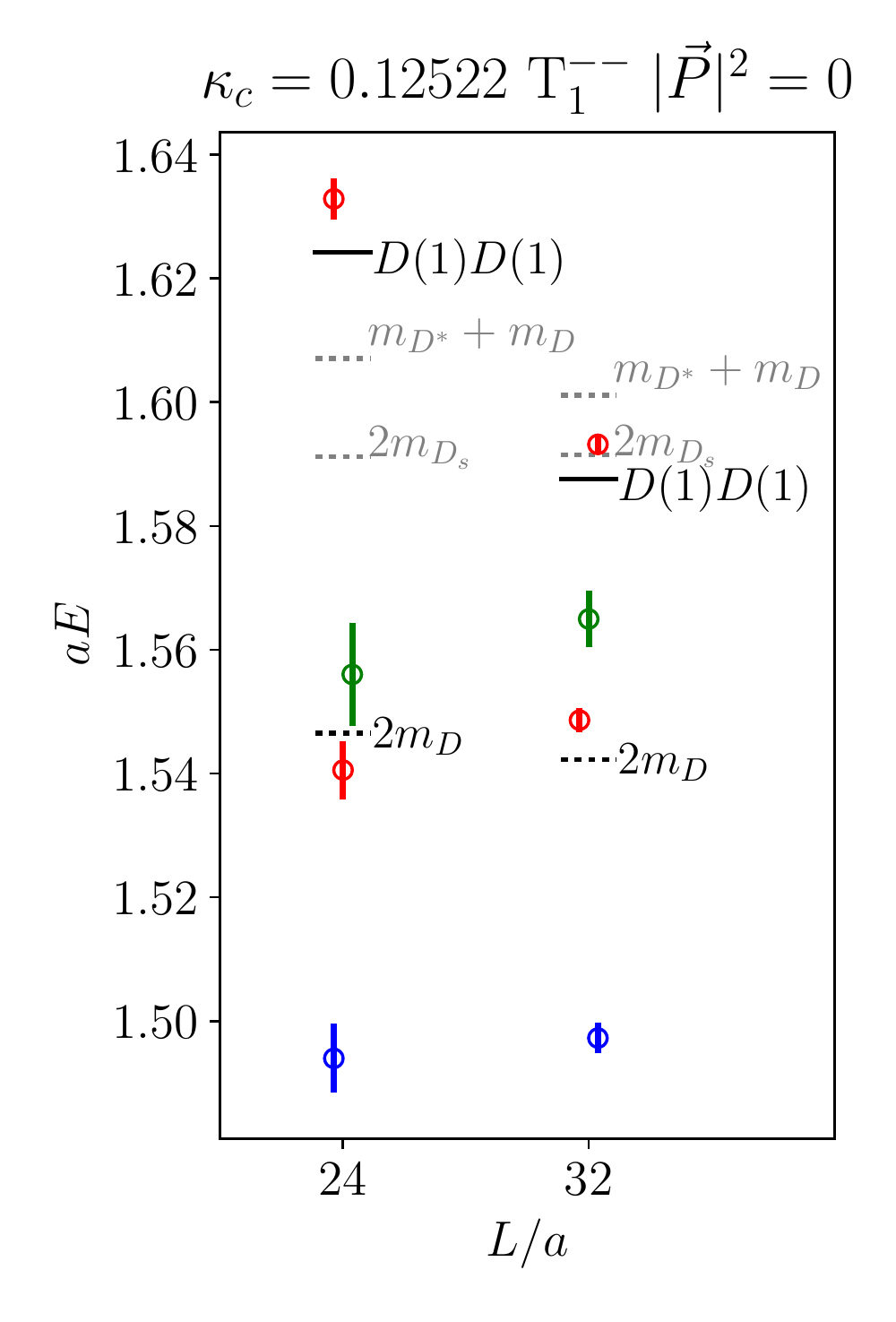}}
 \subfigure{\includegraphics[width=0.235\textwidth]{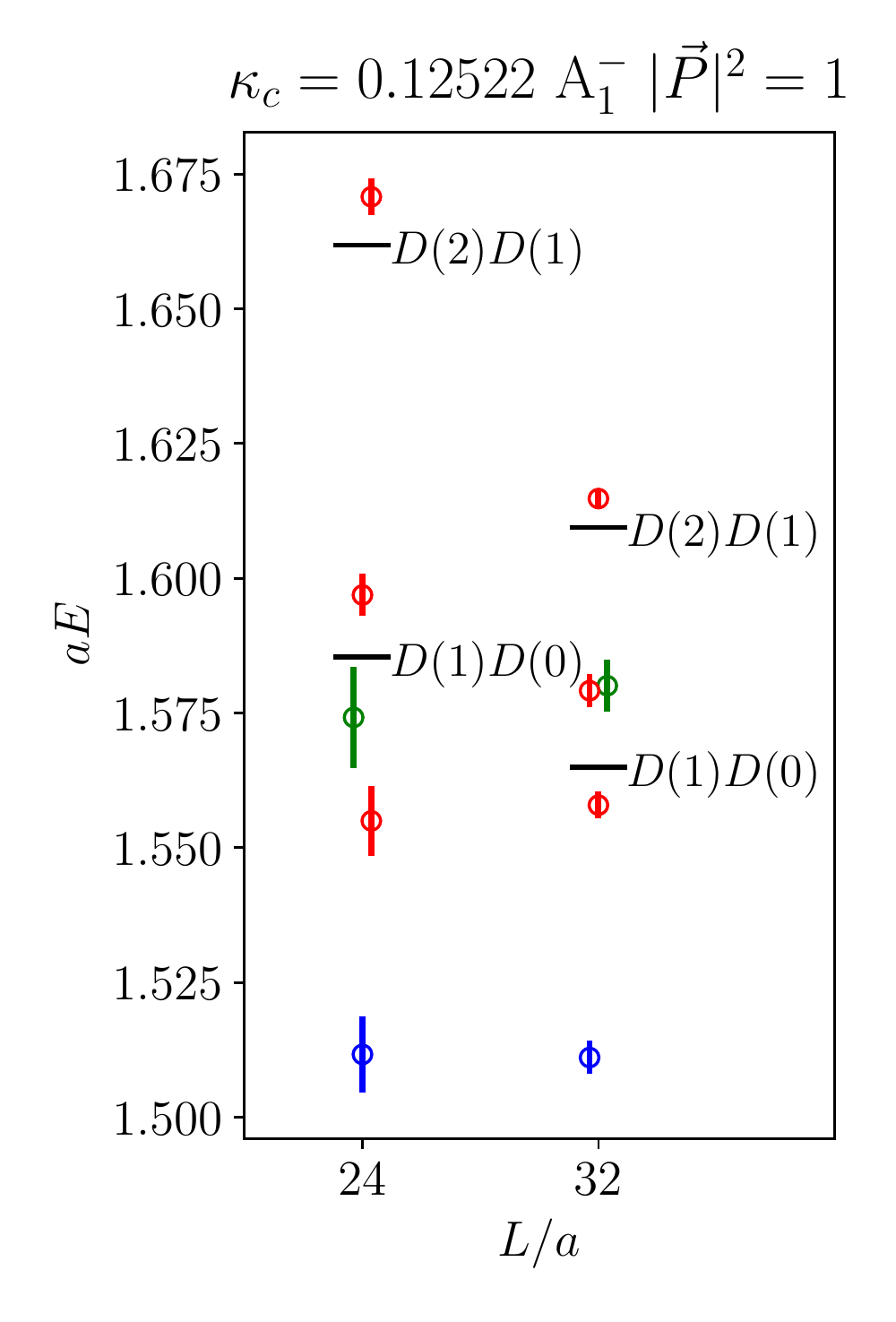}}
 \subfigure{\includegraphics[width=0.235\textwidth]{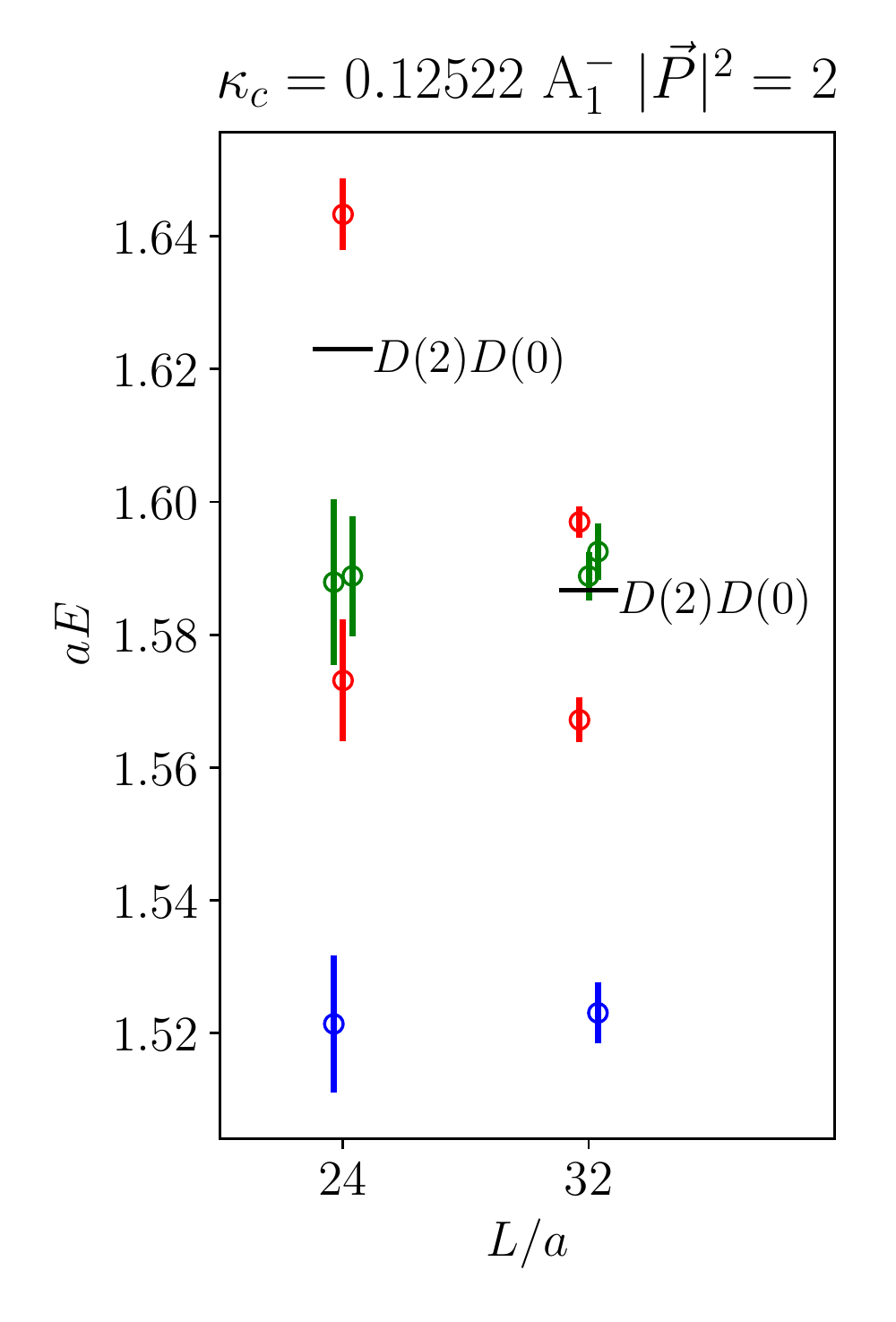}}
 \caption{Energy spectrum $E_n^{\textrm{\textrm{lat}}}$ in various lattice irreducible representations for $\kappa_c=0.12522$ for the U101 ($L = 24$) and H105 ($L = 32$) ensembles. The blue energy levels correspond to the naive $\psi(2S)$ energy levels, while the green points have been identified as having $J^{PC} = 3^{--}$. The red points correspond to the remaining energy levels, which arise from charmonia with $J^{PC} = 1^{--}$ and from $\bar{D}D$. Black solid lines show the energies of the non-interacting $\bar{D}(p_1^2)D(p_2^2)$ lattice energy levels, where $p_{1,2}^2 = 0$, 1 or 2, respectively in units of $(2\pi/L)^2$. The black dashed line in the rest frame indicates the $\bar{D}D$ threshold, while the grey dashed lines refer to the $D_s D_s$ and $\bar{D}D^*$ thresholds. }\label{energy_spectrum_kc0p12522}
\end{figure}

\begin{figure}
\centering
 \subfigure{\includegraphics[width=0.235\textwidth]{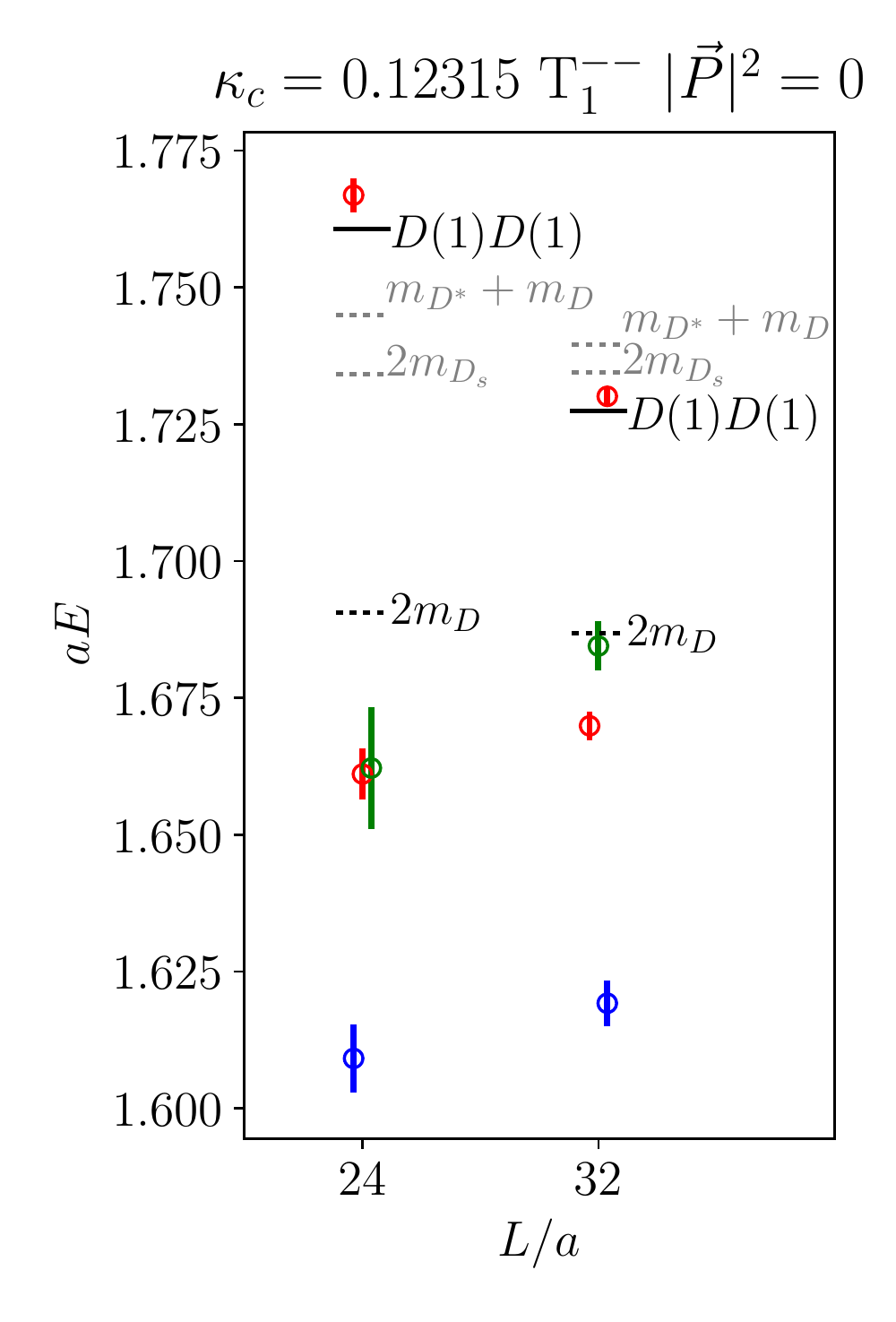}}
 \subfigure{\includegraphics[width=0.235\textwidth]{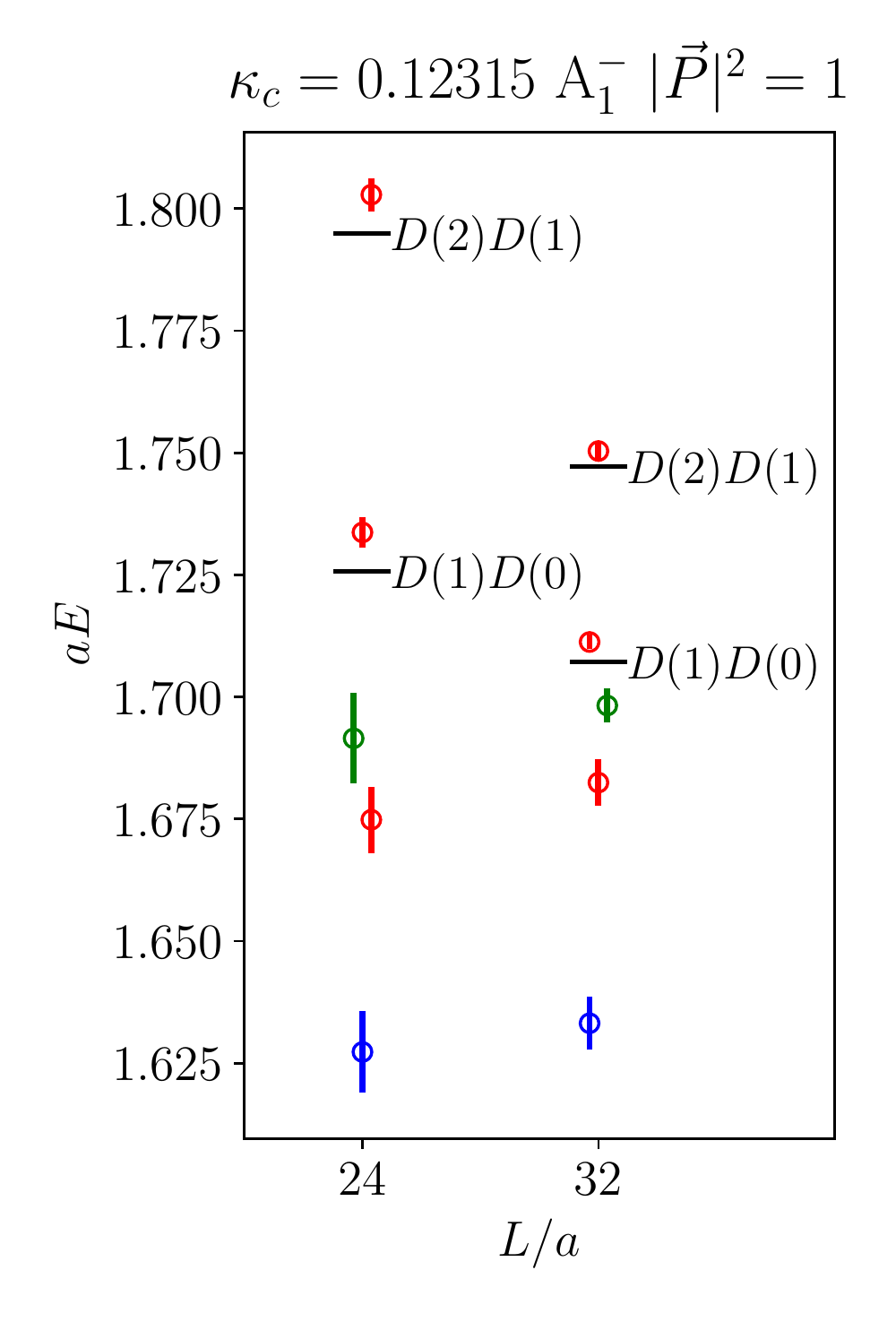}}
 \subfigure{\includegraphics[width=0.235\textwidth]{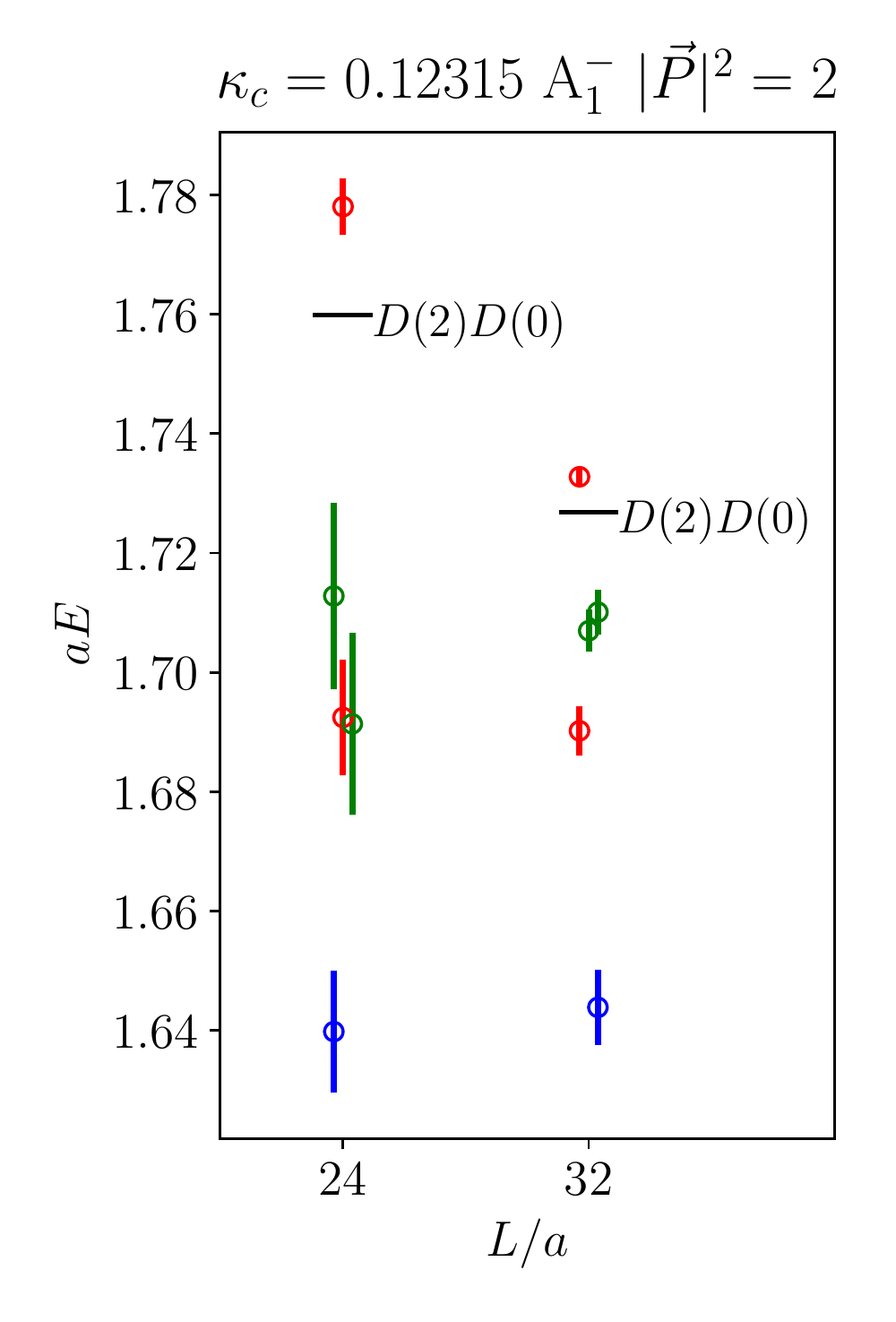}}
 \caption{Energy spectrum $E_n^{\textrm{\textrm{lat}}}$ in various lattice irreducible representations for $\kappa_c=0.12315$, with the same notations and color coding as in Fig.~\ref{energy_spectrum_kc0p12522}.}\label{energy_spectrum_kc0p12315}
\end{figure}

We employ the distillation method to compute our correlation functions \cite{Peardon:2009gh}, {\it i.e.} the propagation of the quarks is computed in the distillation space of the eigenvectors of the 3D Laplacian. As a consequence, we are able to separate the construction of the interpolators from the computation of the light and charm propagators without resorting to storing full propagators. Only at a later time, the quark-lines of the Feynman diagrams are contracted and summed to obtain the entries of the correlation matrix. We employ 90 and 150 eigenvectors of the lattice Laplacian for the U101 and the H105 gauge field ensembles, respectively. We use full distillation for U101. Full distillation is also used on H105 for the charm propagators and the $\phi$-matrices (defined in Eq.~(12) of Ref. \cite{Peardon:2009gh}). The cost of full distillation for the determination of light quark propagators on the H105 ensemble would have been prohibitively expensive, as $150\times 4$ inversions of the Dirac operators on a large volume would have been required for each flavor and each time-source/sink. Therefore, we use stochastic distillation \cite{Morningstar:2011ka} for the light quark propagators on the H105 ensemble only. The perambulator, defined as
\begin{equation}\label{perambulator}
 \tau(t',\mu^\prime,i^\prime;t,\mu,i) = \langle v_{i^\prime}^{\mu^\prime}(t^\prime)| (D^{-1}|v_i^{\mu}(t)\rangle)\,,
\end{equation}
requires to compute the inverse of the Dirac matrix $D$ on a source $|v_i^{\mu}(t)\rangle$ which is zero everywhere except for the timeslice $t'$ and spin index $\nu$, where it is equal to the Laplacian eigenvector $j$. The number of inversions required can be reduced by considering random sources. We introduce $N_s$ complex $Z_2$ stochastic noise vectors 
\begin{equation}
 k_n(t,i,\mu) = \frac{1}{\sqrt{2}}(\pm 1 \pm i)\,,
\end{equation}
using full-time dilution. The stochastic source given to the multigrid solver is constructed for every time-slice $t$ as
\begin{equation}
 |w_n(t)\rangle = \sum_{j,\nu} k_n(t,j,\nu) |v_j^\nu(t)\rangle\,.
\end{equation}
The perambulator can be approximated as
\begin{equation}\label{tau_stoc}
 \tau( t',\mu^\prime,i^\prime;t,\mu,i) = \frac{1}{N_s} \sum_n k_n^\dag(t,i,\mu) \langle v_{i^\prime}^{\mu^\prime}(t^\prime)| (D^{-1}|w_n(t)\rangle)\,,
\end{equation}
which is equal to the exact expression of Eq.~(\ref{perambulator}) in the limit $N_s \rightarrow \infty$, since
\begin{equation}
 \frac{1}{N_s}\sum_{n=1}^{N_s}   k_n^\dag(t,i,\mu) k_n(t,j,\nu)\simeq \delta_{ij}\delta_{\mu\nu}\,.
\end{equation}
Thanks to the use of full-time dilution in our stochastic distillation scheme, there is no bias for contributions to the correlation matrix involving light annihilation diagrams. However, we need to consider two independent sets of random vectors for diagrams involving light quarks propagating at different timeslices. We therefore use two sets of 20 noise vectors, and we average over the two only for light annihilation diagrams. Following this approach, we are able to reduce the number of the expensive inversions of the Dirac operator for the light quarks by a factor ten on our largest volume. At the same time, we retain the advantages of full-dilution for the purely charm contributions and employ full distillation to compute the correlation functions between $\bar cc$-type operators, that need to be estimated precisely for the study of charmonium excited states we are interested in.

\section{Energy spectrum}\label{sec:spec}

\subsection{Determination of the finite volume spectrum}\label{CEOsec}

Many excited energy levels need to be accurately determined on several different volumes and/or in different momentum frames in order to extract the scattering matrix according to L\"uscher's method. The excited energy spectrum $E_n^{\textrm{\textrm{lat}}}$ is extracted from Euclidean time two-point correlation functions 
\begin{equation}
 C_{ij} (t) = \langle O_i(t) O_j^{\dagger}(0) \rangle = \sum_n Z_i^n Z_j^{n\ast} e^{- t E_n^{\textrm{\textrm{lat}}}},
\label{eq:ecmat}
\end{equation}
constructed from a basis of interpolators \{$O_i$\} with the desired quantum numbers. 
Here $Z_i^n=\langle 0| O_i|n\rangle$ refers to the operator state overlap.

Our main purpose is to investigate $\bar{D}D$-meson scattering in partial wave $l=1$. We determine the charmonium spectrum in three lattice irreducible representations that contain $l=1$~(they all also contain partial wave $l=3$):
\begin{align}
\mbox{i)  }  & \Lambda^{PC}=T_1^{--} \mbox{ in the rest frame } |\vec P|^2=0\,, \nonumber \\ 
\mbox{ii)  } & \Lambda^{C}=A_1^{-} \mbox{ in the moving frame } |\vec P|^2=1 \mbox{ and} \nonumber \\
\mbox{iii)  }& \Lambda^{C}=A_1^{-} \mbox{ in the moving frame } |\vec P|^2=2\,, \nonumber 
\end{align}
where each irreducible representation corresponds to a different
inertial frame, with total momentum $\vec{P}$. The values of $|\vec P|^2$ are given in units of
$(2\pi/L)^2$. Note that charge conjugation $C$ is a good quantum
number in all frames, whereas parity $P$ is a good quantum number only
in the rest frame. In order to reliably extract the relevant low
energy spectrum, a large basis of interpolators \{$O_i$\} that
includes single-meson as well as two-meson operators in all the irreps
is required.

For the single-meson interpolators, we follow the procedure proposed in Refs.~\cite{Dudek:2010wm,Thomas:2011rh} and construct all interpolators with up to two gauge covariant derivatives. As a first step in the analysis, the spectrum is extracted using a basis of these single-meson interpolators and, following the spin identification procedure discussed in Ref.~\cite{Padmanath:2018tuc}, the continuum quantum numbers $J^P$ of the levels in the energy region of interest are identified. In this way, we can understand the quantum numbers of the energy levels that are related to a naive $\bar cc$ description that enter into our phase shift analysis. 

The two-meson interpolators are built using the projection %to a given irreducible representation $\Lambda$.
\begin{align}
%O^{\bar D D}(\vec p_1+\vec p_2) &=& \sum_{R\in G(\vec p_1+\vec p_2)} \chi^*_{\Lambda}(R) \{\bar D(R \vec p_1)D(R \vec p_2) \nonumber \\ 
O^{\bar D D}(\vec p_1+\vec p_2) &=& \sum_{R\in G(\vec p_1+\vec p_2)} T^{\Lambda}_{r,r}(R)^* \{\bar D(R \vec p_1)D(R \vec p_2) \nonumber \\ 
  && - \bar D(R \vec p_2)D(R \vec p_1) \},
\label{projection}
\end{align}
where $\vec p_1$ and $\vec p_2$ are the momenta of the scattering particles such that $\vec P = \vec p_1+\vec p_2$, $R$ is a group element of the point symmetry group $G$ of the inertial frame with momentum $\vec P$, $T^{\Lambda}_{r,r}(R)$ is the representation for the group element $R$ in the irrep $\Lambda$ and $r$ is the row of the irrep $\Lambda$. As $\bar DD$ is the predominant decay mode of the $\psi(3770)$ resonance, we include $\bar DD$ interpolators for all momentum combinations within the relevant low energy region. In Table \ref{tab:twomesonops}, we show the two-meson interpolators considered in our calculation in each of the lattice irreps being analyzed. In this study, we neglect the $p$-wave decay $J/\psi\eta$ and in addition the final states including three or more hadrons, such as $J/\psi\pi^+\pi^-$ and $J/\psi\pi^0\pi^0$.

The inclusion of two-meson interpolators in the basis brings additional Wick contractions. We compute all relevant Wick contraction diagrams for the single- and two-meson operators (see Fig. 1 of Ref.~\cite{Lang:2015sba}). We exclude diagrams with disconnected charm quark contractions, that lead to the decay of charmonium into light hadrons. The correlation functions constructed from the single and two meson operators are averaged over the spin and momentum polarization to increase the statistical precision. We also bin our data in blocks of size equal to one, two and four to correctly address the autocorrelation of our configurations. For further details, see Appendix~\ref{app1}.

The extraction of the energy levels proceeds via a variational approach~\cite{Michael:1985ne,Luscher:1990ck,Blossier:2009kd} by solving a generalized eigenvalue problem~(GEVP)
\begin{eqnarray}
% C_{ij}(t) v_n^i(t) = \lambda_n(t) C_{ij}(t_0) v_n^i(t), ~~~ \lambda_n(t)|_{t\rightarrow\infty}\propto e^{-E_n^{\textrm{lat}} t}
 C_{ij}(t) v_n^i(t) &=& \lambda_n(t) C_{ij}(t_0) v_n^i(t), \nonumber \\
 \lambda_n(t)&\propto& e^{-t E_n^{\textrm{\textrm{lat}}} }(1+\mathcal{O}(e^{-\Delta E_n t})) \label{exp_lambda_decay}
\end{eqnarray}
for the correlation matrices in Eq.~(\ref{eq:ecmat}). We set $t_0 = 2$ in our analysis. The eigenvalues $\lambda_n(t)$ are fitted to a single or a double exponential function and the quality of the fits is compared to estimate the best fitting intervals. We remark that increasing $t_0$ leads to consistent results, however, the errors also increase. Corrections to these fit forms arise when using open boundary conditions in the temporal direction and the source or sink timeslice of the correlation function is close to one of the boundaries. However, all our measurements are made in the bulk of the lattice, 28 time slices away from either boundary, and no such corrections are observed. 

\begin{table}[ht]
\begin{tabular}{c|c|c|c}
  \hline
          &  $|\vec P|^2$ & $\Lambda^{(P)C}$ & $O_{\bar DD} = \bar D(p_1^2)D(p_2^2)$        \\\hline
  $O_h$   &  $0$   &  $T_1^{--}$      & $\bar D(1)D(1)$ \\\hline
  $Dic_4$ &  $1$   &  $A_1^{-}$       & $\bar D(1)D(0)$ \\
          &        &                  & $\bar D(2)D(1)$ \\\hline
  $Dic_2$ &  $2$   &  $A_1^{-}$       & $\bar D(2)D(0)$ \\
          &        &                  & $\bar D(3)D(1)$ \\\hline
\end{tabular}
\caption{List of two-meson interpolators $O_{\bar DD}$ used in each lattice irrep $\Lambda^{(P)C}$. For brevity we use the simplified notation $\bar D(p_1^2)D(p_2^2)$, which refers to two-meson interpolators, with good charge conjugation C, built from a $\bar D$ and $D$ meson operator with momentum $\vec{p}_1$ and $\vec{p}_2$, respectively in units of $2\pi/L$. The full expressions for these interpolators are much longer and are omitted for brevity.}\label{tab:twomesonops}
\end{table}

In Figs.~\ref{energy_spectrum_kc0p12522} and \ref{energy_spectrum_kc0p12315}, we present the finite volume excited energy spectrum (above the $J/\psi$) in all the three frames studied for $\kappa_c=0.12522$ and $\kappa_c=0.12315$, respectively.  The statistical uncertainty is smaller on the H105 ensemble, in part due to the larger number of configurations analyzed compared to U101. The color coding of the levels indicates the quantum numbers. Blue levels are related to the $\psi(2S)$ bound state and the green levels are identified to have quantum numbers $J^{PC}=3^{--}$.\footnote{This state appears twice for $\vec P^2=2$ according to \cite{Thomas:2011rh} and Table I of  \cite{Padmanath:2018tuc}: one level has helicity zero and the other  helicity two.}. The remaining levels in red arise from charmonia with $J^{PC}=1^{--}$ and from $\bar DD$ scattering levels shifted due to interaction. We extract up to 4 or 5 excited states for each lattice irrep considered. 

Within the energy range explored, we also observe a $2^{--}$ level in the $A_1^{-} (|\vec P|^2=2)$ spectrum. The helicity two components of the $2^{--}$ state appear in this irrep as can be seen from Table II of \cite{Thomas:2011rh} and Table I of \cite{Padmanath:2018tuc}. Taking the naive energy level for $\kappa_c=0.12522$ and using the continuum dispersion relation, we determine the mass of this state in its rest frame to be 3825(8) MeV, which is consistent with the lowest $2^{--}$ charmonium state observed by Belle \cite{Bhardwaj:2013rmw} and BESIII \cite{Ablikim:2015dlj}. The coupling of this excitation with the rest of the spectrum is expected to be negligible as, in a study of vector channel scattering of two equal mass pseudoscalar mesons, $J^P=2^-$ is an unnatural quantum number and $l=2$ is an irrelevant partial wave. This level is absent when we omit the relevant $2^{--}$ interpolators from the operator basis, while the rest of the spectrum remains unaffected. Hence we exclude this level from the amplitude analysis and also do not show it in the figures.

\begin{table*}
 \begin{tabular}{c|c||c|c|c|c|c}
   & & $aE_D(p^2=0)$ & $aE_D(p^2=1)$ & $aE_D(p^2=2)$ &  $aM_{\textrm{av}}$\\
  \hline
  $\kappa_c=0.12522, ~N_L=24$ & lat & 0.7732(9) & 0.8132(10) & 0.8516(11) & 1.2318(20) \\
   & cont & - & 0.8163(9) & 0.8573(8) & -\\
   \hline
  $\kappa_c=0.12522, ~N_L=32$ & lat & 0.7711(6) & 0.7942(5)  & 0.8166(6)  & 1.2348(13) \\
   & cont & - & 0.7957(9) & 0.8296(8) & -\\
   \hline
 $\kappa_c=0.12315, ~N_L=24$ & lat & 0.8453(13) & 0.8814(10) & 0.9164(65) & 1.3559(20) \\
   & cont & - & 0.8849(13) & 0.9228(12) & -\\
   \hline
 $\kappa_c=0.12315, ~N_L=32$ & lat & 0.8433(6)  & 0.8642(5)  & 0.8844(6)  & 1.3587(14) \\
   & cont & - & 0.8659(8) & 0.8878(8) & -\\
   \hline
 \end{tabular}
 \caption{Lattice energies of $D$ meson for different momenta $p^2$ compared to the continuum values from Eq.~(\ref{dispcont}). Above $p^2$ are given in units of $(2\pi/L)^2$. The spin-averaged charmonium mass of Eq.~(\ref{mav}) is also given.}\label{tab:disp_rel}
\end{table*}

The thresholds for the channels $D_s D_s$ and $\bar{D} D^*$, which are omitted in our simulations, are shown in Fig.~\ref{energy_spectrum_kc0p12522} and \ref{energy_spectrum_kc0p12315}. These thresholds are lower than some of the employed energy levels on the $N_L = 24$ ensemble, however we expect that the effect of the omitted channels is very small because they appear in $p$-wave and they  open at relatively high energy. Note that there are no close-by two-particle levels of the omitted channels because they appear in $p$-wave. The influence of the omitted channels on the results from the $N_L=32$ ensemble is expected to be negligible\footnote{We have verified that masses and couplings of the extracted charmonia from the $N_L=32$ ensemble alone are compatible with the ones based on combined fits of all volumes.}.

\subsection{Approach to dispersion relations and charm-quark discretization errors}\label{approach}

The lattice energy levels presented in Figs. \ref{energy_spectrum_kc0p12522} and \ref{energy_spectrum_kc0p12315} are inputs to the quantization condition given in Eq.~(\ref{eq:det}) (realized using the TwoHadronsInBox package of Ref.~\cite{Morningstar:2017spu}), which is based on continuum dispersion relations. However, lattice energy levels are subject to discretization effects and can deviate from continuum expectations. Such lattice artifacts are larger in magnitude for charm than for the light quarks $u,d$ and $s$, since, in dimensionless units, $1\sim am_c\gg am_{u/d,s}$ even on many currently available lattices. At finite lattice spacing $a$, there are non-negligible deviations of the dispersion relation $E^{\textrm{lat}}_D(p)$ of the $D$-meson from its continuum counterpart
\begin{equation}\label{dispcont}
 E^{\textrm{cont}}_D(\vec p)=\sqrt{m_D^2+|\vec p|^2}\,.
\end{equation}
As is evident from Table \ref{tab:disp_rel}, there are non-negligible deviations from $E^{\textrm{cont}}_D(\vec p)$ particularly for the smaller physical volume~(for which the lattice momentum is larger), and hence care is required when extracting the scattering matrix in the $\bar DD$ channel. In this section we describe our approach to this issue.

First, we determine the energy shift of each interacting eigenstate with total momentum $\vec P$ with respect to the nearest non-interacting state $\bar D(\vec p_1)D(\vec p_2)$  
\begin{equation}\label{delE}
(\Delta E)_s=(E^{\textrm{lat}})_s-(E^{\textrm{lat}}_{D(\vec p_1)})_s-(E^{\textrm{lat}}_{D(\vec p_2)})_s\,,
\end{equation}
where $\vec p_{1,2}=\vec n_{1,2} \tfrac{2\pi}{L}$ and $\vec p_1+\vec p_2=\vec P$.  All three energies on the r.h.s. are extracted from their corresponding eigenvalues of the GEVP, separately. Here, $(E^{\textrm{lat}})_s$ is the energy of the interacting $\bar DD$ system, presented in Section~\ref{CEOsec}, while $(E^{\textrm{lat}}_{D(\vec p_1)})_s$ is the  energy of a single $D$ meson with momentum $\vec p_1$ measured on the lattice. All are extracted on a given bootstrap or jackknife sample $s$. If the scattering matrix is equal to the identity, the phase shift is equal to zero and hence the energy shifts in Eq.~(\ref{delE}) should be equal to zero. Given that the quantization condition is based on the continuum dispersion relation,\footnote{The L\"uscher $\mathcal{Z}$-functions, that enter the box matrix $B$ in the quantization condition, have poles at energies $E=E^{\textrm{cont}}_{D(\vec p_1)}+E^{\textrm{cont}}_{D(\vec p_2)}$ with $\vec p_{1,2}=\vec n_{1,2} \tfrac{2\pi}{L}$ and $\vec p_1 + \vec p_2=\vec P$, while $\tan\delta$ in the L\"uscher-type relations is a sum of terms with $\mathcal{Z}$-functions in the denominator. Therefore $\delta=0$ at those energies $E$. } we ensure this important constraint is satisfied by using the energies  
\begin{equation}\label{Ecalc}
(E^{\textrm{calc}})_s=(\Delta E^{\textrm{lat}})_s+\left(E^{\textrm{cont}}_{D(\vec p_1)}\right)_s+\left(E^{\textrm{cont}}_{D(\vec p_2)}\right)_s
\end{equation}
as input to Eq.~(\ref{eq:det}). The energies $(E^{\textrm{cont}}_{D(\vec p_1)})_s$ and $(E^{\textrm{cont}}_{D(\vec p_2)})_s$ are computed from Eq.~(\ref{dispcont}) using the lattice momenta $\vec p_{1,2}$ and the $D$-meson mass at rest~(also determined from the bootstrap or jackknife sample $s$). Note that, the energies $E^{\textrm{calc}}$ are equal to $E^{\textrm{lat}}$ in the continuum limit $a\rightarrow0$ by construction (Eqs. (\ref{delE}) and (\ref{Ecalc})). Furthermore, the above $E^{\textrm{calc}}$ also clearly provides a zero phase-shift when $\Delta E$ in Eq.~(\ref{delE}) is zero, even if non-negligible heavy-quark discretization effects are present in our simulation. 

\section{Results for charmonia with $J^{PC}=1^{--}$ and $3^{--}$}\label{sec:res}

\begin{figure}
  \subfigure[~$\kappa_c = 0.12522$]{\includegraphics[width=.47\textwidth]{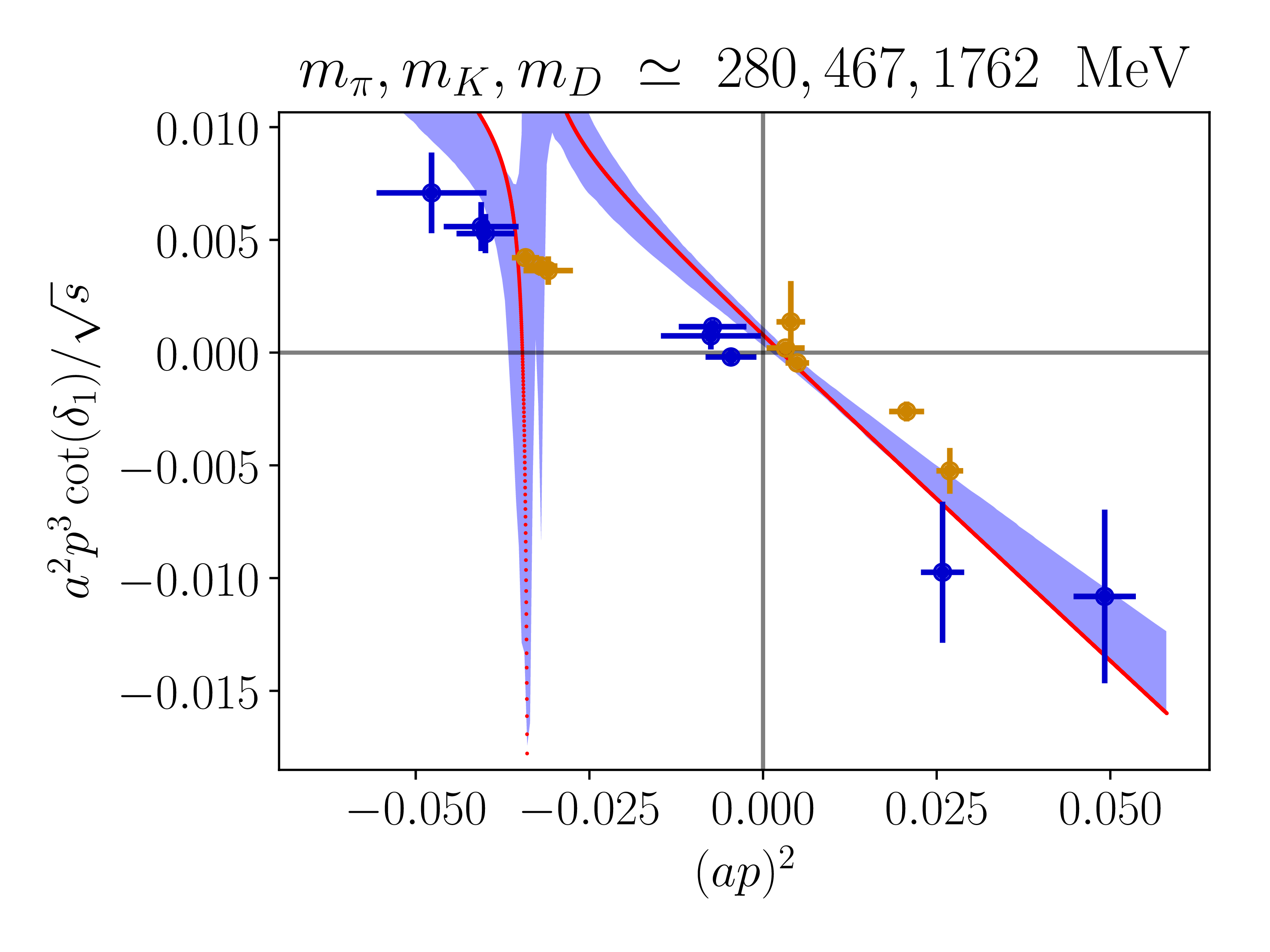}\label{double_pole_fit_12522_vector_channel}}
  \subfigure[~$\kappa_c = 0.12315$]{\includegraphics[width=.47\textwidth]{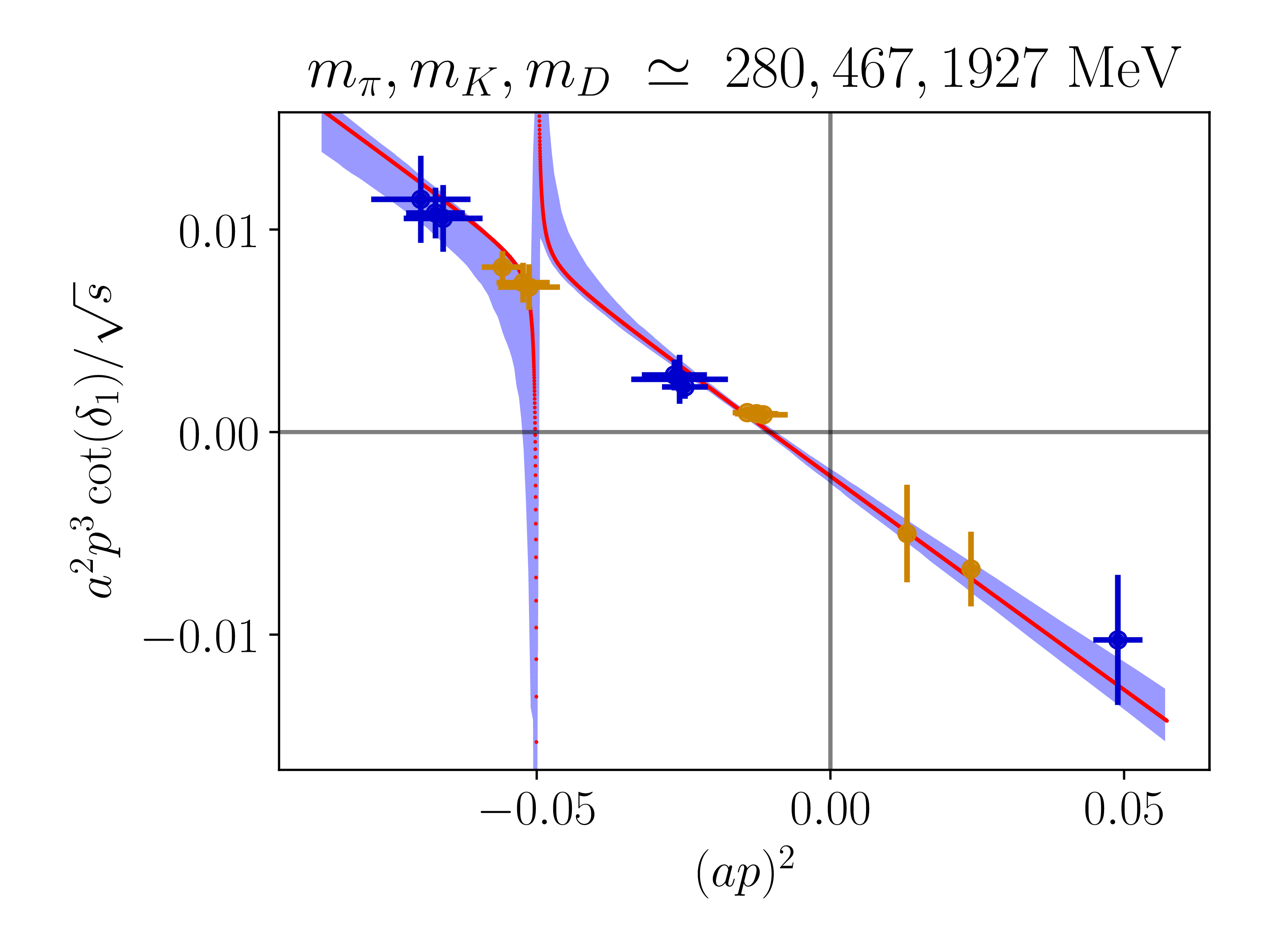}\label{double_pole_fit_12315_vector_channel}}
  \caption{$p^3\cot{\delta}/\sqrt{s}$ in the vector channel for the two different $\kappa_c$. The red curve represents a ``double pole'' fit. Only data points obtained from energy levels arising from charmonia with  $J^{PC} = 1^{--}$ and from $\bar{D} D$ are shown in the figures. Furthermore, we omit results for $p^3 \cot(\delta)/\sqrt{s}$ with errors larger than 0.004 for clarity. The $V=32^3$~($V=24^3$) data are indicated by orange~(blue) points. Finally, the violet error band is determined from the central 68 \% of the bootstrap distribution of the fitted curve. The distribution is in some cases asymmetric and not centered on the red curve.}\label{double_pole_fits_vector}
\end{figure}

The ground state in the vector channel is the $J/\psi$, while the first excited state, the $\psi(2S)$, lies below the open charm decay threshold. Beyond these two states, there is a relatively narrow resonance, the $\psi(3770)$ with $\Gamma^{\textrm{exp}}\approx 27$ MeV, and a broader resonance, the $\psi(4040)$, both which can decay to $\bar{D}D$ in $p$-wave. 

The near-threshold charmonium resonance $\psi(3770)$ is the main target in this investigation. In particular, we aim to extract the pole position of the $\psi(3770)$ and its coupling to $\bar{D}D$, which is related to the $\psi(3770)$ decay width. The study of the vector channel provides a benchmark for our simulations enabling us to assess systematic uncertainties which will also arise when considering more complex cases where several decay channels are involved~(and for which the experimental situation is unclear). We assume that the $\psi(3770)$ resonance is only coupled with the $\bar{D}D$ scattering channel and hence its parameters can be completely determined from the elastic
$\bar{D}D$ scattering amplitude. Experimentally, the $\psi(3770)$ decays into $\bar{D}D$ with a branching ratio of $93 ~_{+8}^{-9}~ \%$ \cite{Tanabashi:2018oca}. In this work, we neglect the effects from other hadronic decay modes, such as $J/\psi\eta$, $J/\psi\pi^0\pi^0$ and $J/\psi\pi^+\pi^-$, which collectively have a branching ratio below 0.2\%. We also neglect decays into light hadrons through charm-annihilation.

Our second aim is to study the lowest $3^{--}$ charmonium resonance. A candidate for this state referred to as the $X(3842)$\footnote{The LHCb collaboration has not identified the quantum numbers of this state but assumes $J^{PC}=3^{--}$ as the mass and width of the $X(3842)$ fit model expectations.} was recently discovered by LHCb in $\bar{D}D$ decay and has width $\Gamma^{\textrm{exp}}\approx 2.8$ MeV \cite{Aaij:2019evc}.

\subsection{Fits of the phase shift}\label{sec:fits}

\begin{figure}
  \subfigure[~$\kappa_c = 0.12522$]{\includegraphics[width=.47\textwidth]{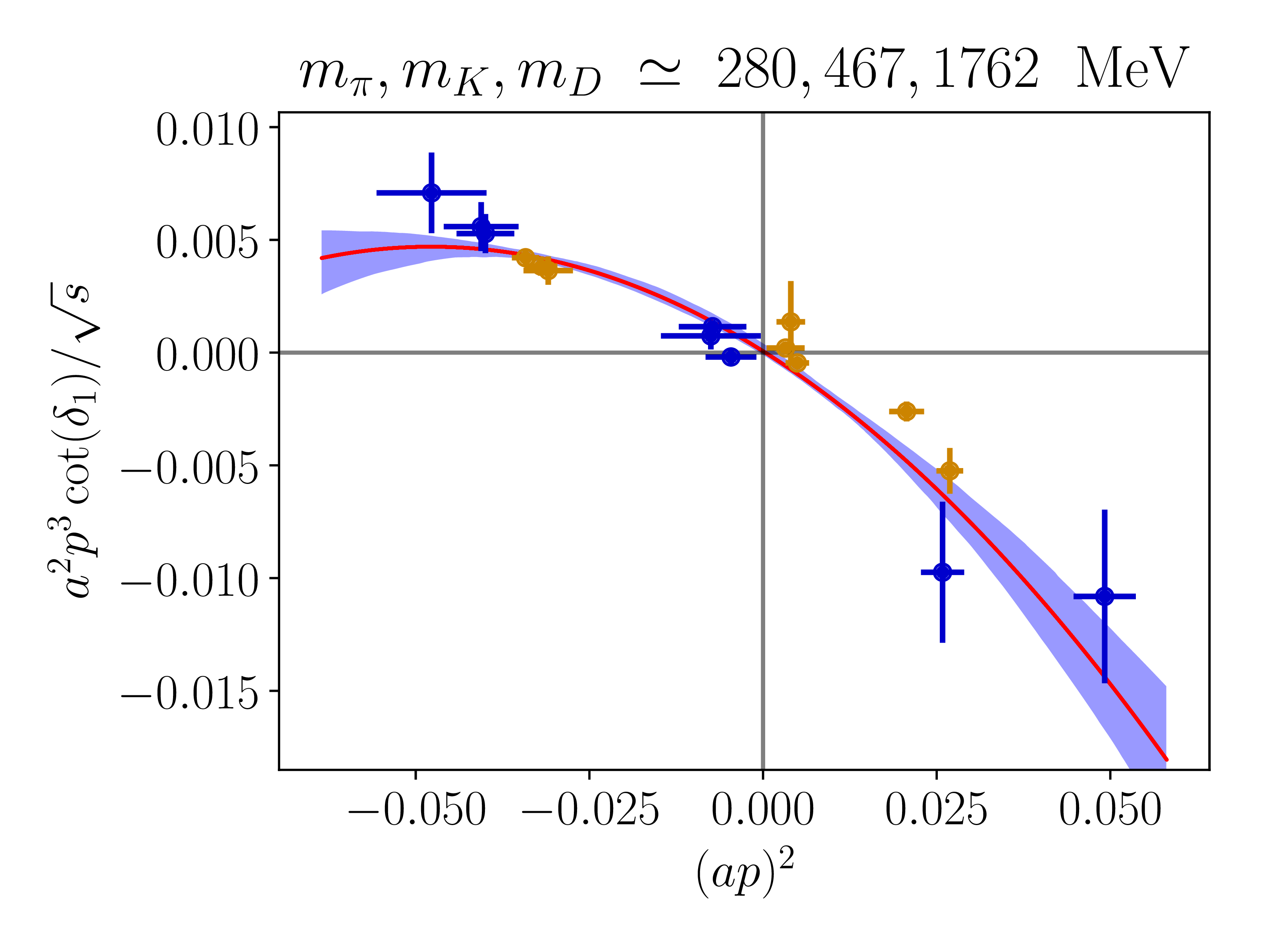}\label{quadratic_fit_12522_vector_channel}}
  \subfigure[~$\kappa_c = 0.12315$]{\includegraphics[width=.47\textwidth]{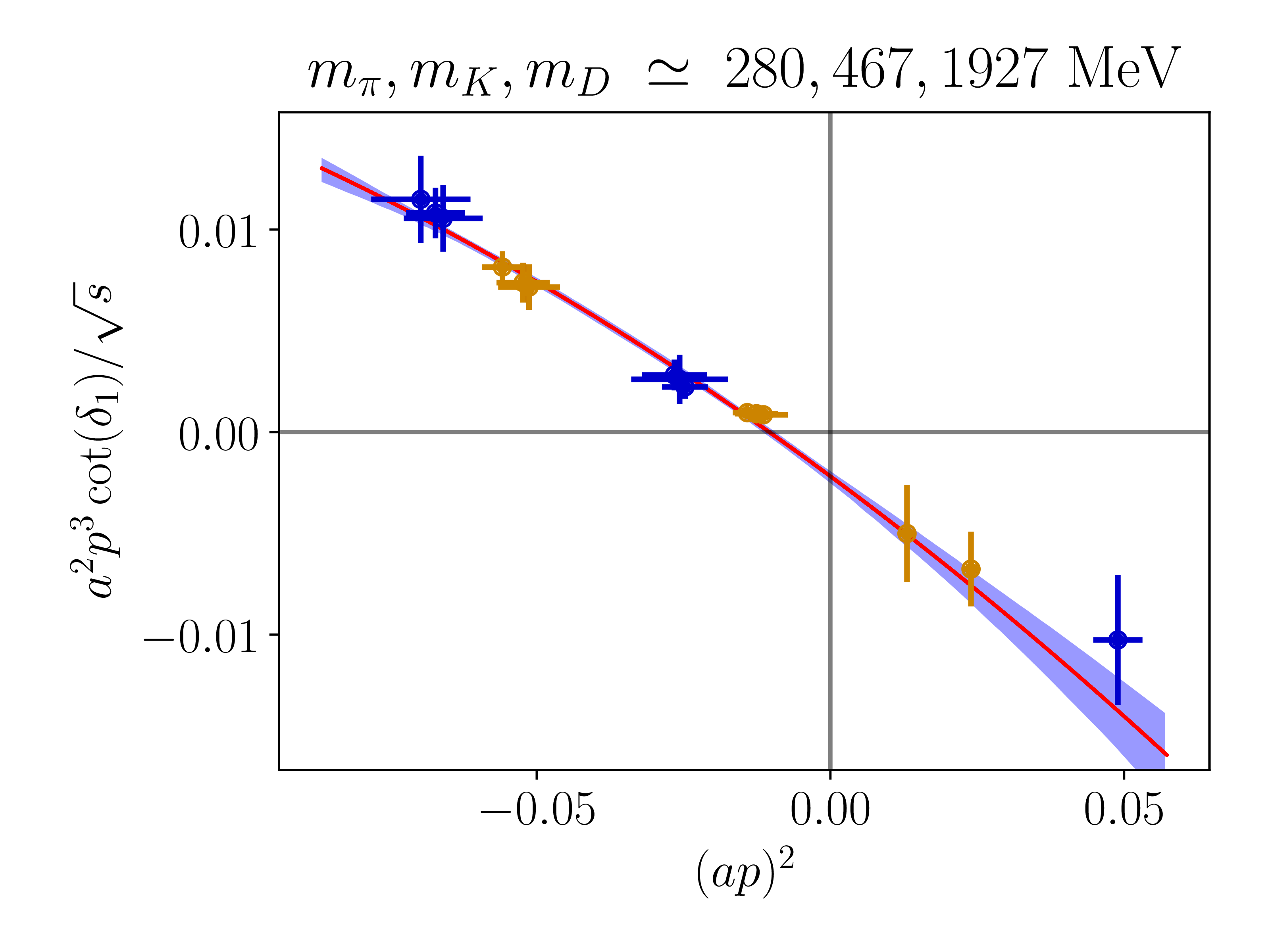}\label{quadratic_fit_12315_vector_channel}}
  \caption{$p^3\cot{\delta}/\sqrt{s}$ in the vector channel for the two different $\kappa_c$. The red curve represents a ``quadratic'' fit to the data. Color coding and conventions are as in Fig. \ref{double_pole_fits_vector}.}
\end{figure}

We perform a combined fit of the energy levels calculated on the U101 and H105 ensembles using bootstrapping for the determination of the uncertainties~(see Appendix~\ref{app1} for details). The ground state level in each frame, corresponding to $J/\psi$, is excluded from our fits. We include the energy level corresponding to the $\psi(2S)$ to use the maximal information from our spectrum and to correctly describe the spectrum in the vicinity of the threshold. The fitting forms we have explored for the vector channel are the ``double-pole''
\begin{equation}
 \frac{p^3 \cot(\delta_1)}{\sqrt{s}} = \left(\frac{G_1^2}{m_1^2 - s} + \frac{G_2^2}{m_2^2 - s}\right)^{-1}\,,\label{eq:doublefitform}
\end{equation}
inspired by two states in the region of interest, the $\psi(2S)$ and the $\psi(3770)$\footnote{Here $\cot(\delta_1) = 0$ at $s = m_1^2$ and $s=m_2^2$. Between these two energies, $\cot(\delta_1)$ has a pole and $T$ has a zero at $s=\frac{G_2^2 m_1^2 + G_1^2 m_2^2}{G_1^2+G_2^2}$. The presence of a pole in $\cot(\delta_1)$ allows to fulfill easily the consistency check in Sec.~\ref{sec:consistency_check} and the requirement that the parametrization should not predict unobserved energy levels.};
and the ``quadratic'' form
\begin{equation}
 \frac{p^3 \cot(\delta_1)}{\sqrt{s}} = A + B s + C s^2\,,
\end{equation}
where $p^2 = s/4 - m_D^2$. Other parameterizations were investigated. A linear form did not describe the data well; a triple-pole form would require additional data at large momenta to enable a third pole to be resolved; a double pole form with an additional constant is described in Appendix \ref{app3}.

Although many  partial wave amplitudes in principle contribute to the finite volume spectrum in any given irrep starting at the $\bar{D}D$ threshold, the effect of higher partial waves should be small, unless there is a resonance contributing to a higher partial wave in the energy region of interest. The spin-identified finite volume spectrum for charmonium from our previous investigation \cite{Padmanath:2018tuc} indicates a low lying $J^{PC}=3^{--}$ resonance, that  contributes to the $l=3$ partial wave. In order to investigate the effects of such a low-lying $l=3$ partial wave excitation on the rest of the discrete spectrum, we parametrize it with a simple 
Breit-Wigner form given by
\begin{equation}
 \frac{p^7 \cot(\delta_3)}{\sqrt{s}} = \frac{m_3^2 - s}{g_3^2}\,.\label{del3BW}
\end{equation}

\begin{widetext}

Our fits include all the fourteen excited energy levels we have been able to determine on the H105 and the U101 ensembles. For both $\kappa_c$, the fits are more constrained by the data from the H105 ensemble. The ``double pole'' fit is presented for the $1^{--}$ channel in Fig.~\ref{double_pole_fit_12522_vector_channel} for $\kappa_c=0.12522$ and in Fig.~\ref{double_pole_fit_12315_vector_channel} for $\kappa_c=0.12315$. For this parametrization~(and Eq.~(\ref{del3BW})), we have been able to fit all our data with
\begin{equation}\label{T1MM_3MM_double_pole_kc12522}
 \frac{(ap)^{2l+1} \cot(\delta)}{\sqrt{a^2 s}} = \left\{
\begin{array}{ll}
      \left(\frac{[0.63(33)]^2}{[1.4966(30)]^2 -a^2 s} + \frac{[3.69(37)]^2}{[1.5457(32)]^2 - a^2 s}\right)^{-1} & ~~~l = 1\\
      \frac{[1.568(11)]^2 - a^2 s}{[0.07(3)]^2} & ~~~l = 3 \\
\end{array} 
\right.%\,,
\end{equation}
for $\kappa_c = 0.12522$ with a $\chi^2/\textrm{d.o.f.}=0.5$, while we have 
\begin{equation}\label{T1MM_3MM_double_pole_kc12315}
 \frac{(ap)^{2l+1} \cot(\delta)}{\sqrt{a^2 s}} = \left\{
\begin{array}{ll}
      \left(\frac{[0.40(50)]^2}{[1.63(1)]^2 -a^2 s} + \frac{[4.15(13)]^2}{[1.6745(17)]^2 - a^2 s}\right)^{-1} & ~~~l = 1\\
      \frac{[1.6883(85)]^2 - a^2 s}{[0.025(17)]^2} & ~~~ l = 3 \\
\end{array} 
\right.\,,
\end{equation}
for $\kappa_c = 0.12315$ with a $\chi^2/\textrm{d.o.f.}=0.3$.

The quadratic fit is presented for the $1^{--}$ channel in Fig.~\ref{quadratic_fit_12522_vector_channel} for $\kappa_c=0.12522$ and in Fig.~\ref{quadratic_fit_12315_vector_channel} for $\kappa_c=0.12315$. Compared to the double pole fit form, the minimum of the correlated $\chi^2$ is more unstable across different bootstrapping resamples, resulting in an asymmetric distribution of the fitted parameters with long tails. We therefore quote asymmetric uncertainties for this parametrization. The fitted quadratic functional form for $\kappa_c=0.12522$ is equal to
\begin{equation}\label{T1MM_3MM_quadratic_kc12522}
 \frac{(ap)^{2l+1} \cot(\delta)}{\sqrt{a^2 s}} = \left\{
\begin{array}{ll}
      {\scriptstyle -0.601({}^{+200}_{-240}) + 0.554({}^{+200}_{-170}) (a^2 s) - 0.1267({}^{+380}_{-440}) (a^2 s)^2} &~~~l = 1\\[\bigskipamount]
      \frac{[1.5639({}^{+83}_{-14})]^2 - a^2 s}{[0.058({}^{+94}_{-55})]^2} & ~~~l = 3 \\
\end{array} 
\right.%\,,
\end{equation}
with a $\chi^2/\textrm{d.o.f.} = 1.32$, while for $\kappa_c=0.12315$
\begin{equation}\label{T1MM_3MM_quadratic_kc12315}
 \frac{(ap)^{2l+1} \cot(\delta)}{\sqrt{a^2 s}} = \left\{
\begin{array}{ll}
      {\scriptstyle -0.08({}^{+11}_{-13}) + 0.107({}^{+100}_{-80}) (a^2 s) - 0.028({}^{+15}_{-17}) (a^2 s)^2} &~~~l = 1\\[\bigskipamount]
      \frac{[1.6869({}^{+33}_{-32})]^2 - a^2 s}{[0.042({}^{+53}_{-22})]^2} & ~~~l = 3 \\
\end{array} 
\right.%\,,
\end{equation}
with a $\chi^2/\textrm{d.o.f.} = 1.15$. 

%\end{strip}
\end{widetext}

\begin{figure}
\centering
 \subfigure[]{\includegraphics[width=0.155\textwidth]{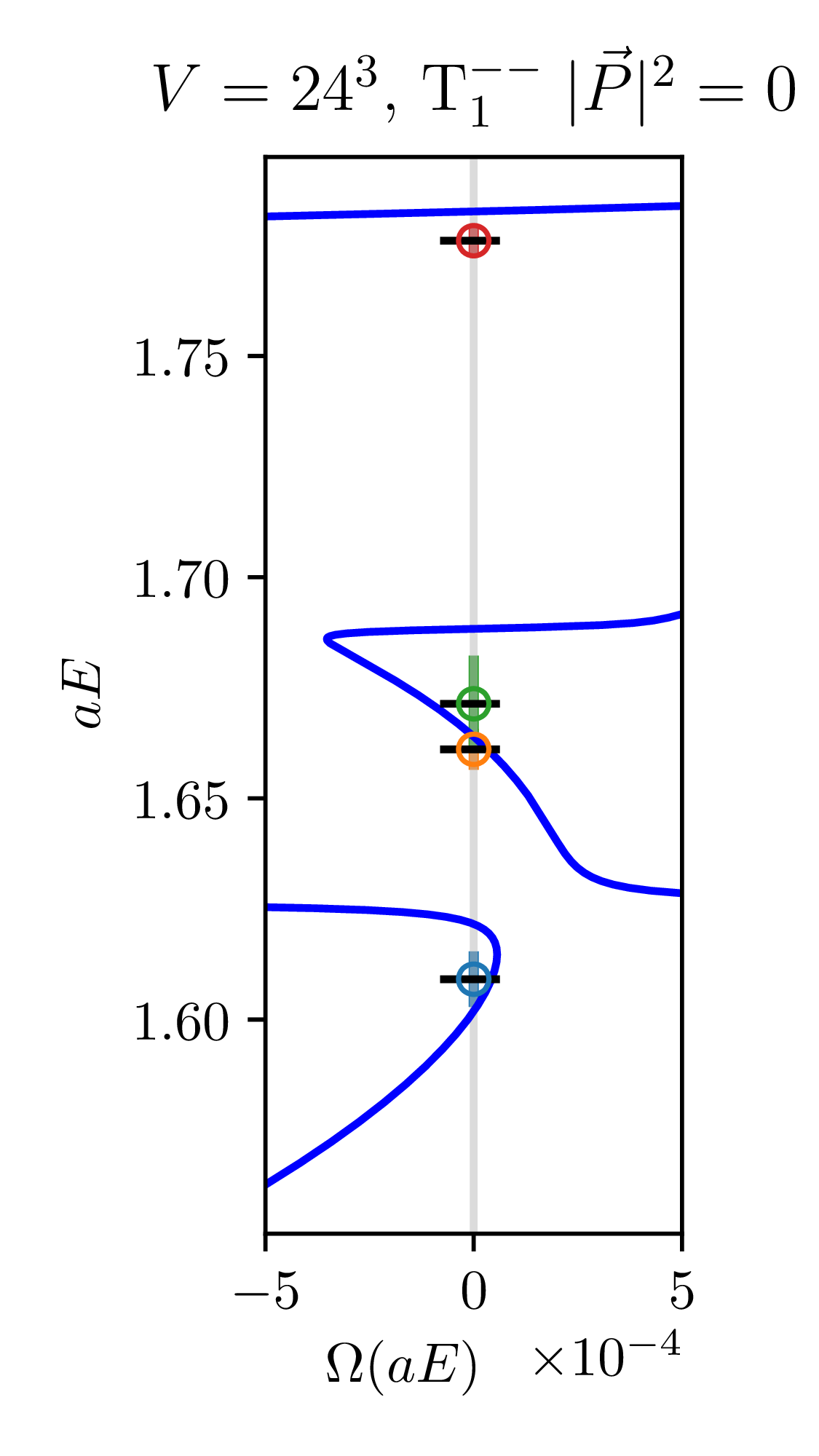}}
 \subfigure[]{\includegraphics[width=0.155\textwidth]{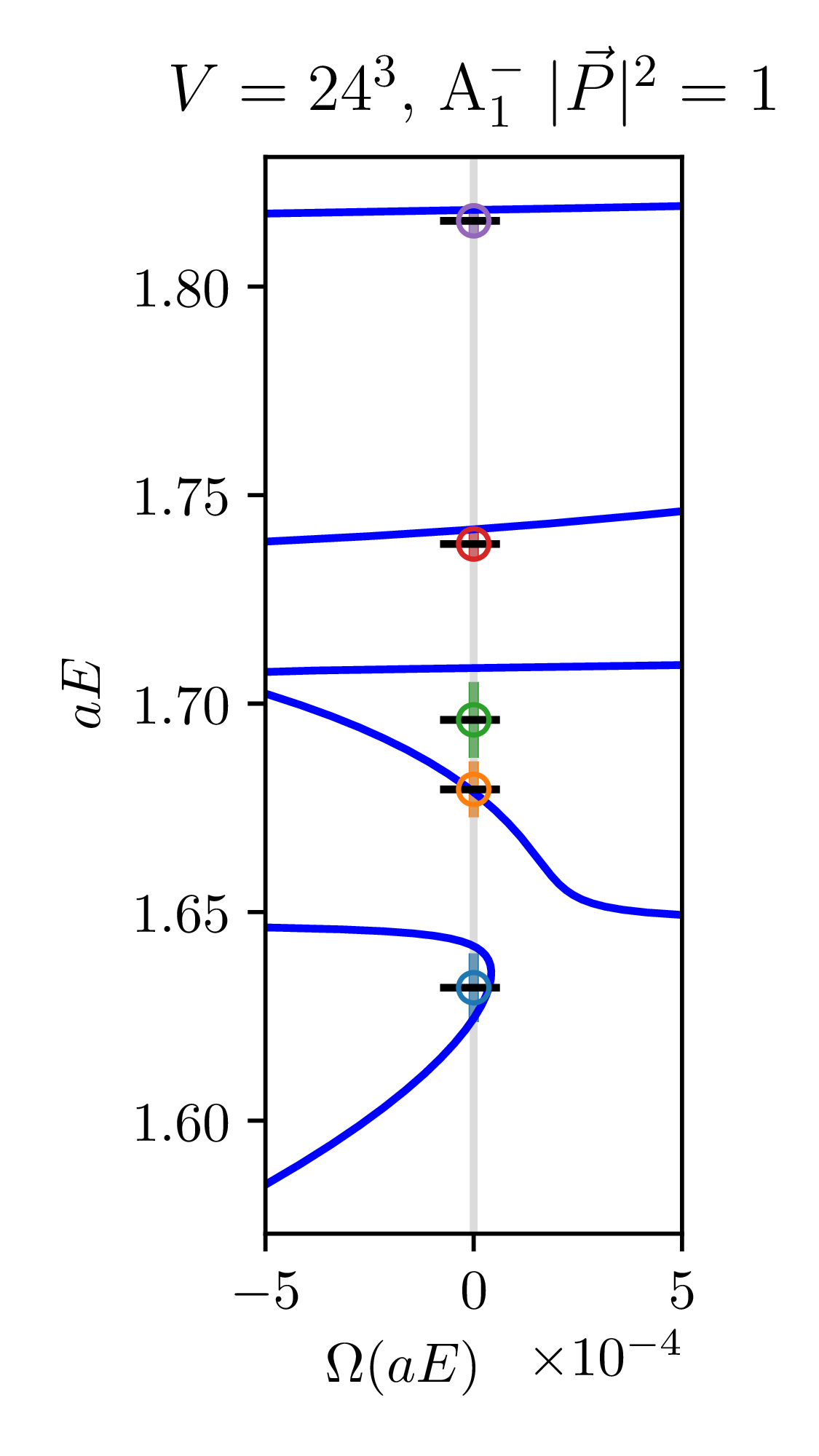}}
 \subfigure[]{\includegraphics[width=0.155\textwidth]{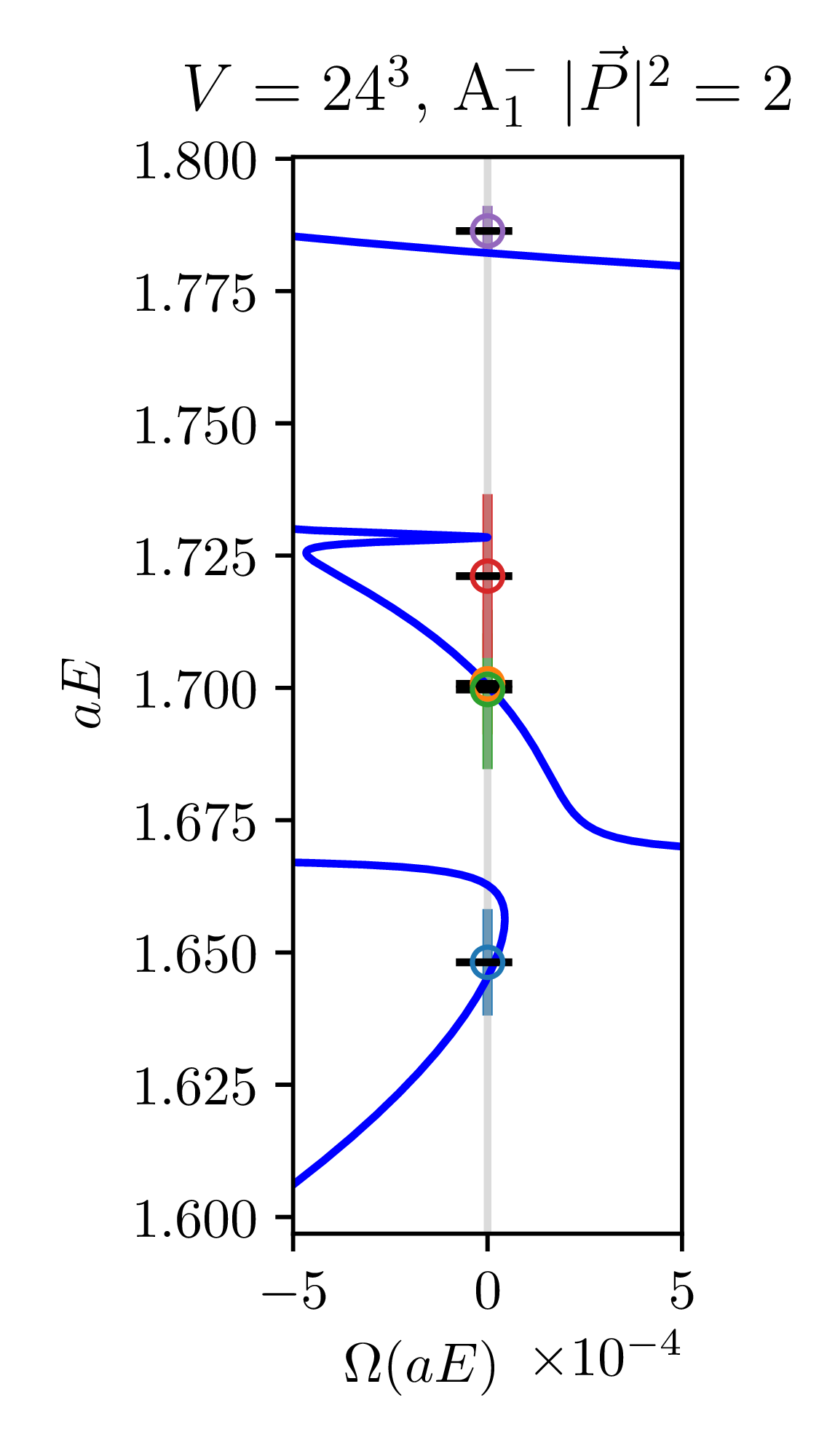}}
 \subfigure[]{\includegraphics[width=0.155\textwidth]{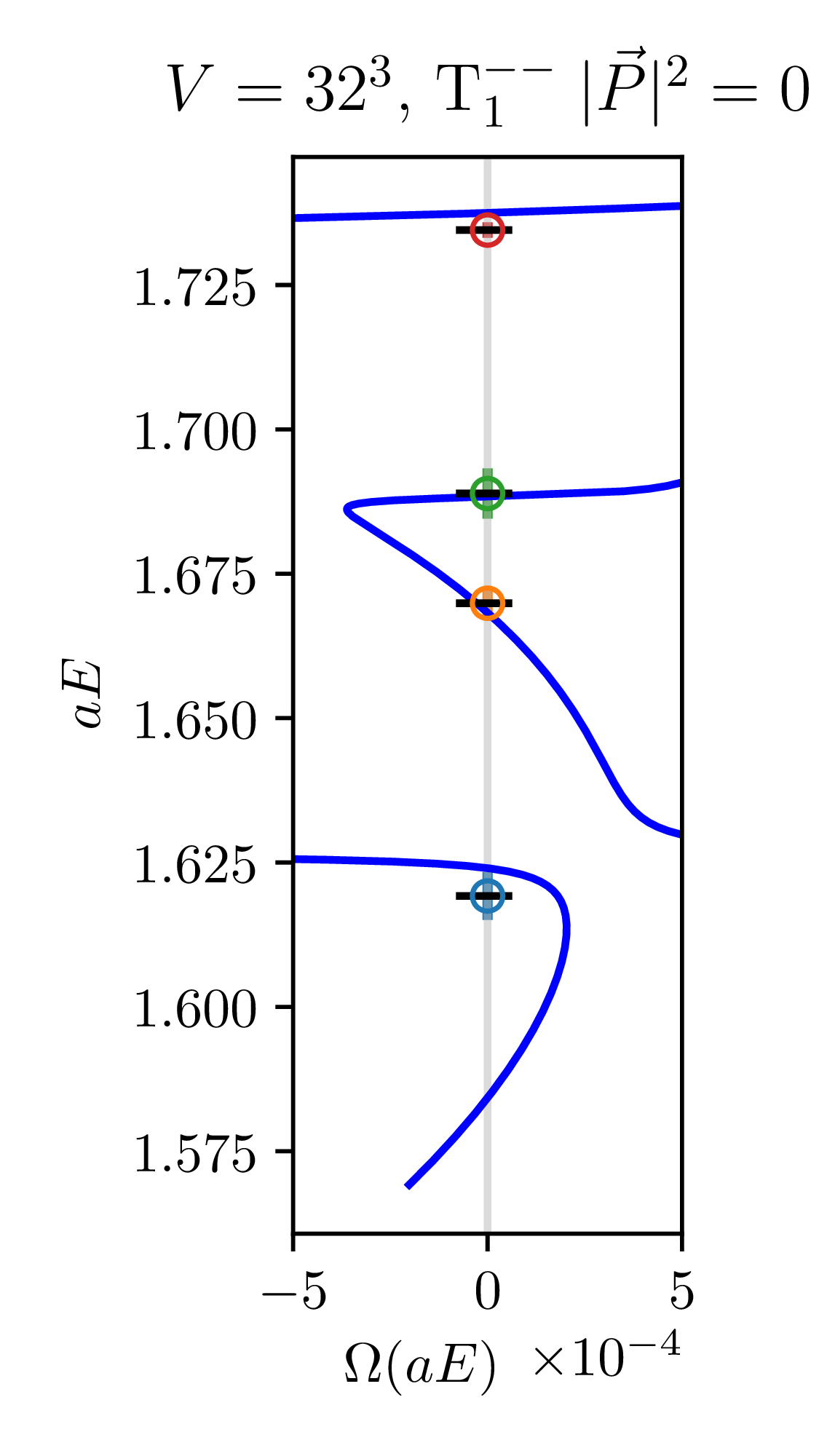}}
 \subfigure[]{\includegraphics[width=0.155\textwidth]{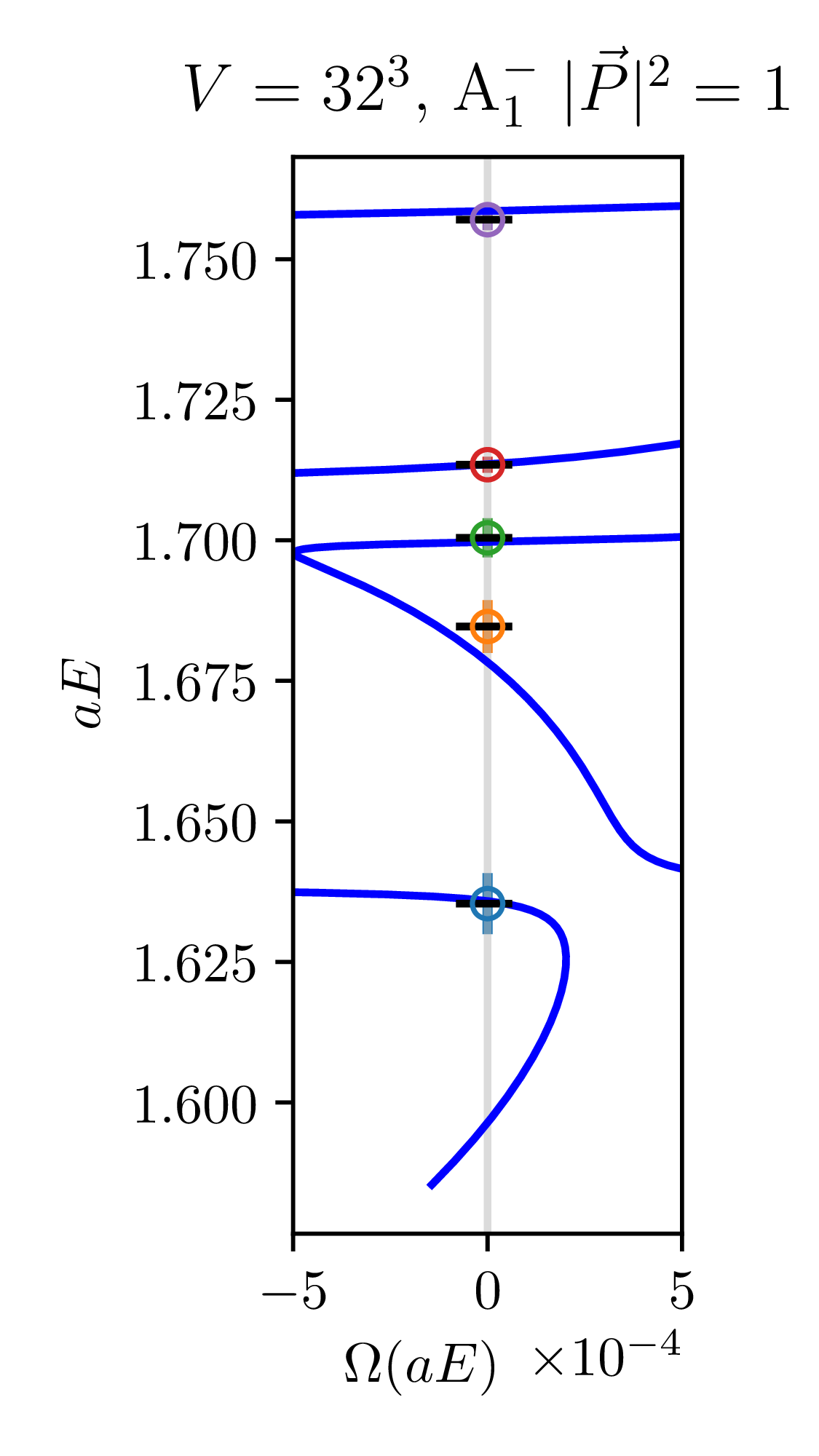}}
 \subfigure[]{\includegraphics[width=0.155\textwidth]{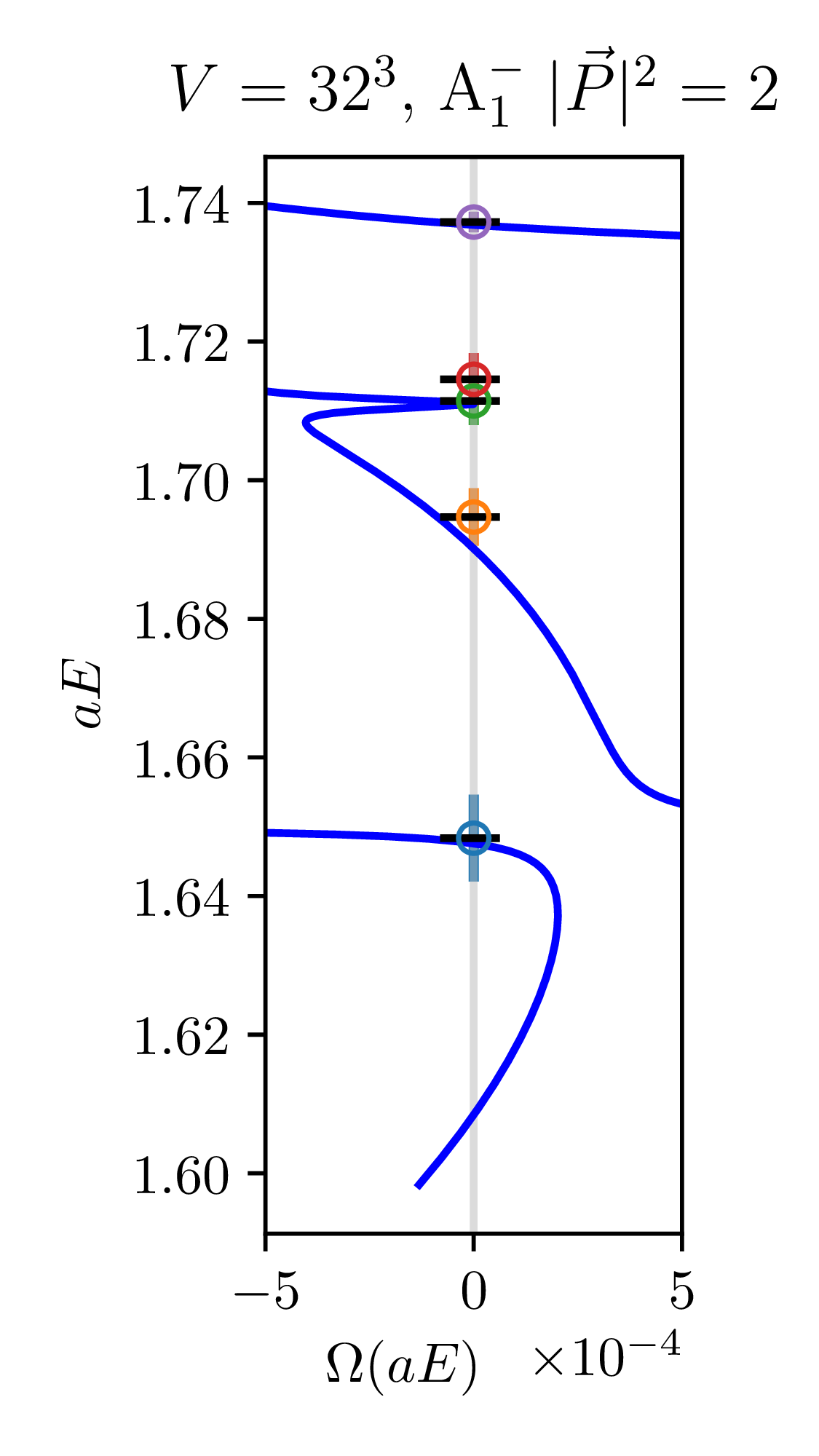}}
 \caption{$\Omega$-function in the three lattice irreducible representations in various frames for the vector channel resulting from the fit of all energy levels using the ``double pole'' fit ansatz for the phase shift for $\kappa_c = 0.12315$. The horizontal black lines help to identify the exact position of the energy levels. In particular, thicker lines are visible if two energy levels are close one another.}\label{Omega_T1MM_3MM_double_pole_kinv_kc0p12315}
\end{figure}

\begin{figure}
\centering
 \subfigure[]{\includegraphics[width=0.155\textwidth]{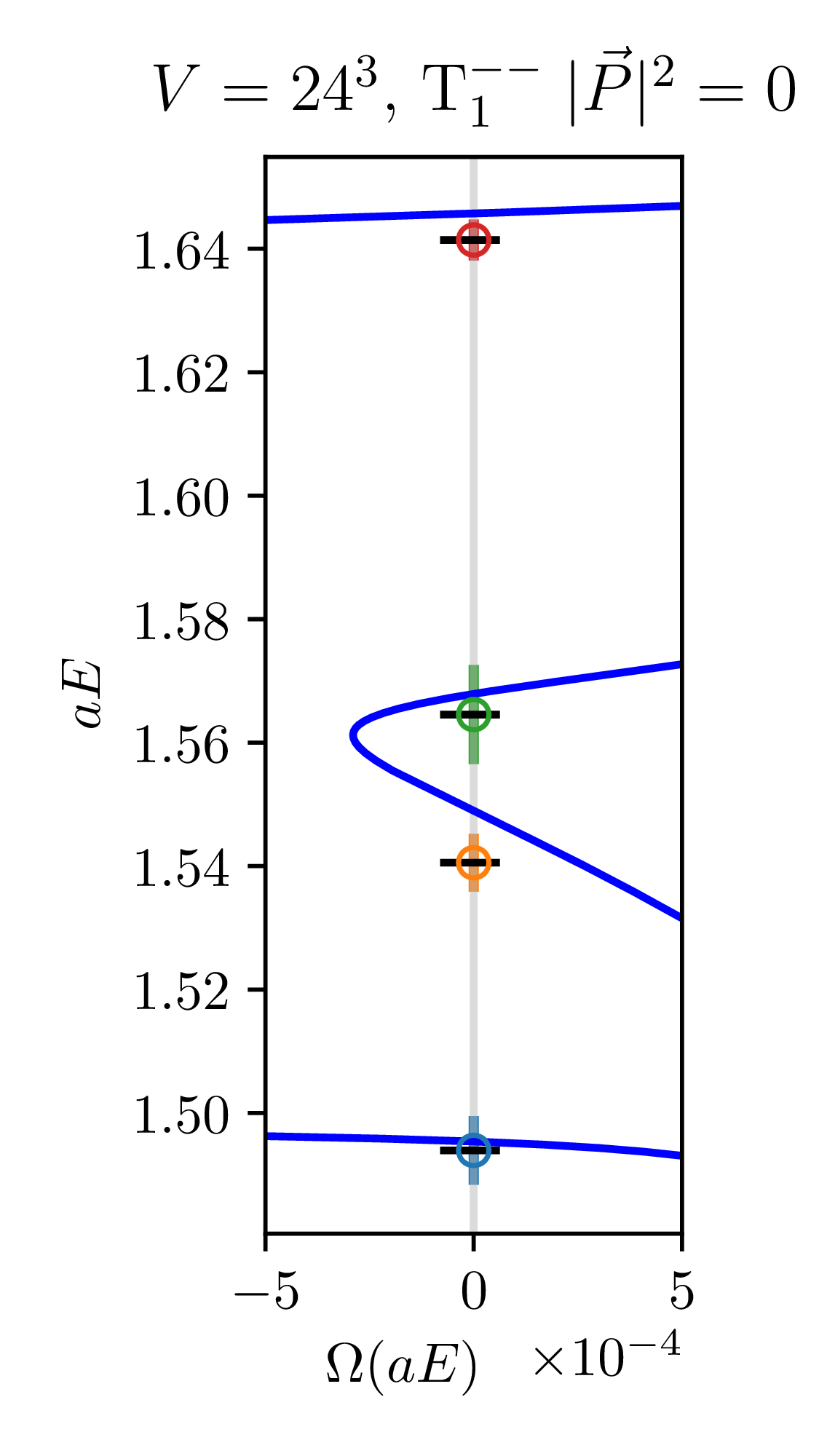}}
 \subfigure[]{\includegraphics[width=0.155\textwidth]{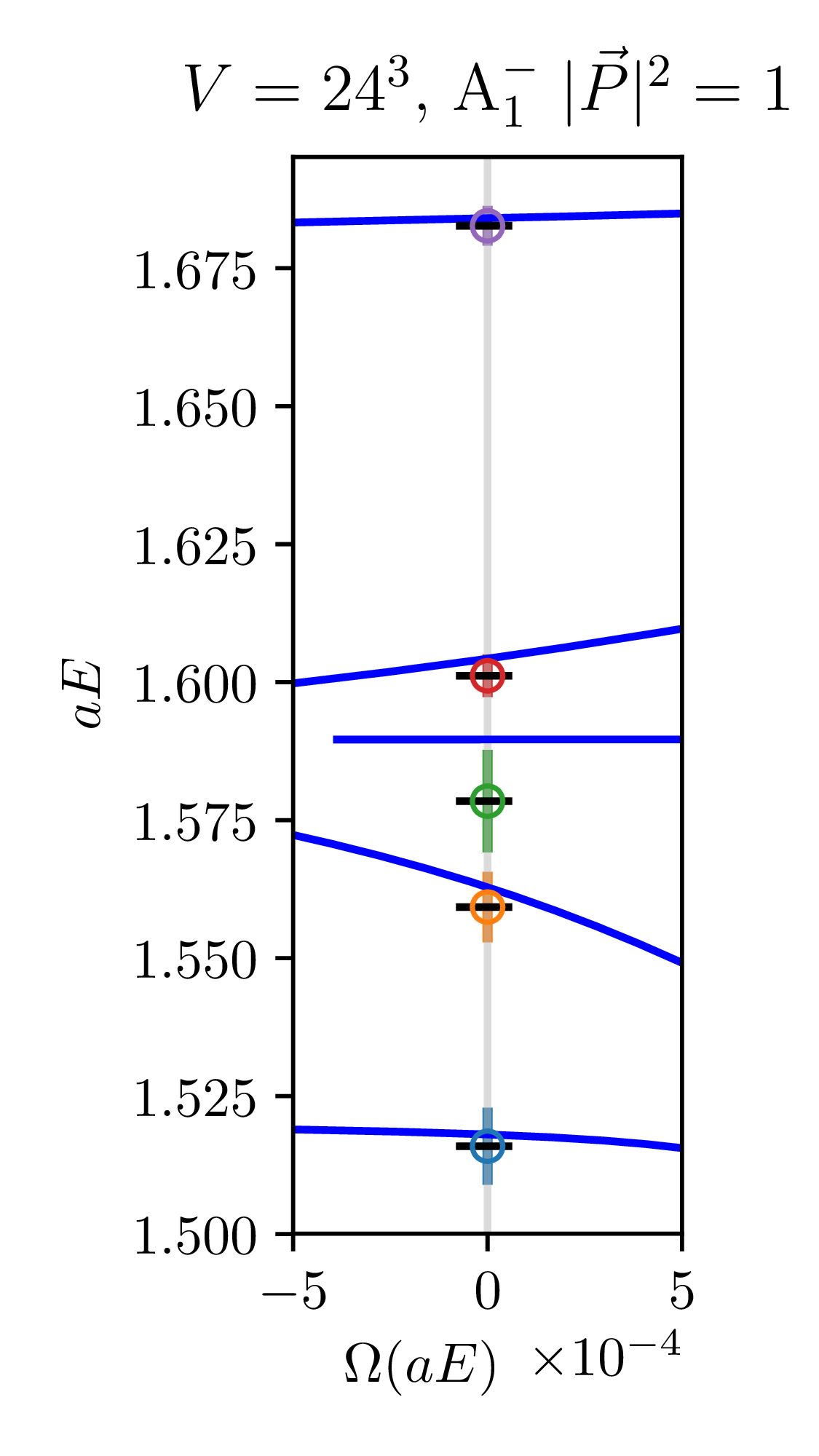}}
 \subfigure[]{\includegraphics[width=0.155\textwidth]{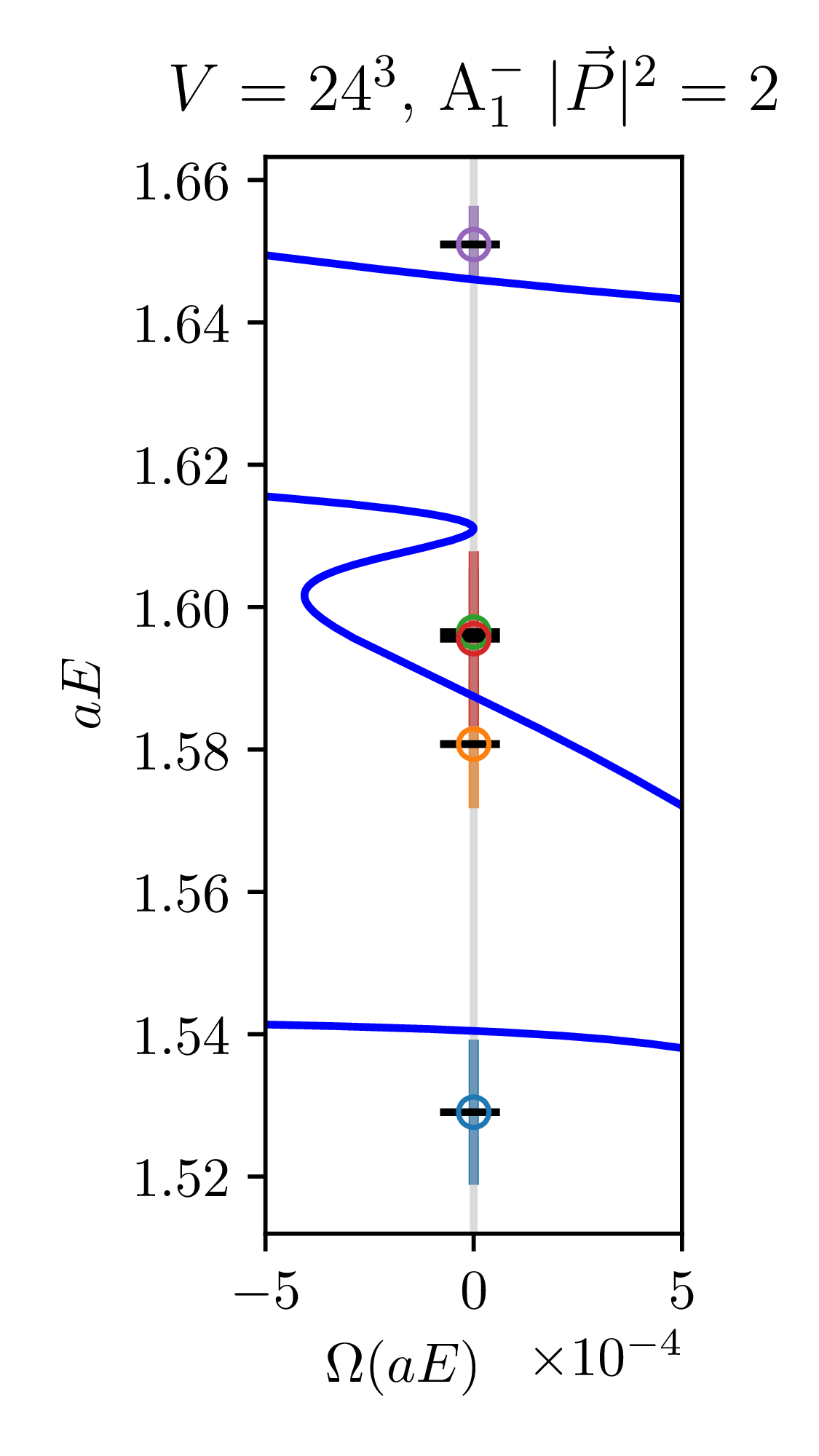}}
 \subfigure[]{\includegraphics[width=0.155\textwidth]{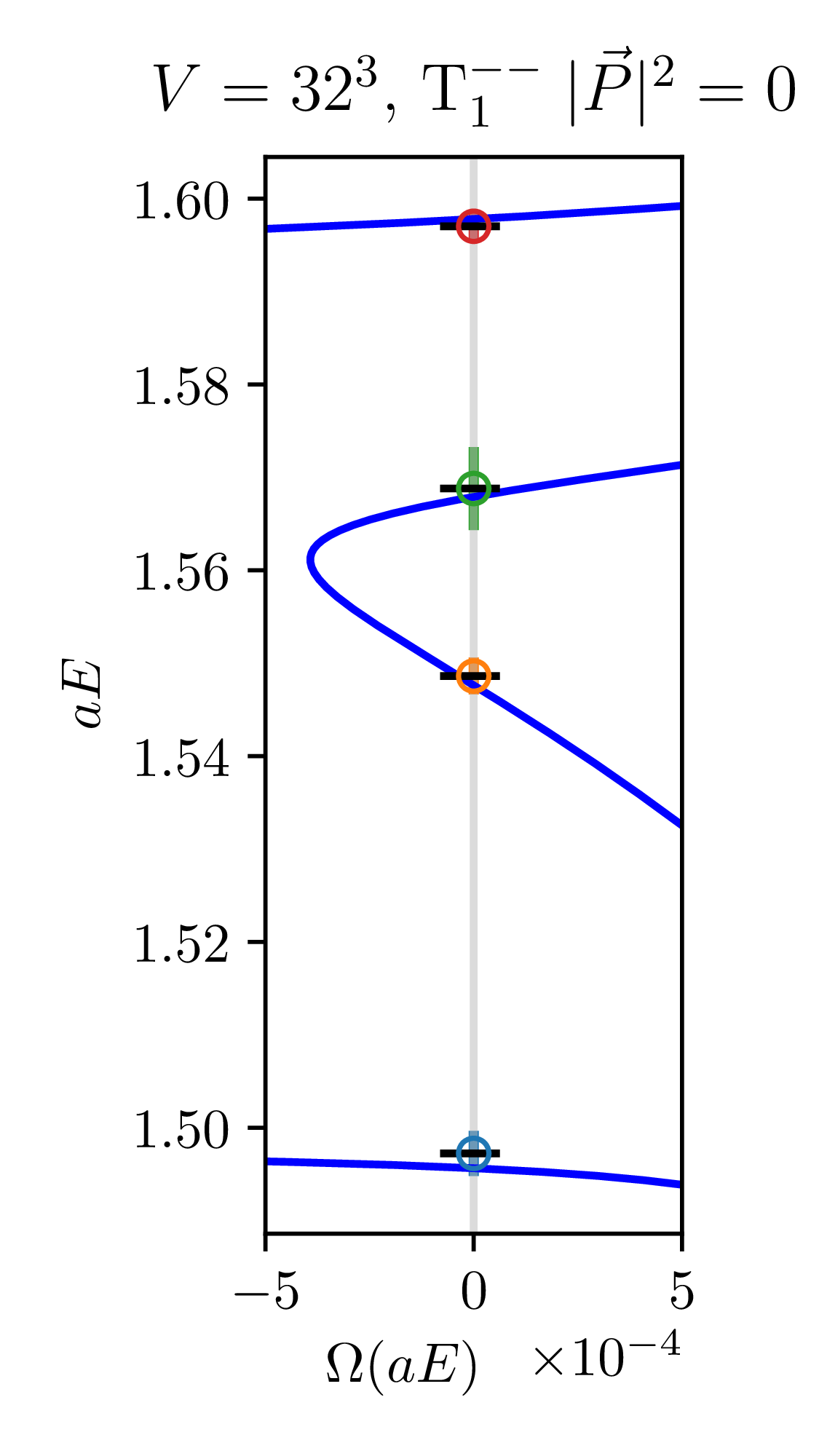}}
 \subfigure[]{\includegraphics[width=0.155\textwidth]{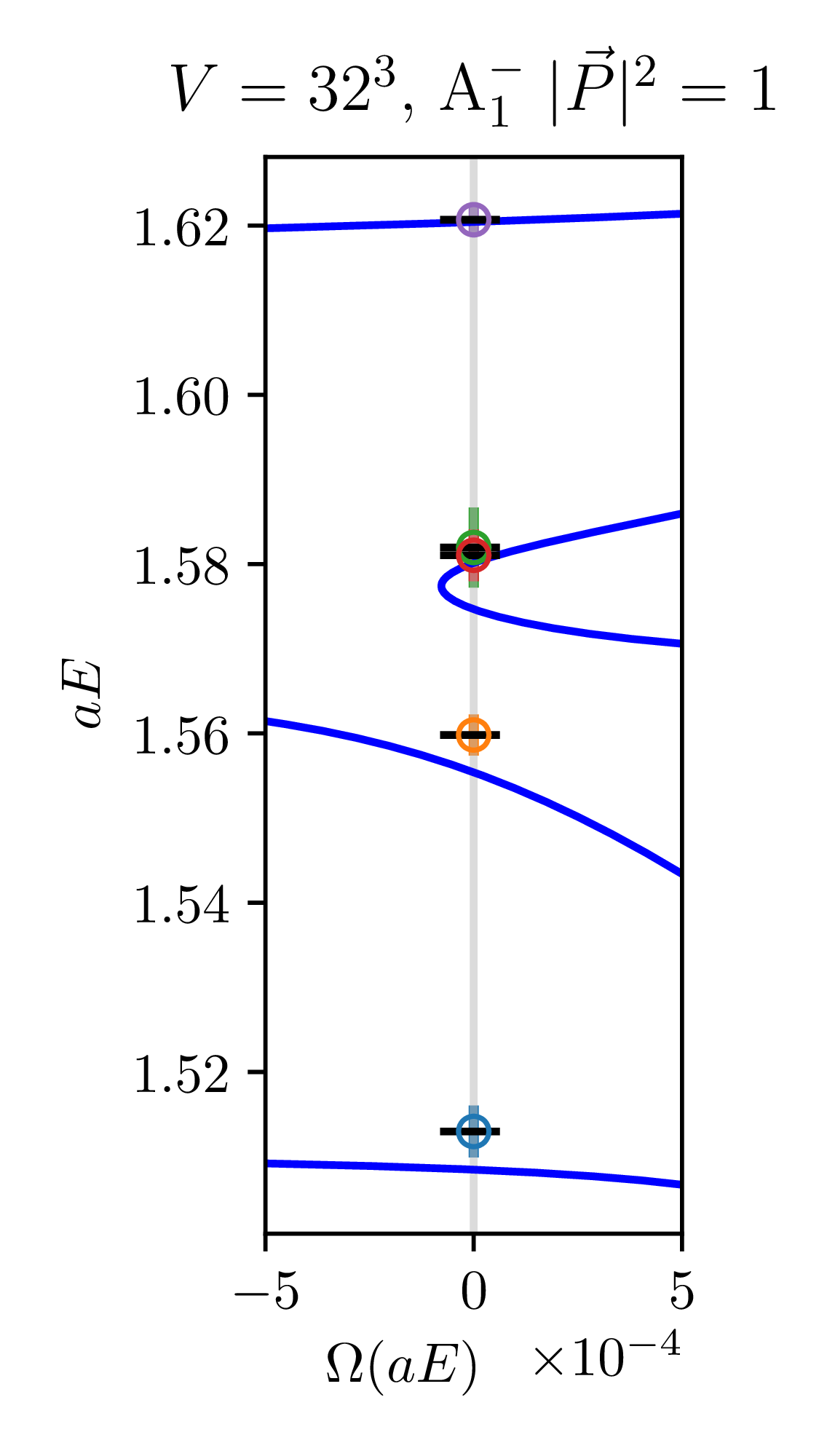}}
 \subfigure[]{\includegraphics[width=0.155\textwidth]{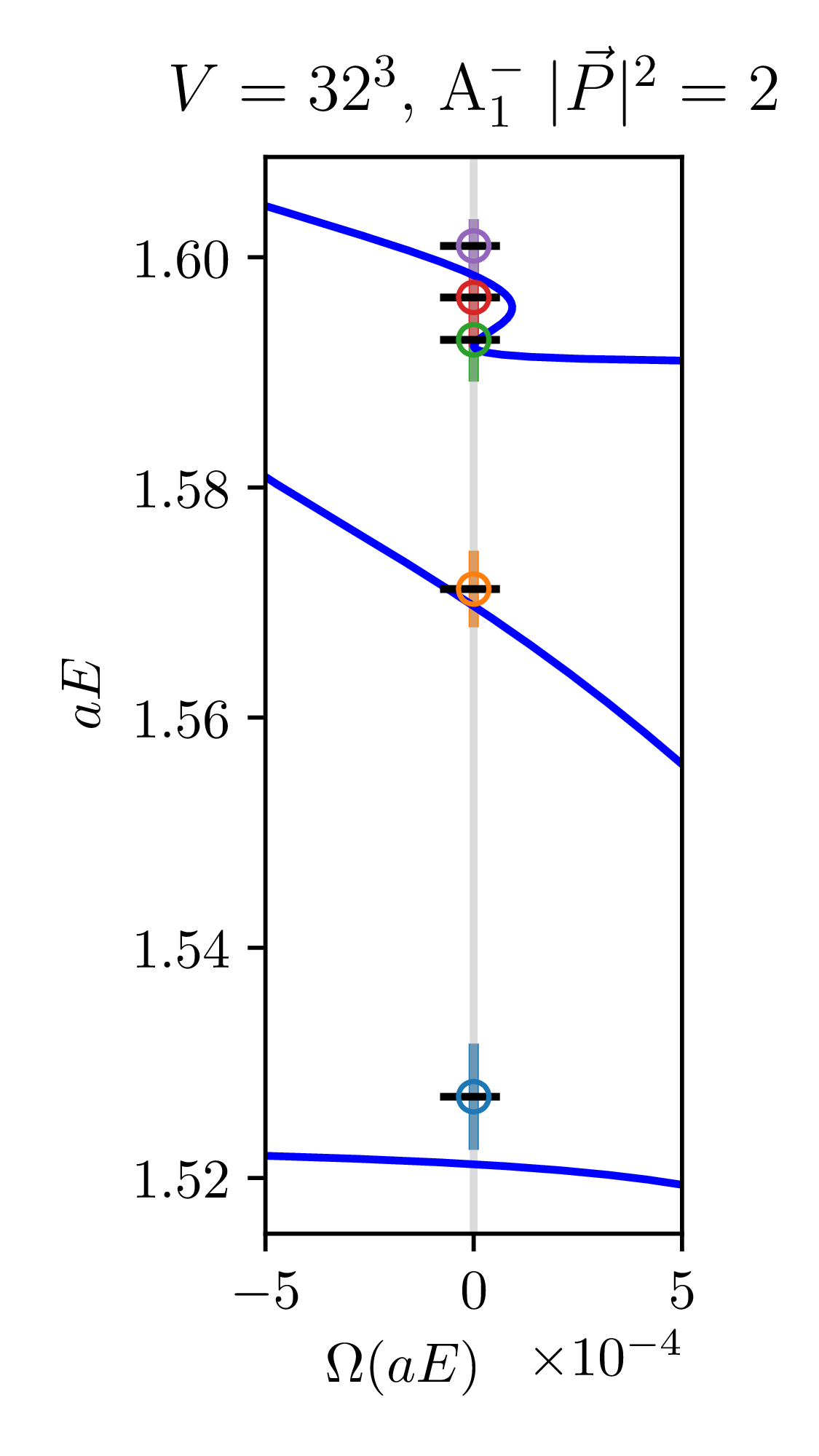}}
 \caption{$\Omega$-function in the three lattice irreducible representations in various frames for the vector channel resulting from the fit of all energy levels using the ``double pole'' fit ansatz for the phase shift for $\kappa_c = 0.12522$.}\label{Omega_T1MM_3MM_double_pole_kinv_kc0p12522}
\end{figure}

\begin{figure}
\centering
 \subfigure[]{\includegraphics[width=0.155\textwidth]{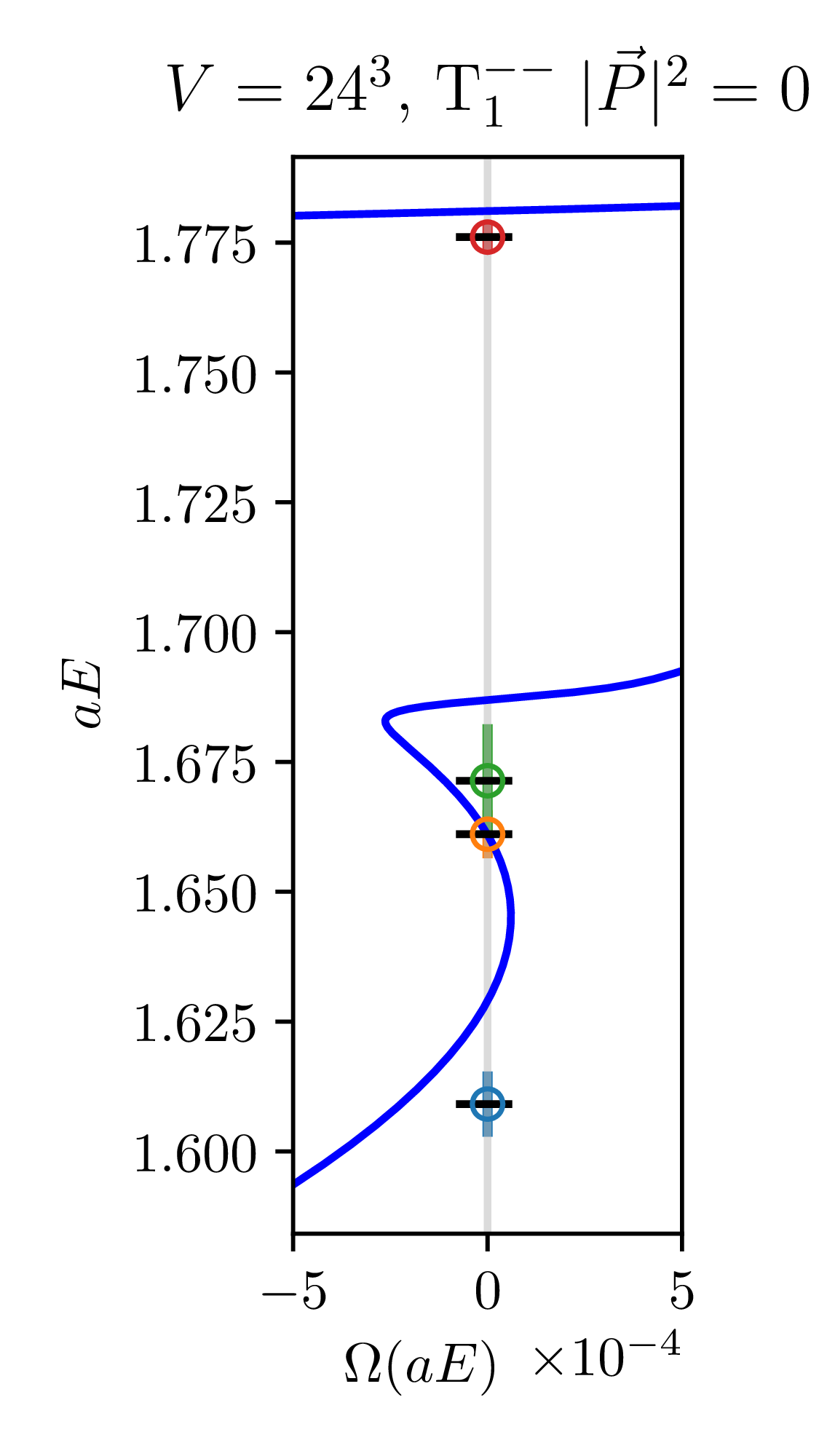}}
 \subfigure[]{\includegraphics[width=0.155\textwidth]{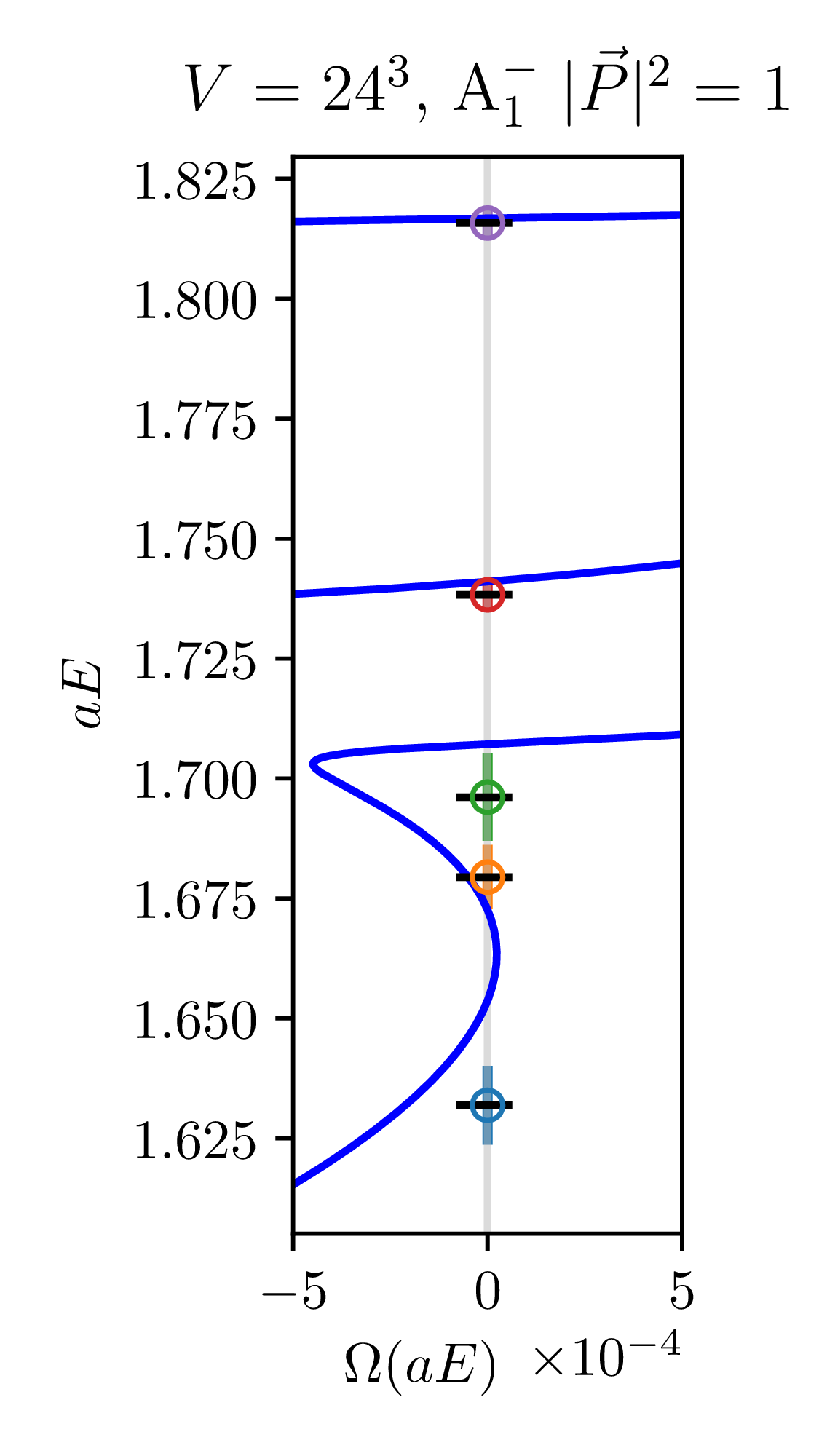}}
 \subfigure[]{\includegraphics[width=0.155\textwidth]{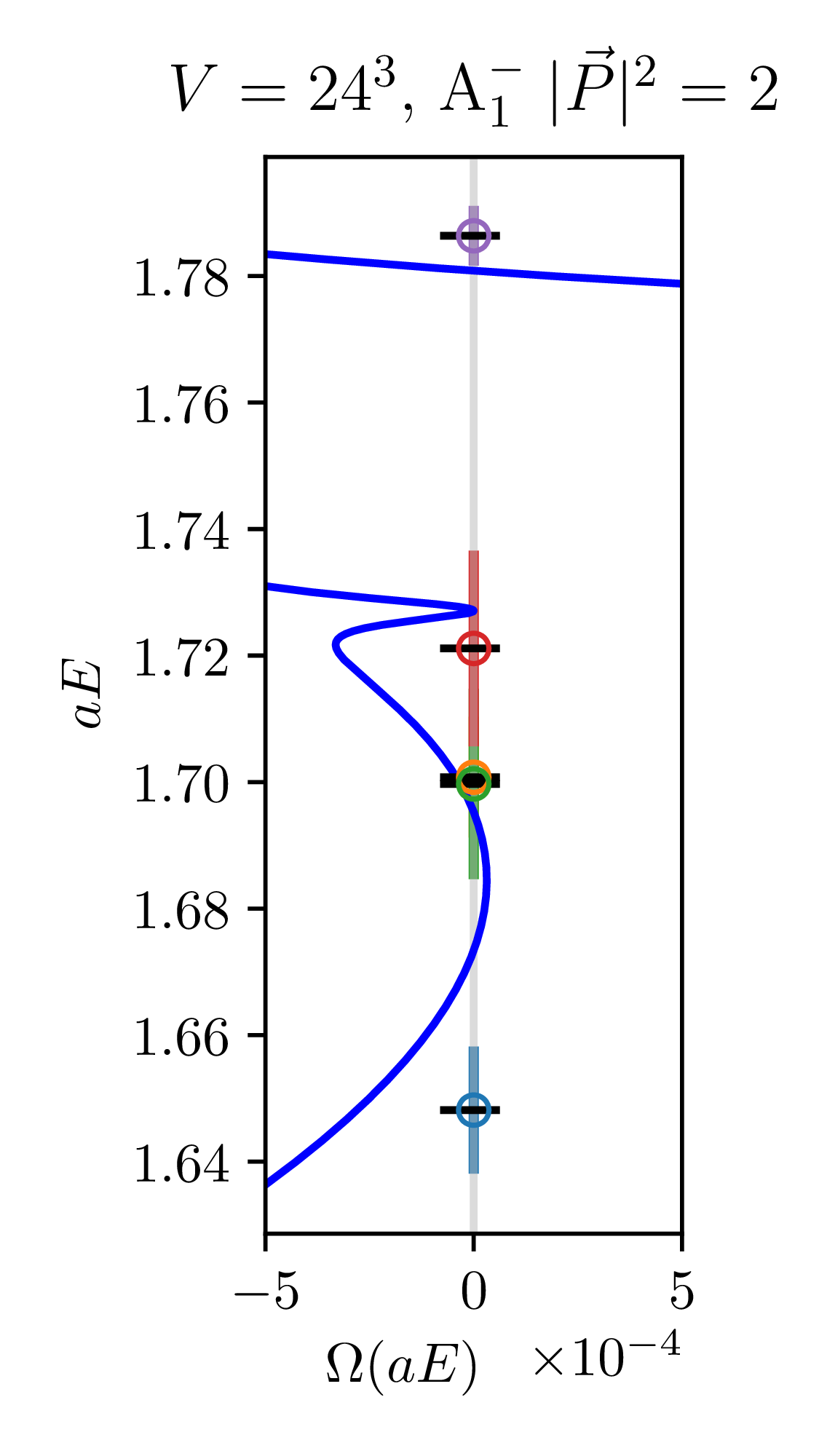}}
 \subfigure[]{\includegraphics[width=0.155\textwidth]{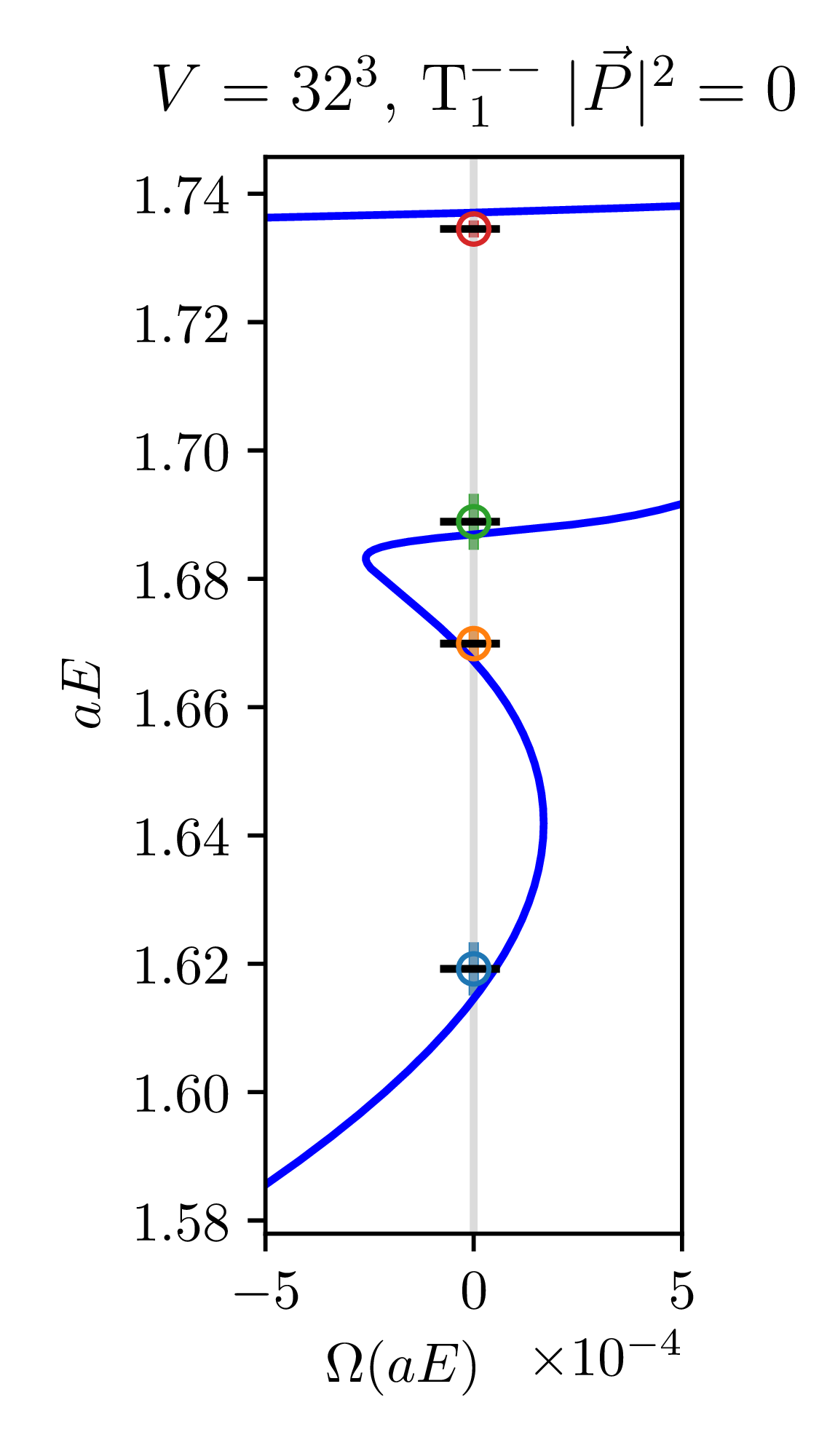}}
 \subfigure[]{\includegraphics[width=0.155\textwidth]{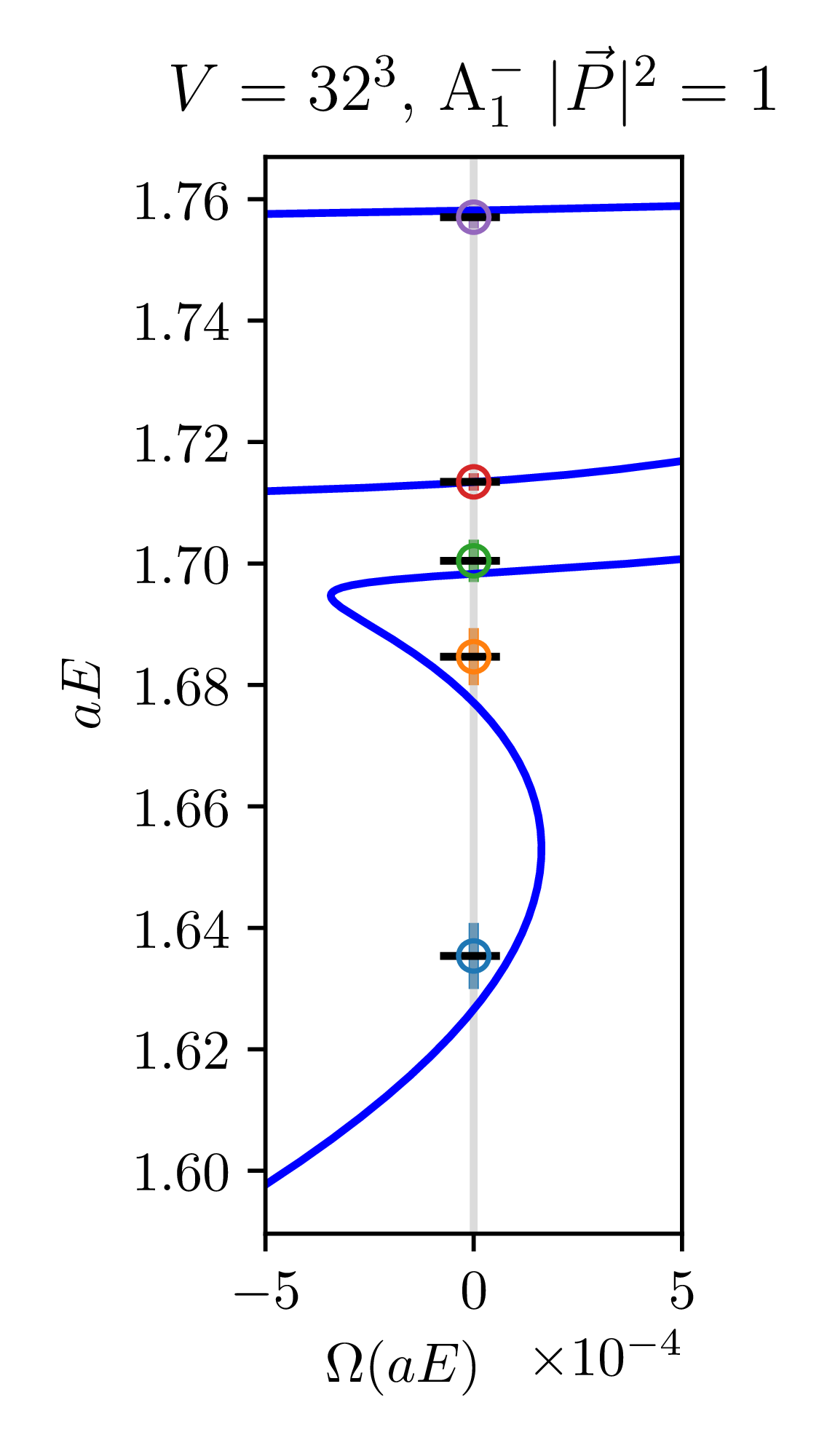}}
 \subfigure[]{\includegraphics[width=0.155\textwidth]{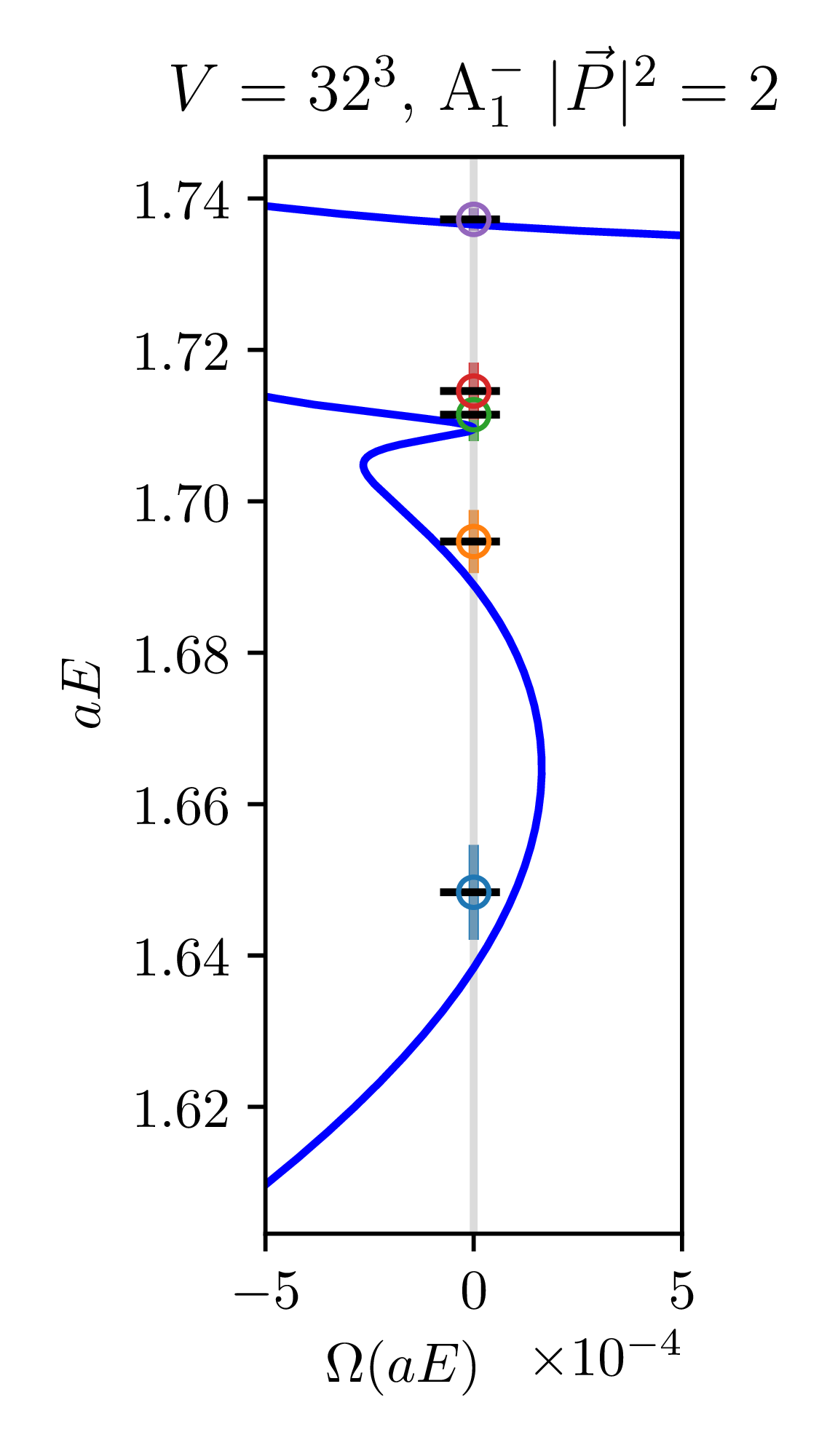}}
 \caption{$\Omega$-function in the three lattice irreducible representations in various frames for the vector channel resulting from the fit of all energy levels using the ``quadratic'' fit ansatz for the phase shift for $\kappa_c = 0.12315$.}\label{Omega_T1MM_3MM_quadratic_kinv_kc0p12315}
\end{figure}

\begin{figure}
\centering
 \subfigure[]{\includegraphics[width=0.155\textwidth]{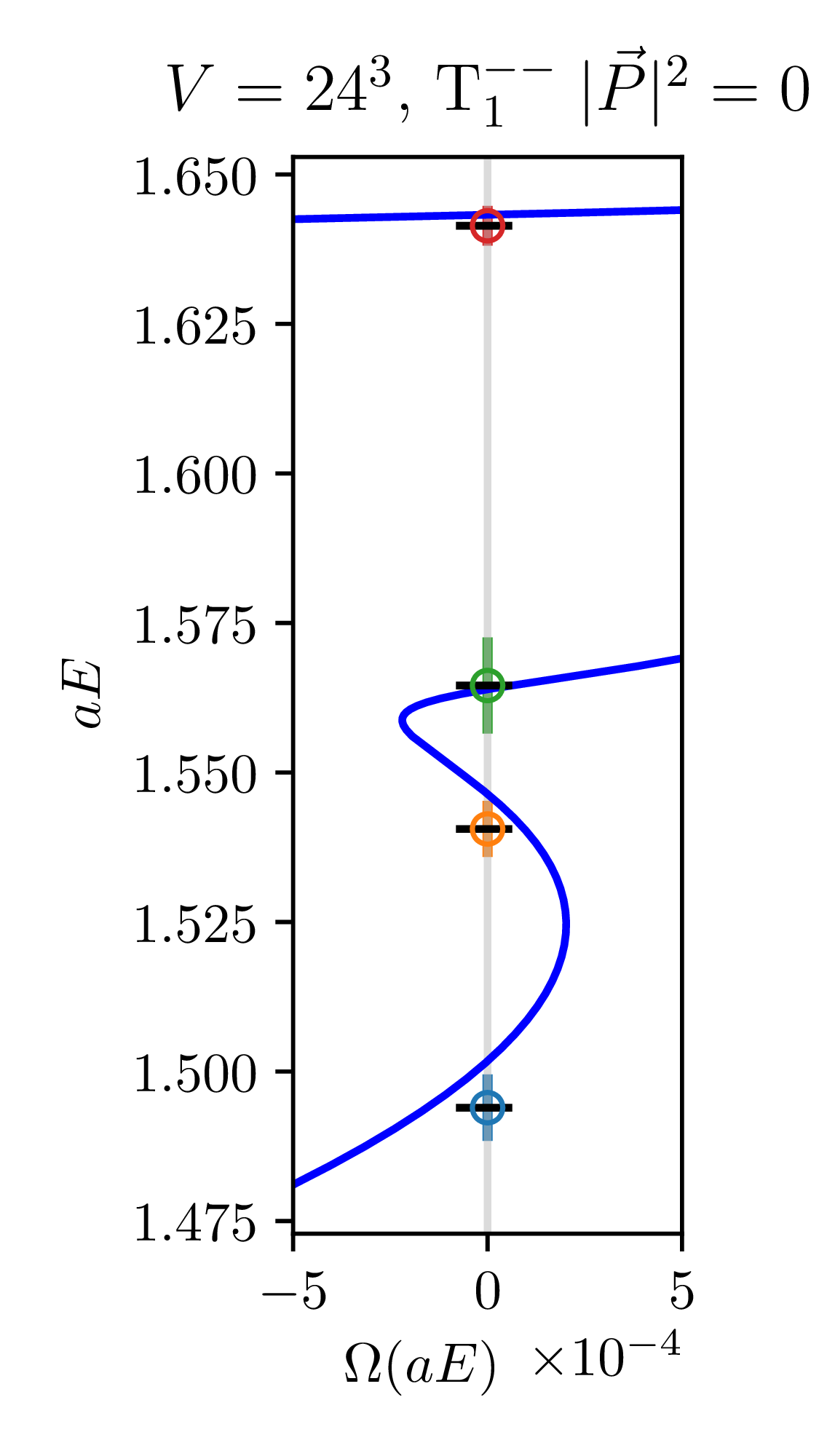}}
 \subfigure[]{\includegraphics[width=0.155\textwidth]{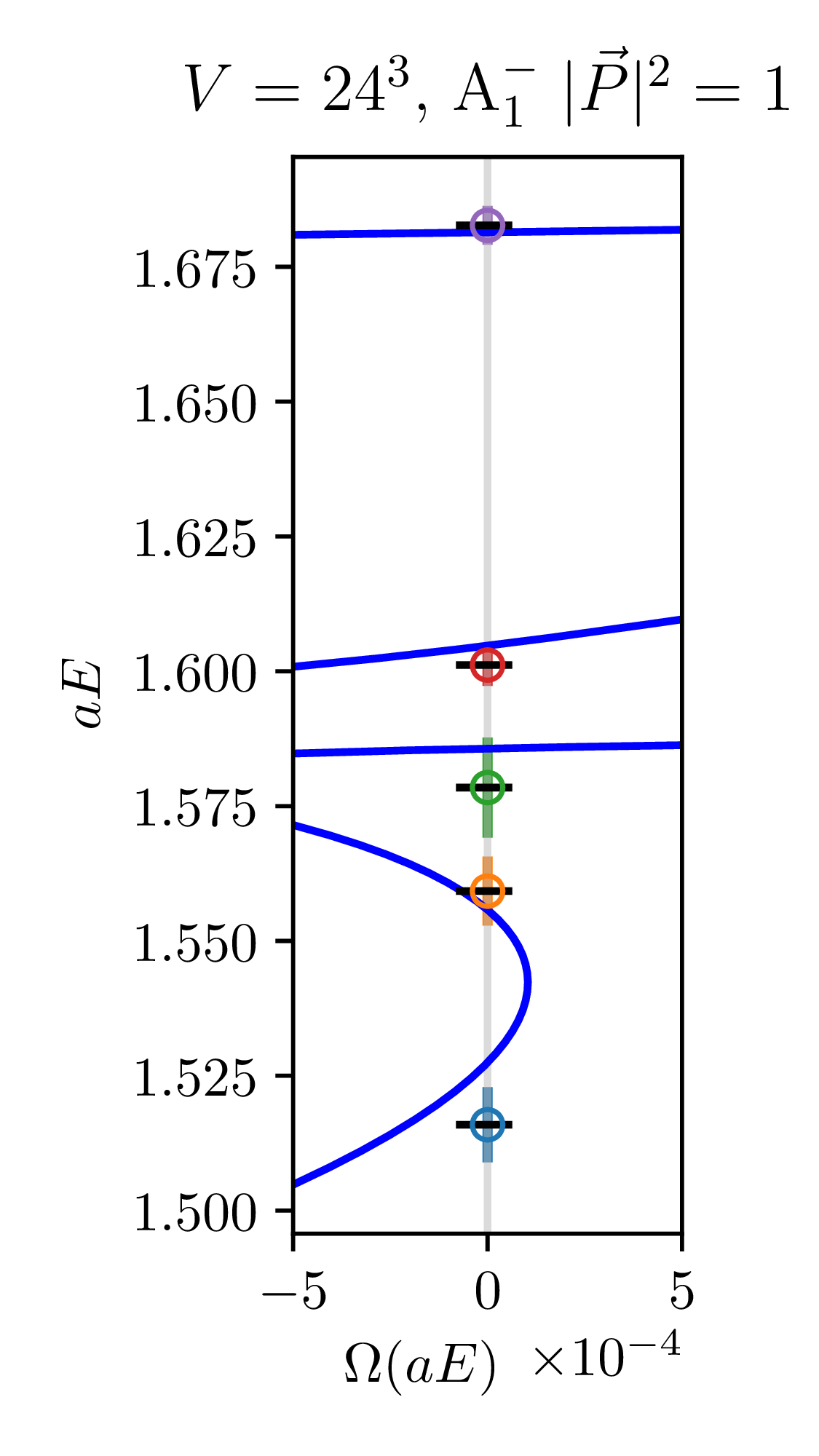}}
 \subfigure[]{\includegraphics[width=0.155\textwidth]{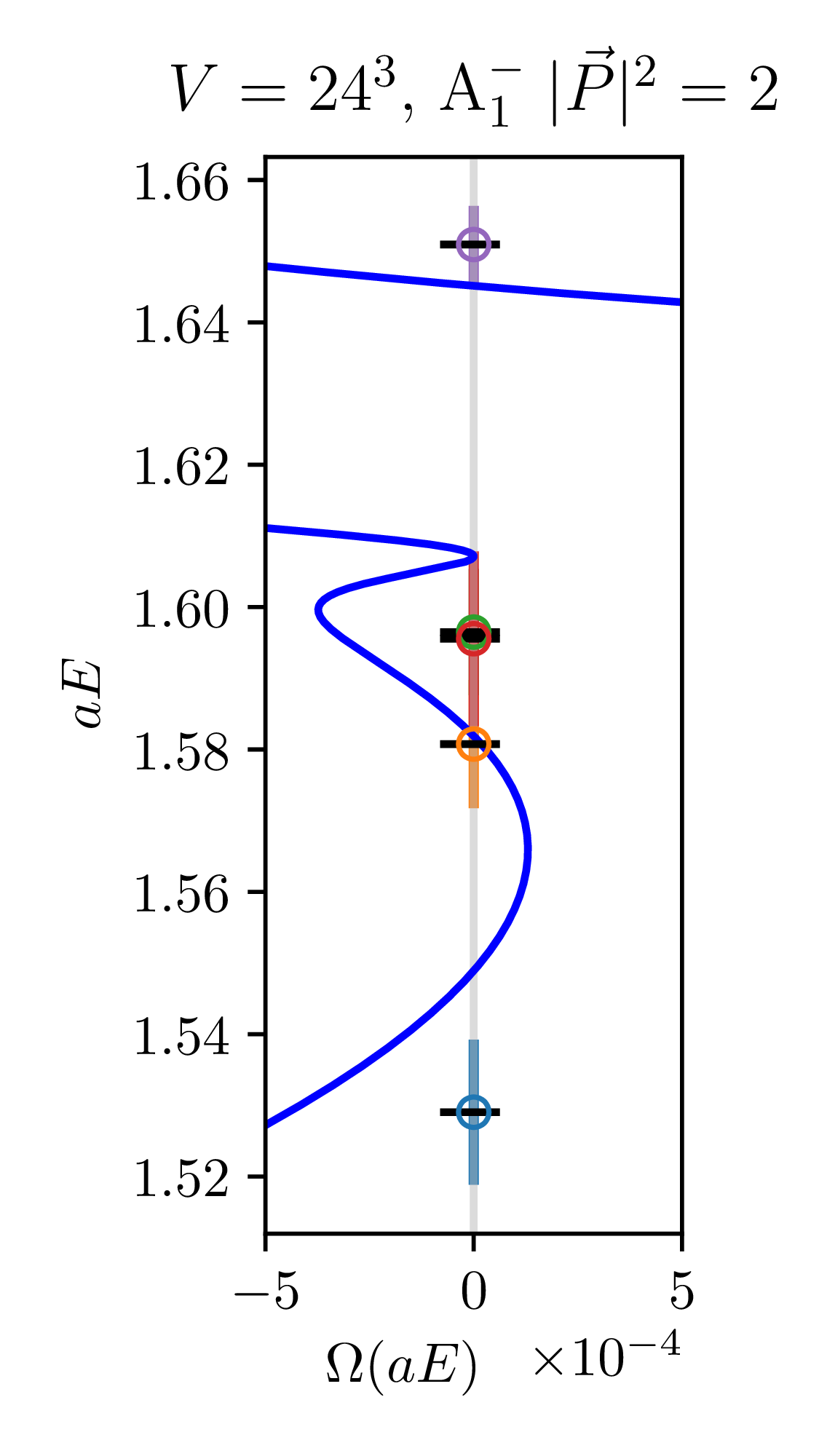}}
 \subfigure[]{\includegraphics[width=0.155\textwidth]{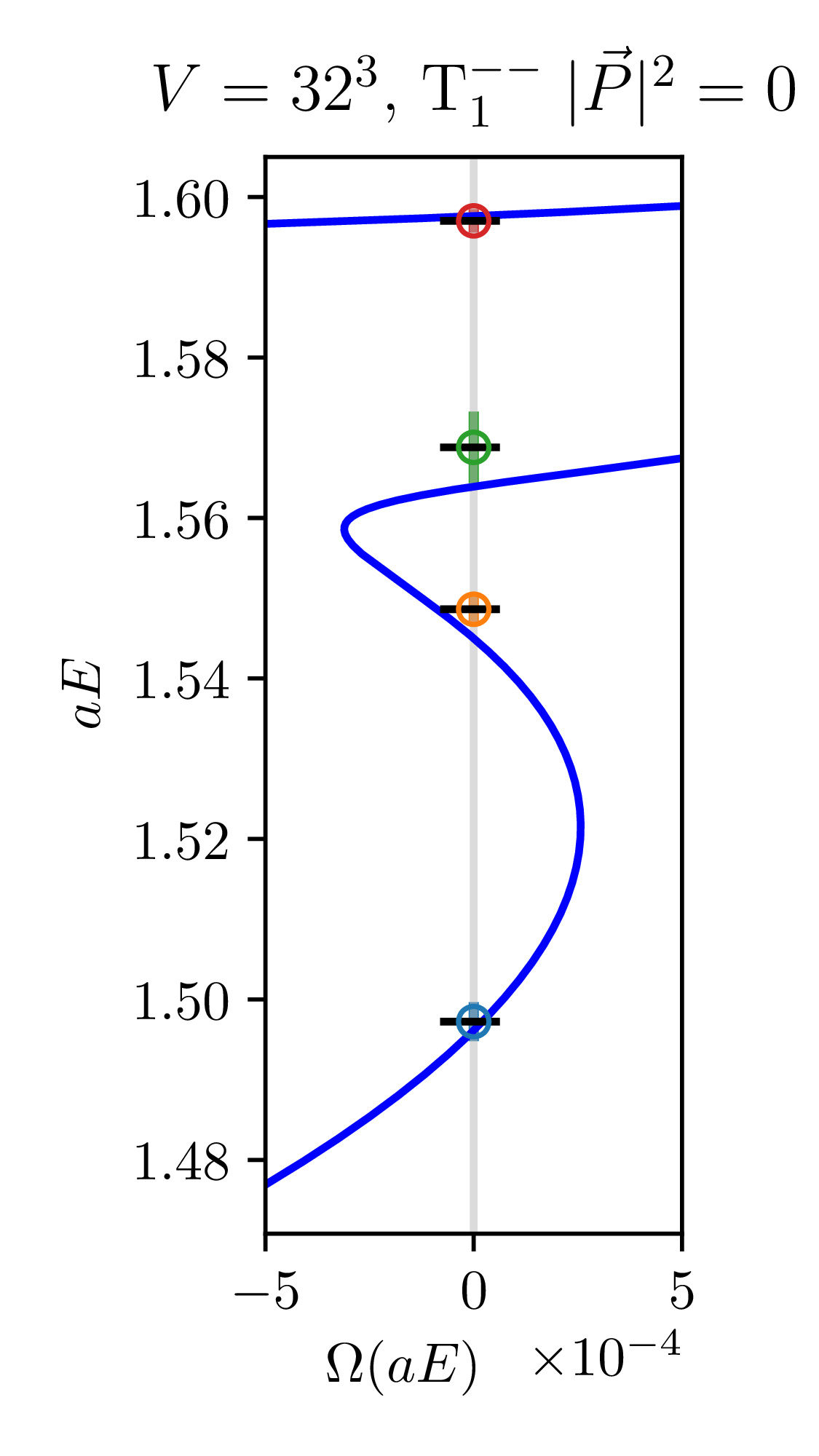}}
 \subfigure[]{\includegraphics[width=0.155\textwidth]{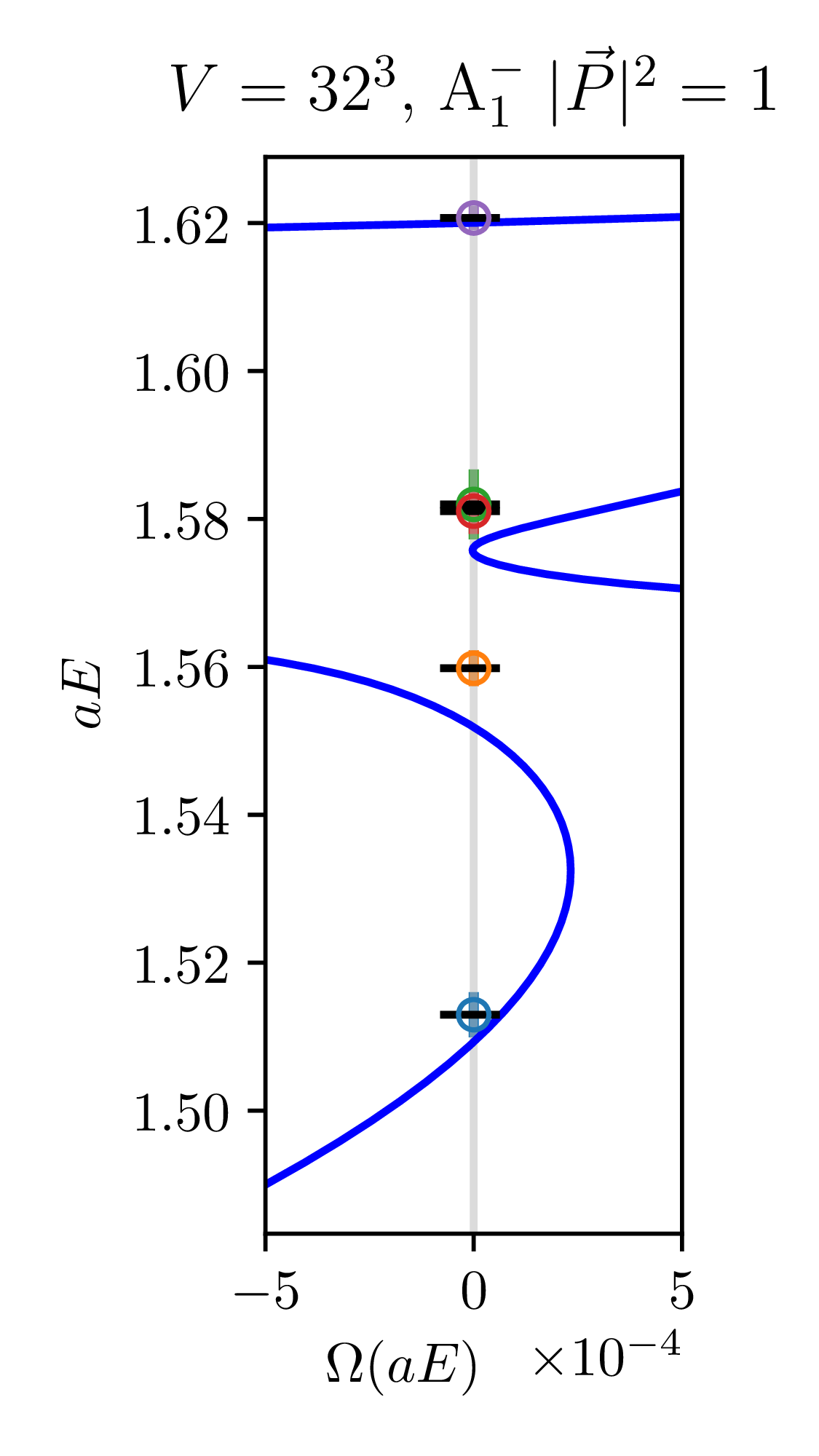}}
 \subfigure[]{\includegraphics[width=0.155\textwidth]{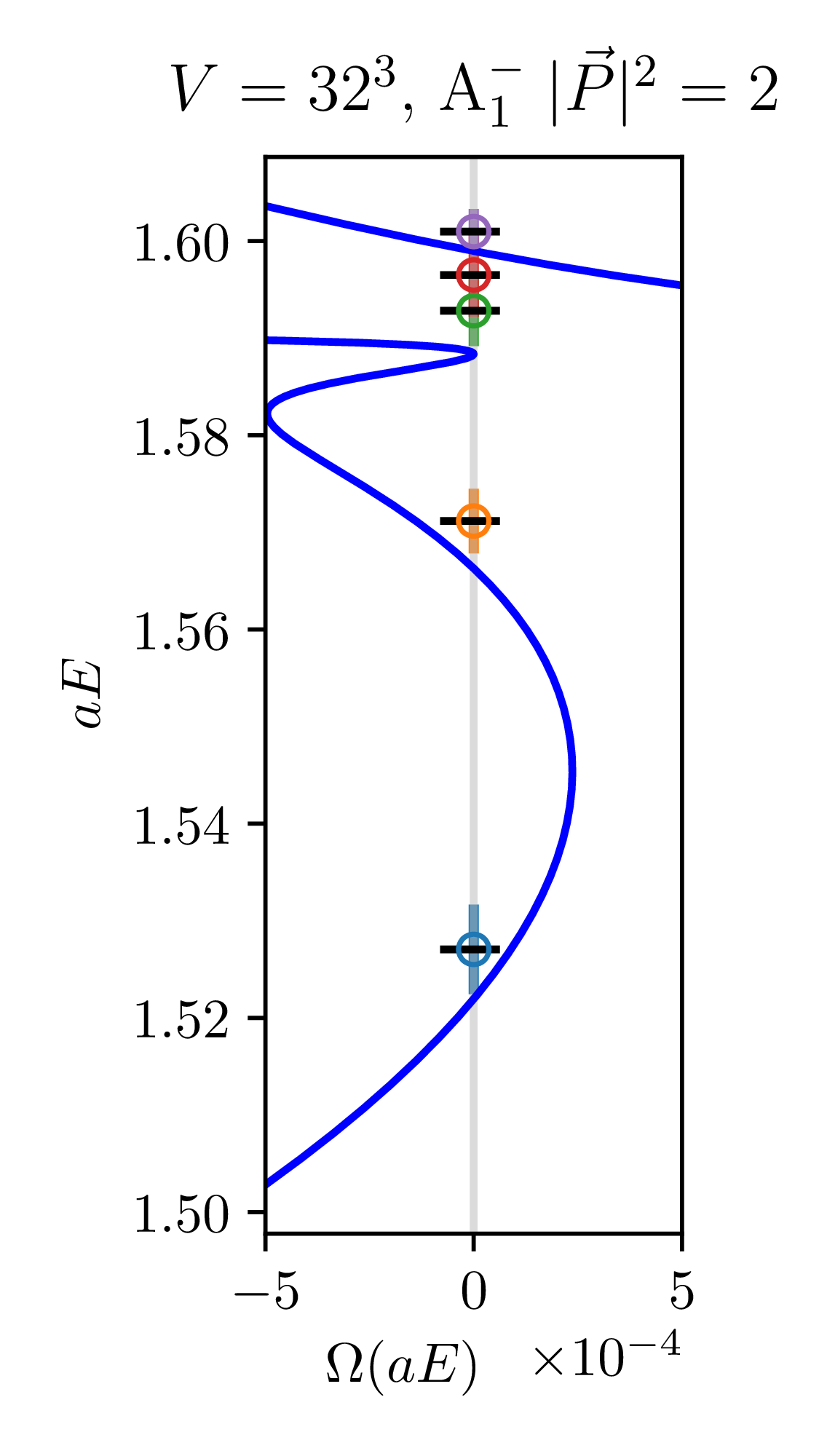}}
 \caption{$\Omega$-function in the three lattice irreducible representations in various frames for the vector channel resulting from the fit of all energy levels using the ``quadratic'' fit ansatz for the phase shift for $\kappa_c = 0.12522$.}\label{Omega_T1MM_3MM_quadratic_kinv_kc0p12522}
\end{figure}

\begin{figure}
  \subfigure[~double pole]{\includegraphics[width=.33\textwidth]{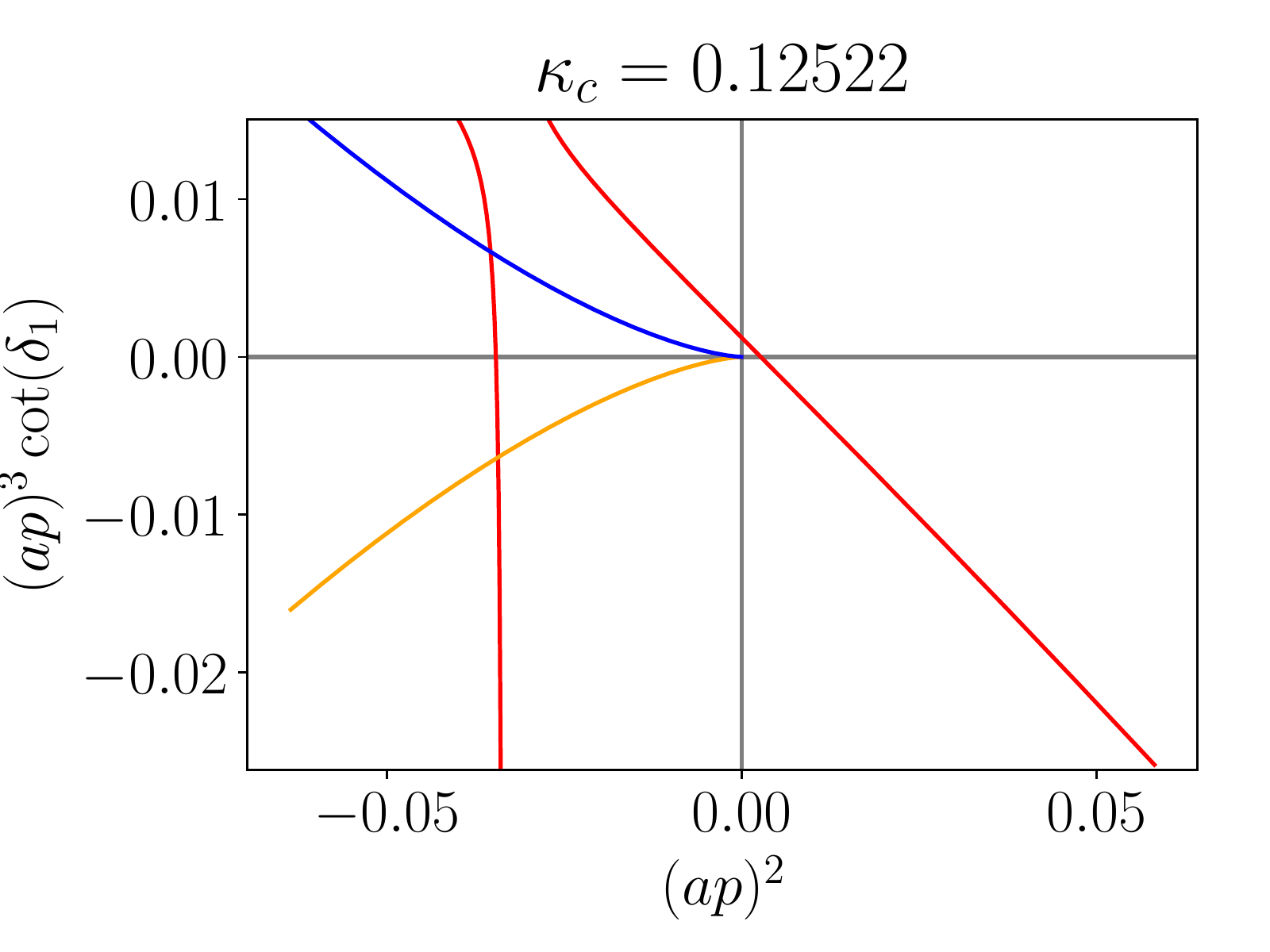}\label{vector_channel_crossing_poles_kc12522}}
  \subfigure[~double pole]{\includegraphics[width=.33\textwidth]{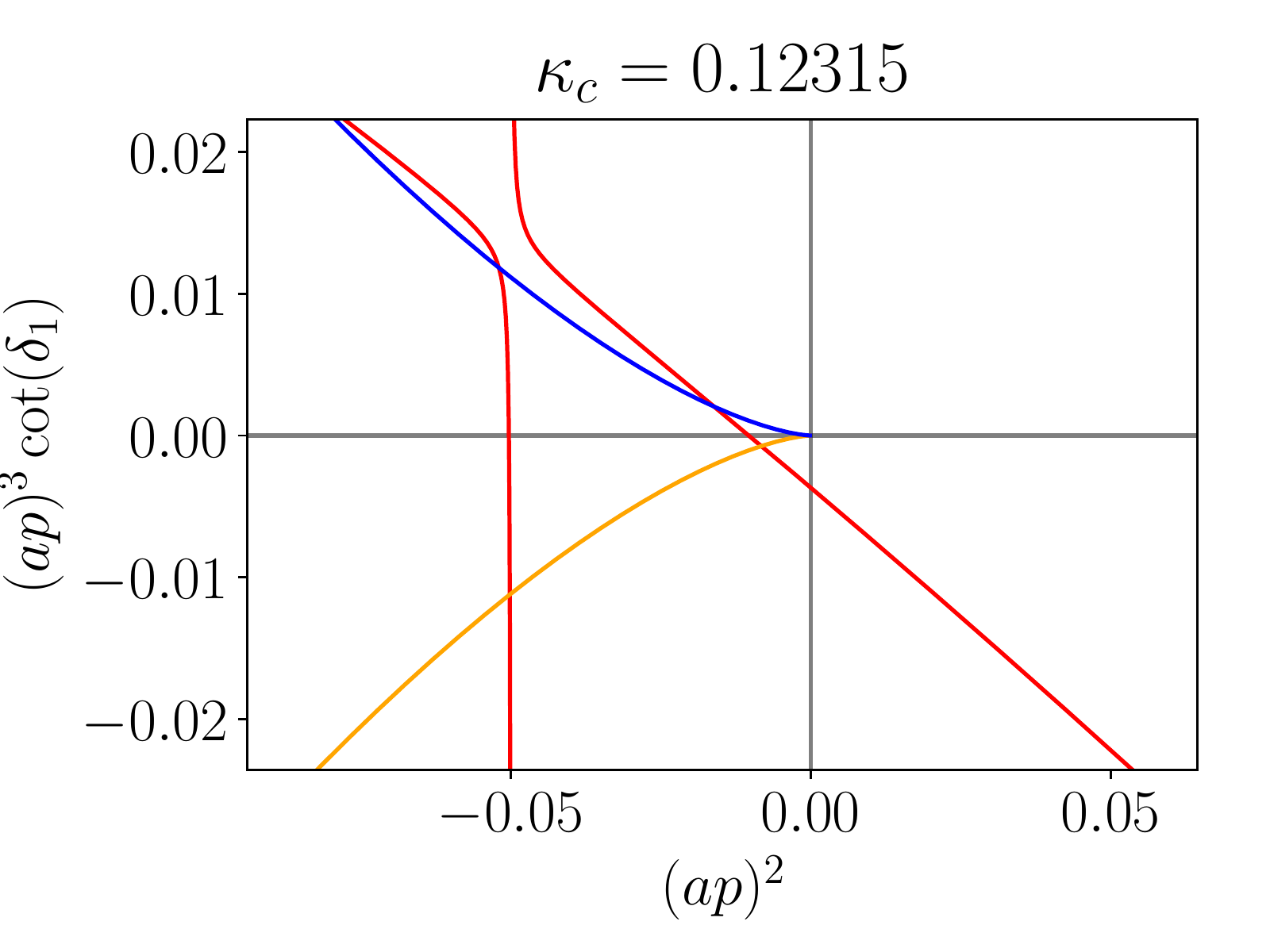}\label{vector_channel_crossing_poles_kc12315}}
  \subfigure[~quadratic]{\includegraphics[width=.33\textwidth]{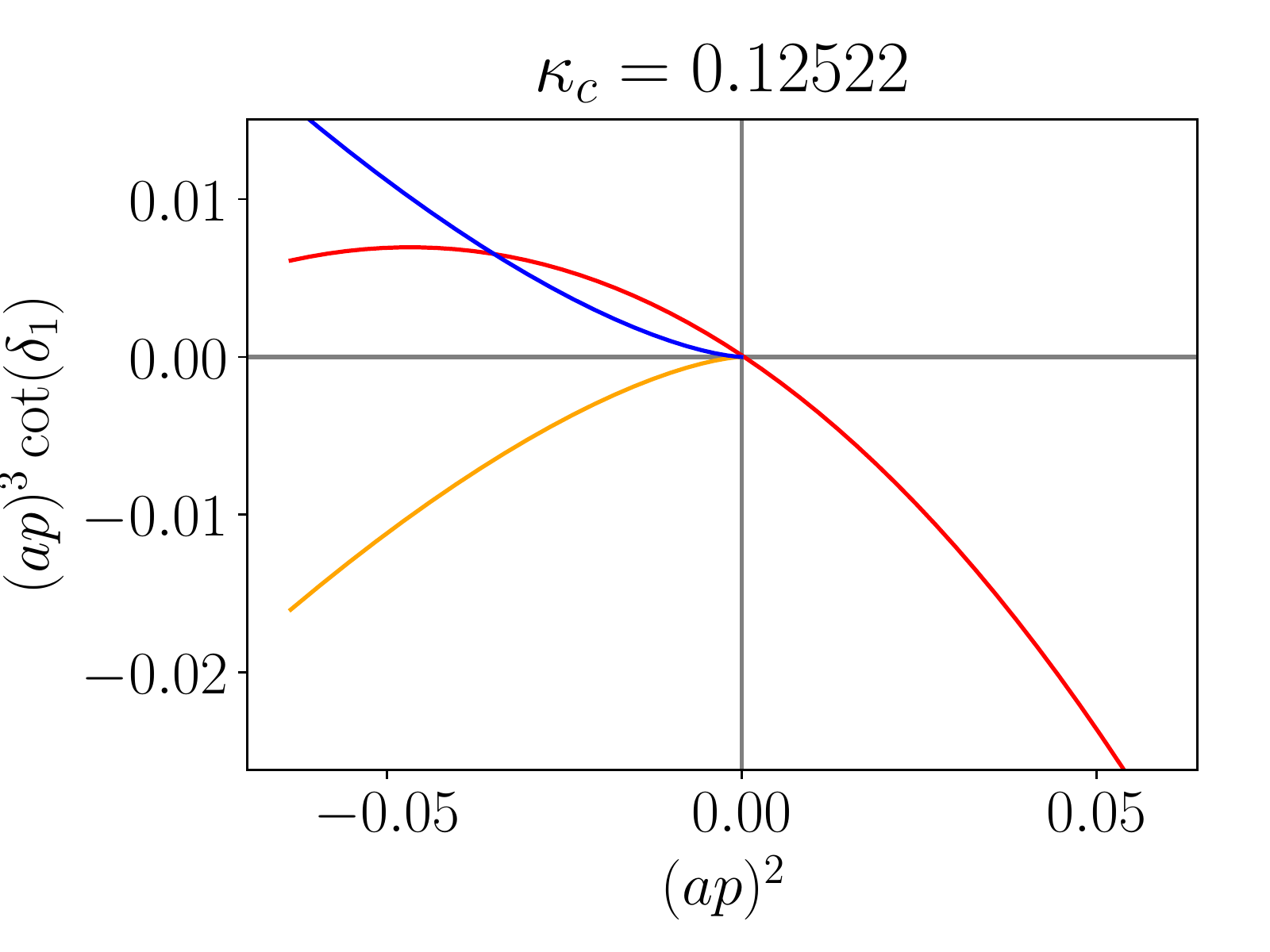}\label{vector_channel_crossing_quadratic_kc12522}}
  \subfigure[~quadratic]{\includegraphics[width=.33\textwidth]{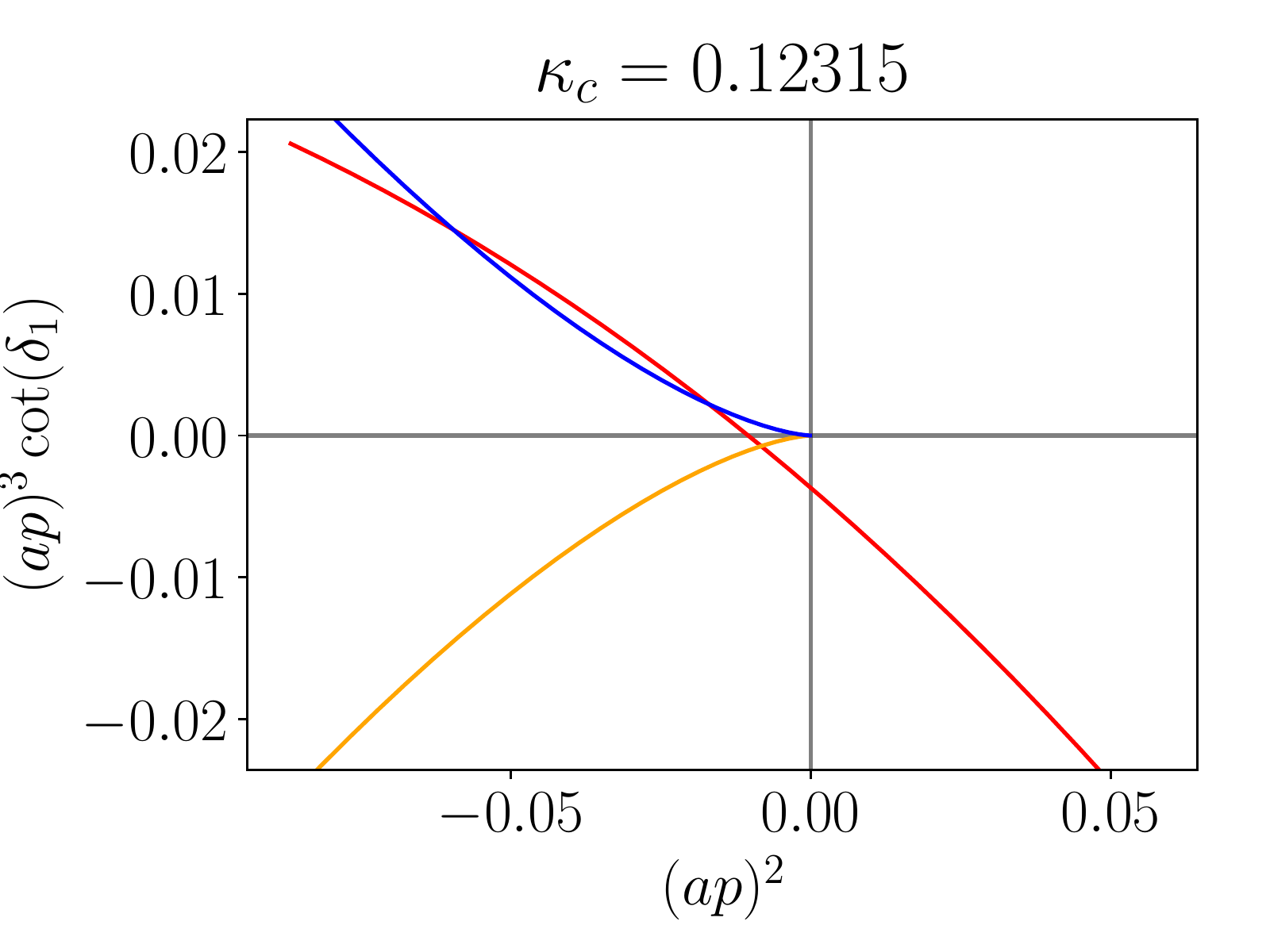}\label{vector_channel_crossing_quadratic_kc12315}}
  \caption{$(ap)^3 \cot(\delta_1)$ (red curves) for the double pole and quadratic fits together with curves representing the bound state~(\ref{bs1}) and virtual bound state~(\ref{bs2}) conditions. The intersection of the red and blue curves corresponds to a bound state, while the intersection of the red and orange curves corresponds to a virtual bound state.}\label{crossings}
\end{figure}

The quality of our fits for the combined channel case is presented in Figs.~\ref{Omega_T1MM_3MM_double_pole_kinv_kc0p12315}, \ref{Omega_T1MM_3MM_double_pole_kinv_kc0p12522}, \ref{Omega_T1MM_3MM_quadratic_kinv_kc0p12315} and \ref{Omega_T1MM_3MM_quadratic_kinv_kc0p12522}, in terms of the $\Omega$ function, 
\begin{equation}\label{omega_function}
\Omega(\mu,A(s)) = \frac{\det(A)}{\det((\mu^2 + AA^\dag)^{\frac{1}{2}})}\,,
\end{equation}
as introduced in Ref.~\cite{Morningstar:2017spu} to minimize the $\chi^2$ in the ``determinant residual method''. Here the matrix $A(s)$ is the argument of the determinant in Eq.~(\ref{eq:det}) and $\mu=8$ is a regularization parameter, see Appendix~\ref{app1} for further details on the choice of $\mu$. Each crossing of the $\Omega$ function with the vertical axis $\Omega=0$ represents a predicted energy level from our parametrization, which can be compared with our observed energy spectrum. The number of energy levels is indeed equal to the number of solutions of $\Omega=0$ in  the energy region between $E\simeq m_{\psi(2S)}$ (the blue energy level) and the highest level. In particular, there are two approximately degenerate energy levels when the fitted $\Omega$-function has a double zero~(this occurs for  $3^{--}$  when $P^2 = 2$).  We remark that our scattering amplitude captures the energy dependence only in the energy region where we have data and is not  applicable outside of this region. There is an additional crossing of $\Omega=0$ in certain irreps for $\kappa_c = 0.12315$ in Fig.~\ref{Omega_T1MM_3MM_double_pole_kinv_kc0p12315}, however, it occurs for $E<m_{\psi(2S)}$. The amplitude is not constrained and is, therefore, not applicable there. \footnote{This additional crossing would not occur if the slope of $p^3 \cot \delta_1/\sqrt{s}$ was slightly bigger (in terms of the absolute value) in the energy region $E<m_{\psi(2S)}$ of Fig.~\ref{vector_channel_crossing_poles_kc12315} (see also Appendix~\ref{app3} for details).}

The parameters for the $3^{--}$ resonance are quite unstable across the bootstrap samples, resulting in large uncertainties, especially for the coupling $g_3$. To check the stability of the fits in the vector channel, we have also performed an alternative analysis: the levels related to $J^{PC} = 3^{--}$ are identified (following the methods utilized in Ref.~\cite{Padmanath:2018tuc}) and excluded from the fits. The resulting couplings and masses for the $1^{--}$ resonance are found to be compatible with the values quoted above and we can therefore conclude that the influence of the $3^{--}$ resonance on the $\psi(3770)$ is negligible in our simulation. 

\subsection{Bound states, virtual bound states and resonances}\label{sec:polesdef}

Experiments and lattice QCD determine the scattering amplitudes $t_l$ for real energies. The physical interpretation in terms of (virtual) bound states and resonances is however conventionally performed by looking at poles of the $t_l(s)$ in Eq.~(\ref{eq:TKSmatrices}), continued to the complex $s$-plane 
\begin{align}\label{complex-plane}
&t_{l}(s)=\frac{1}{\rho~\cot(\delta_l)-i~\rho}\,,  \\
& \rho=\frac{2p}{\sqrt{s}} = \sqrt{1-4\tfrac{m_D^2}{s}}\,.  \nonumber
 \end{align}
The two Riemann sheets I and II, from the square root branch cut are defined to have Im$(\rho)>0$ and Im$(\rho)<0$, respectively\footnote{The square-root branch cut for $\rho$ is chosen in the conventional sense, such that it runs from the threshold $s=4 m_D^2$ to $+\infty$ along the positive real axis. We neglect the effects from the left hand cut and from the freedom in choosing the real part of $i\rho$ in Eq.~(\ref{complex-plane}) (such as using a Chew-Mandelstam phase space), either of which results in pole structure deep below the energy region being studied.}. The product
 \begin{equation}\label{rhocotdelta}
\rho\cot\delta_l=\frac{2}{p^{2l}}~\biggl[\frac{p^{2l+1}~\cot\delta_l}{\sqrt{s}}\biggr]
\end{equation}
is an analytic function of $s$ only, where our lattice results for $p^{2l+1}\cot\delta_l/\sqrt{s}$ were expressed in terms of $s$ in Section \ref{sec:fits}. 

The scattering amplitude $t_l$ in Eq.~(\ref{complex-plane}) has a pole at a given $p$ if $\cot\delta(p) = i$. A bound state appears as a pole of the $t_l$ matrix at $p = i \kappa_B$, {\it i.e.} on the real axis below the scattering threshold on the first Riemann sheet. A virtual state appears as a pole at $p =  -i\kappa_B$, {\it i.e.} on the real axis on the second Riemann sheet. In both cases $\kappa_B$ denotes a real positive number. It is convenient to express these pole conditions in terms of the quantity $p^{2l+1} \cot\delta$, which is considered in the Section \ref{sec:fits} for $l=1,3$. Such poles appear if the conditions
\begin{eqnarray}
p^{2l+1}\cot(\delta_l)&=&-(p^2)^l \sqrt{-p^2} ~~ \mathrm {(bound\ state)} \label{bs1}\\
p^{2l+1}\cot(\delta_l)&=&(p^2)^l \sqrt{-p^2} ~~ \mathrm{(virtual\ bound\ state) \label{bs2}}
\end{eqnarray}
are fulfilled. A resonance corresponds instead to a pair of poles away from the real axis above threshold on the second sheet, influencing the physical region in the first sheet.

\subsection{Consistency check for a bound state in partial wave $l$}\label{sec:consistency_check}

The $S(p)=e^{2i\delta(p)}$ matrix for a bound state has a pole at $p=i\kappa_B$, as discussed in the previous section. The analyticity of the $S(p)$ matrix near the real energy axes implies the sign of the pre-factor in front of the pole to be 
\begin{equation}
\label{sanity1} S(p)=-\frac{(-1)^l ~i~\beta^2}{p-i\kappa_B}\,,
\end{equation}
where $\beta^2$ is a real and positive number. This functional form is derived from the general form of the solutions for $S(k)$ continued to complex $p$ in Ref. \cite{SITENKO} (see Eq.~(7.60)) and applied for $s$-wave in, for example, in Ref. \cite{Iritani:2017rlk}. Equation (\ref{sanity1}) must be satisfied for all bound states, but it does not apply to virtual bound states. We have explicitly verified that the relation (\ref{sanity1}) with positive $\beta^2$ is satisfied for $s$-wave and $p$-wave bound states in 3D non-relativistic quantum mechanics for various forms of the potential (square well, Yukawa, Wood-Saxon, and several other forms of the type $V(r) = \frac{1}{r^n} \exp{(- A r^m)}$).

Let us  express the condition in Eq.~(\ref{sanity1}) in terms of $p^{2l+1}\cot(\delta_l)$ in Eq.~(\ref{bs1}) which was fitted from our data. For this purpose we consider the dependence of the quantity 
\begin{equation}\label{quantitycheck}
 p^{2l+1}\cot(\delta_l)-[-(p^2)^l \sqrt{-p^2}]
\end{equation}
on $p^2$ near the bound state pole, where it equals zero in Eq.~(\ref{bs1}). Inserting in Eq.~(\ref{quantitycheck}) $\cot\delta=i(S+1)/(S-1)$ with $S$ from Eq.~(\ref{sanity1}), and expressing $p$ with $p^2$ as $p=i|p| =i \sqrt{-p^2}$, we obtain
 \begin{align}&p^{2l+1}\cot(\delta)-[-(p^2)^l \sqrt{-p^2}]\nonumber\\
 =&-\frac{2\;(-1)^l \; (-p^2)^l\; (\kappa_B\;\sqrt{-p^2}+p^2)}{(-1)^l \;\beta^2 -\kappa_B+\sqrt{-p^2}}\,.
 \end{align}
Taking the derivative of the previous expression with respect to $p^2$ and evaluating it at $p^2=-\kappa_B^2$, we get
 \begin{equation}\label{sanity2}
 \frac{d}{d\;p^2} \biggl(p^{2l+1}\cot(\delta)-[-(p^2)^l \sqrt{-p^2}]\biggr)\bigg\vert_{p^2=-\kappa_B^2} =-\frac{\kappa_B^{2l}}{\beta^2}<0\,.
 \end{equation}
The above condition implies that the slope of $p^{2l+1}\cot\delta$ in terms of $p^2$ has to be smaller\footnote{Note that this applies for the values of these quantities, not for absolute values of these quantities.} than the slope of $-(p^2)^l \sqrt{-p^2}$  at the position of the bound state in partial wave $l$, while it does not apply for virtual bound states. Note that the position of the bound state is where these two curves cross. The condition for $s$-wave was referred to as a sanity-check in Ref. \cite{Iritani:2017rlk}. 
 
Let us now turn to the fitted phase shifts in Section~\ref{sec:fits} and investigate which results satisfy the condition of Eq.~(\ref{sanity1}) or  equivalently the consistency check of Eq.~(\ref{sanity2}). The slope of $p^{2l+1}\cot(\delta)$ (the red curves in Fig.~\ref{crossings}) has to be smaller than the slope of  $-(p^2)^l \sqrt{-p^2}$ (the blue curve in Fig.~\ref{crossings}) where these two curves cross. This  is satisfied  for the double-pole fits, while it is not satisfied for the quadratic fit as can be seen from Fig.~\ref{crossings}. Note that all these slopes are negative, and in this case the absolute value of slope for $p^{2l+1}\cot(\delta)$ has to be larger than the absolute value of slope for $-(p^2)^l \sqrt{-p^2}$.

\subsection{Bound states in the vector channel, and mass and coupling of the $\psi(3770)$}

The physical parameters for bound states and resonances are based on our fits of $D\bar D$ scattering in p-wave. The first step is to ensure that our fitted parametrizations fulfill all the physical requirements, and in particular the consistency check given by the condition in Eq.~(\ref{sanity2}). The double-pole fits in Section~\ref{sec:fits} satisfy this condition, while the quadratic fits do not, as explained in the previous subsection. Therefore, in the following, in terms of the physics conclusions we focus on the double-pole fits.

The absolute value of the resulting amplitude $t_{l=1}(s)$~(Eqs.~(\ref{complex-plane}) and (\ref{rhocotdelta})) in the complex energy plane is shown in Figs.~\ref{fig:poles12522} and \ref{fig:poles12315} for $\kappa_c=0.12522$ and $0.12315$, respectively. The locations of these poles on both Riemann sheets are collected in Fig. \ref{fig:poles_locations_2D}. These poles are related to (virtual) bound states and resonances according to the general criteria from Section \ref{sec:polesdef}. The final masses of various states will be quoted according to
 \begin{equation}\label{m}
 m=m^{\textrm{lat}}-M_{\textrm{av}}^{\textrm{lat}}+M_{\textrm{av}}^{\textrm{exp}}\,,
\end{equation}
where the mass splitting $[m-M_{\textrm{av}}]^{\textrm{lat}}$ with respect to $M_{\textrm{av}}$ of Eq.~(\ref{mav}) is determined from the lattice. Figure \ref{fig:states} and Table \ref{tab:states} summarize our results for various states, which are detailed below:
\begin{itemize}
\item $\bold{\kappa_c=0.12522}$ ($m_D = 1762(2)$ MeV): For our lighter charm quark mass there is a complex conjugated pair of poles in the second sheet corresponding to the $\psi(3770)$ resonance, see Fig.~\ref{complex_plane_vector_channel_kc12522_II}. Figure \ref{complex_plane_vector_channel_kc12522_I} shows a single pole of the $t$-matrix on the real axis in the first Riemann sheet, corresponding to the bound state $\psi(2S)$. In addition, there is a virtual bound state on the real axis on the second sheet.

\begin{figure}
  \subfigure[~I]{\includegraphics[width=.47\textwidth]{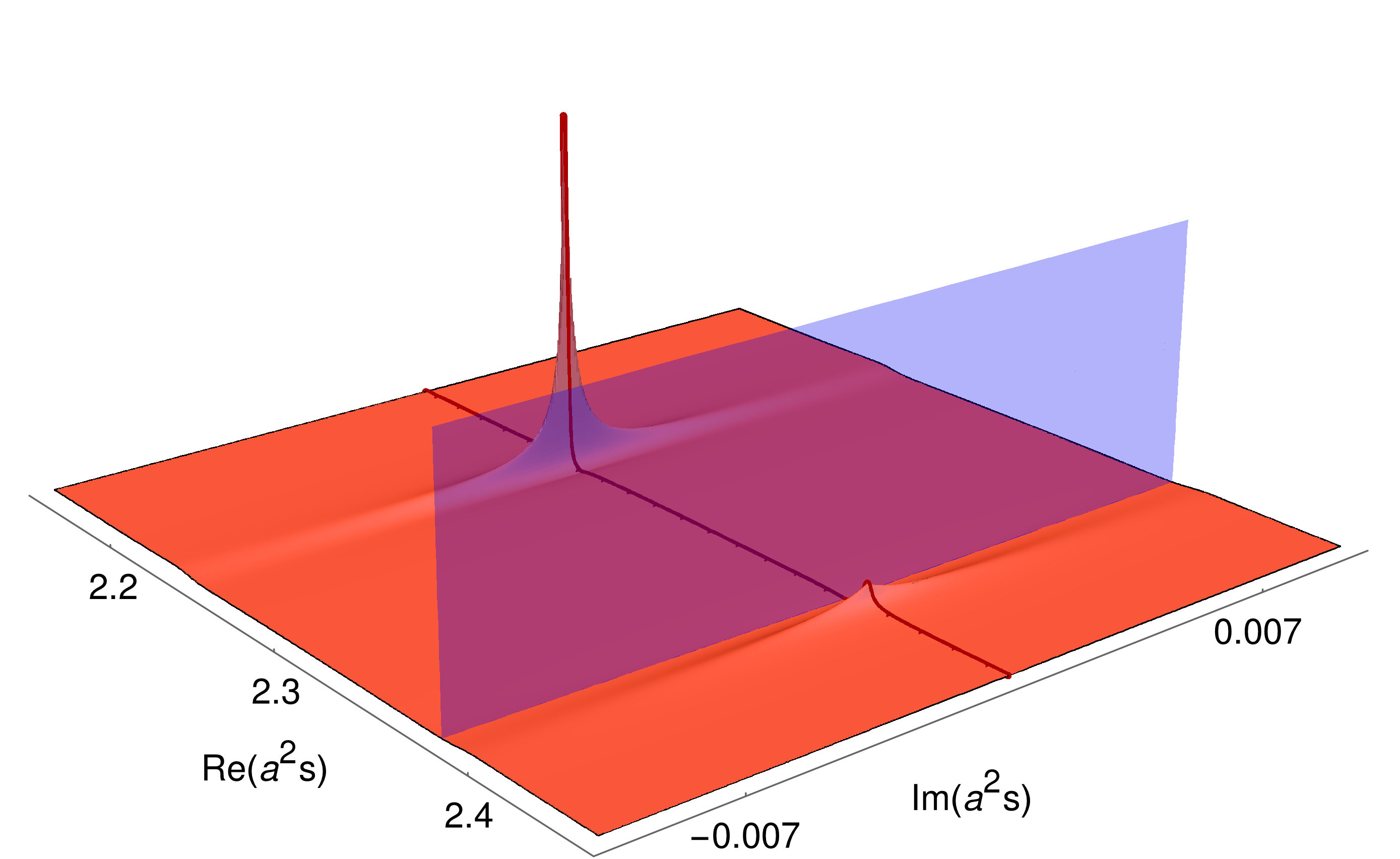}\label{complex_plane_vector_channel_kc12522_I}}
  \subfigure[~II]{\includegraphics[width=.47\textwidth]{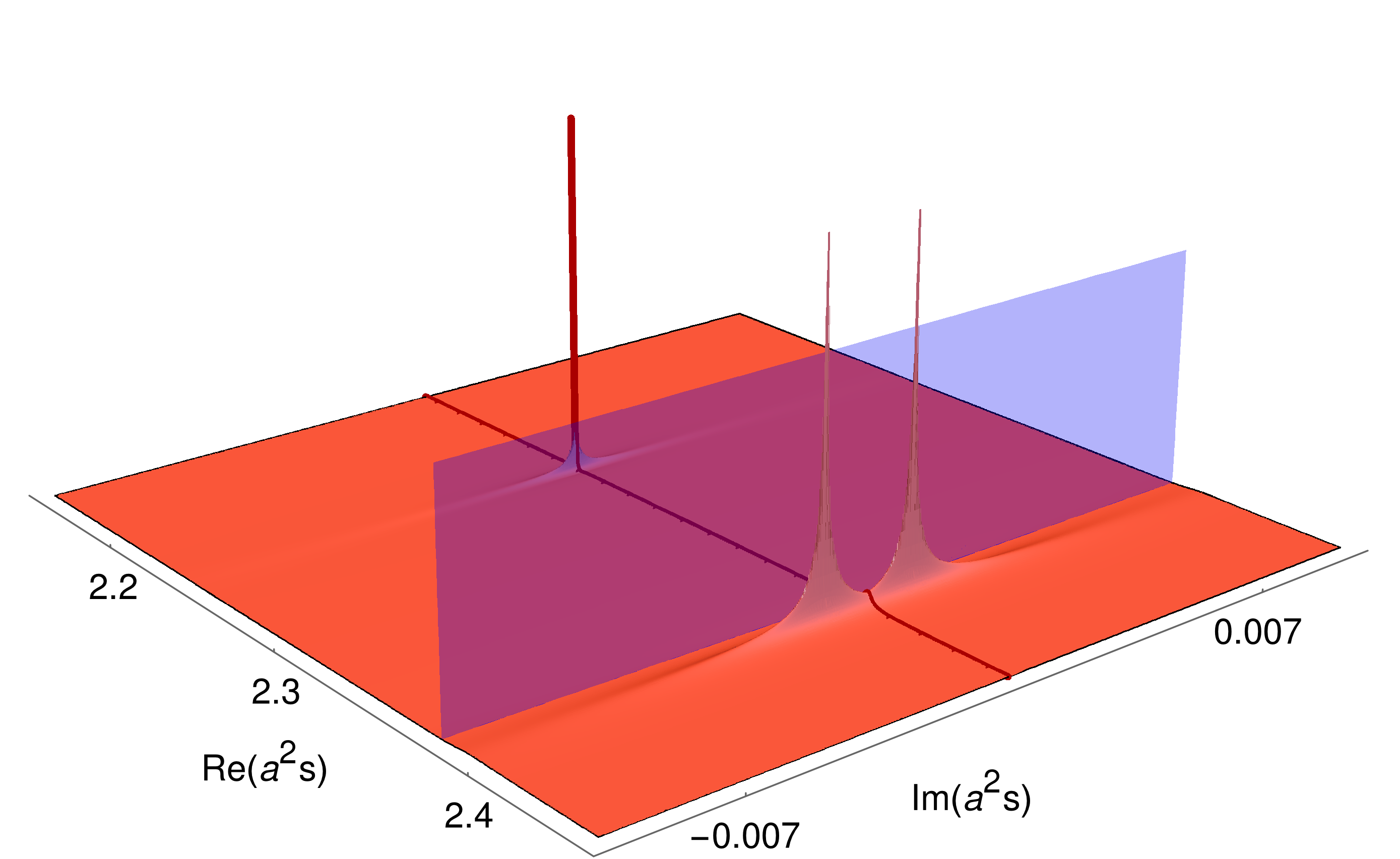}\label{complex_plane_vector_channel_kc12522_II}}
  \caption{The amplitude modulus $|t_{l=1}|$ of Eq.~(\ref{eq:TKSmatrices}) for $\bar DD$ scattering in the vector channel plotted in the complex energy plane for $\kappa_c=0.12522$. The bound state ($\psi(2S)$) is the pole on the real axis of the first Riemann sheet, while the resonance ($\psi(3770)$) appears on the second Riemann sheet as poles off the real axis above the scattering threshold. In the same position, a shoulder appears on the first sheet above threshold.}\label{fig:poles12522}
\end{figure}

\begin{figure}
  \subfigure[~I]{\includegraphics[width=.47\textwidth]{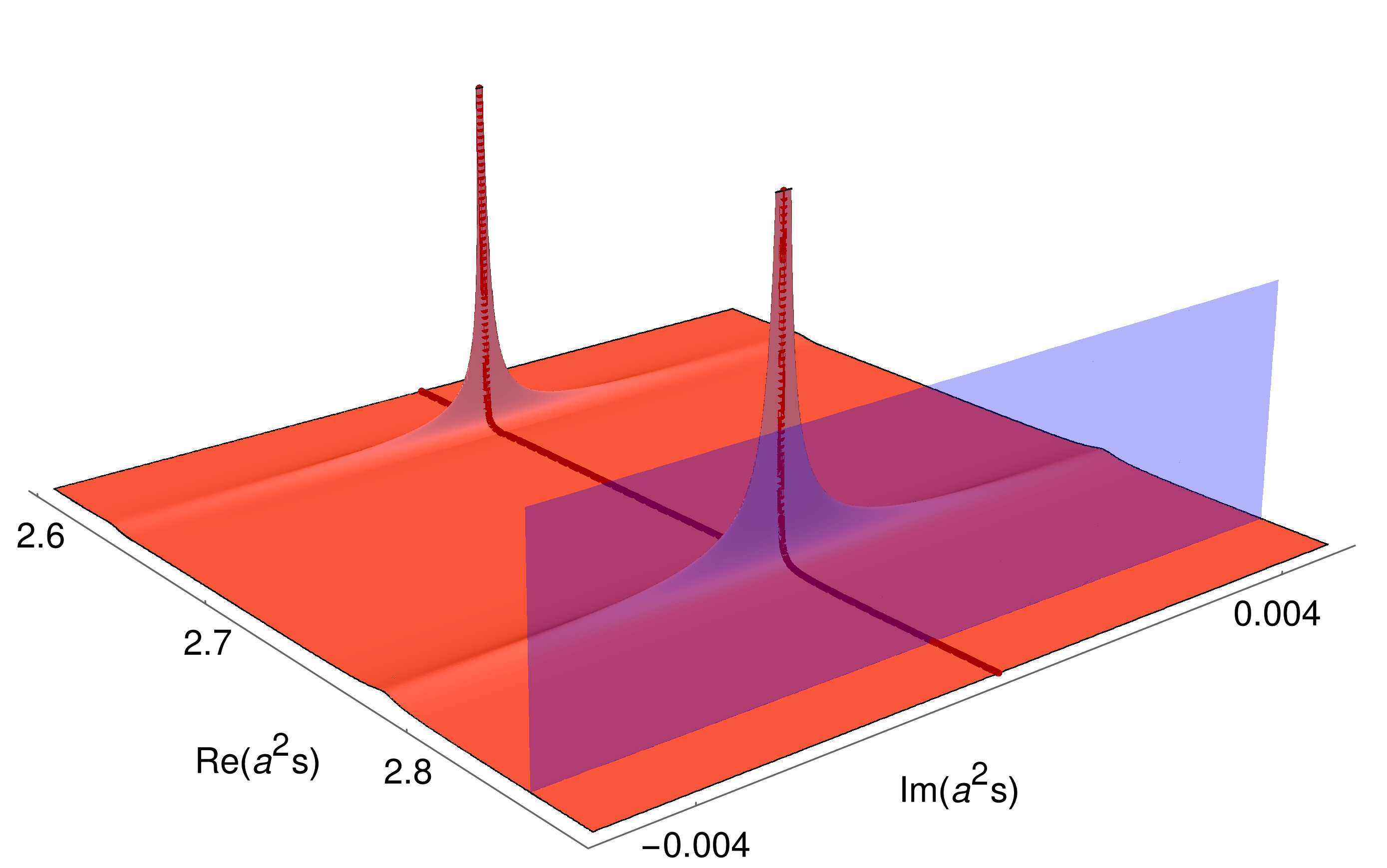}\label{complex_plane_vector_channel_kc12315_I}}
  \subfigure[~II]{\includegraphics[width=.47\textwidth]{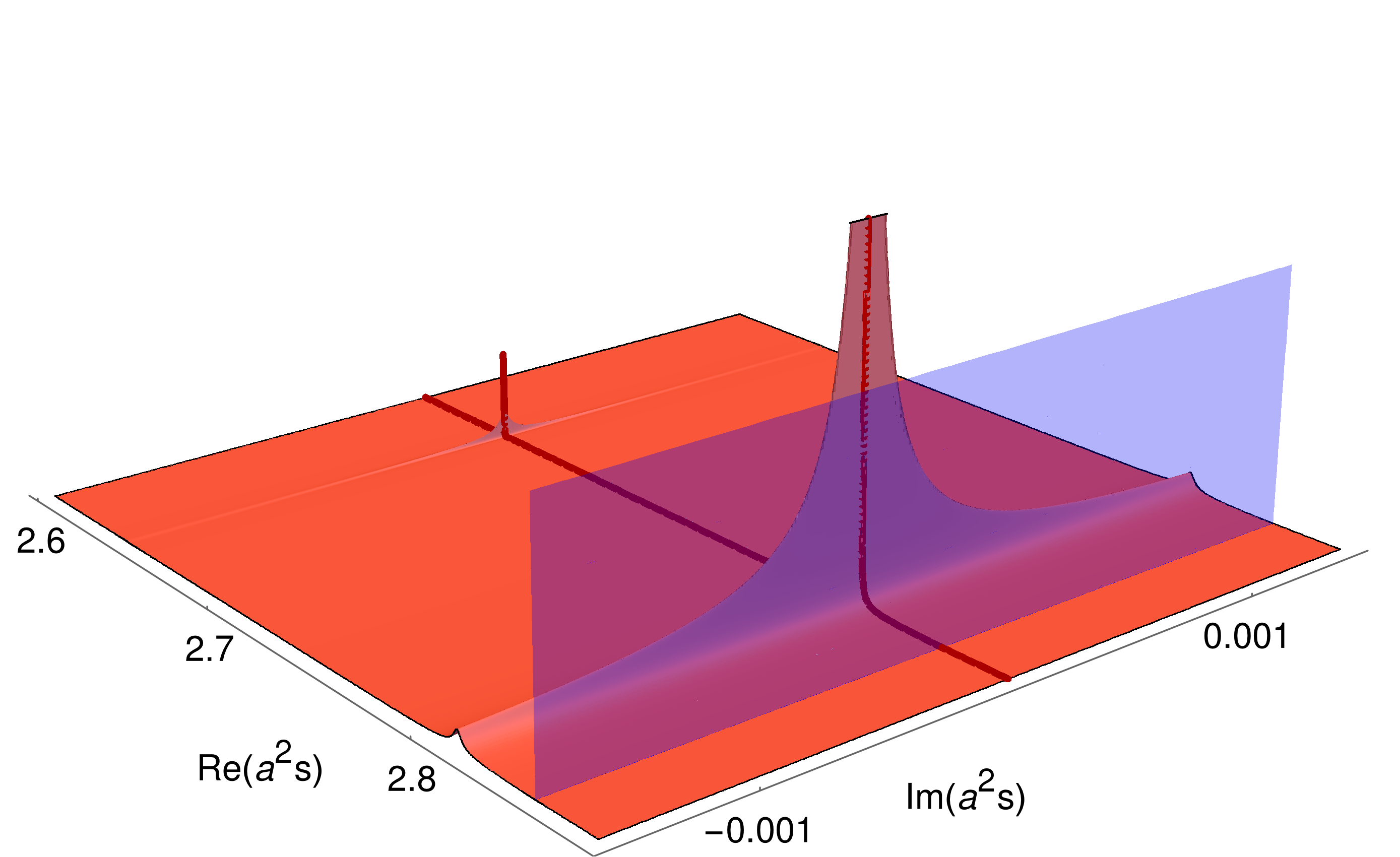}\label{complex_plane_vector_channel_kc12315_II}}
  \caption{The amplitude modulus $|t_{l=1}|$ of Eq.~(\ref{eq:TKSmatrices}) for $\bar DD$ scattering in the vector channel plotted in the complex energy plane for $\kappa_c=0.12315$. The semitransparent vertical plane represents the position of the $\bar DD$ threshold.}\label{fig:poles12315}
\end{figure}

The parameters for the $\psi(3770)$ resonance  are extracted from the linearized (Breit-Wigner) behavior of Eq.~(\ref{T1MM_3MM_double_pole_kc12522}) in the resonance region
\begin{equation}\label{bw}
\frac{p^3\cot\delta_1}{\sqrt{s}}\vert_{s\simeq m^2}=\frac{6\pi}{g^2}(m^2-s)~,\quad \Gamma= \frac{g^2 ~p^3 }{6\pi~s}.
\end{equation}
The resonance width $\Gamma$ can not be directly compared to experiment  since  the phase space (proportional to $p^3$) for our unphysical $D$ meson mass is different to that in experiment.  Therefore we extract the  dimensionless coupling $g = 16.0({}^{+2.1}_{-0.2})$ that parametrizes the $\psi(3770)\to\bar DD$ width. This coupling agrees within errors with  the experimental value $g^{\textrm{exp}}=18.7(9)$ obtained from   $\Gamma^{\textrm{exp}}$. The resonance mass equals $m_2$ of Eq.~(\ref{eq:doublefitform}). The fitted value is given in Eq.~(\ref{T1MM_3MM_double_pole_kc12522}) and~(utilizing  Eq.~(\ref{m})) this yields a final mass $m_2 = 3780(7)$~MeV, which is consistent with the experimental value of $3773.13(35)$~MeV.

\begin{figure}
  \includegraphics[width=.47\textwidth]{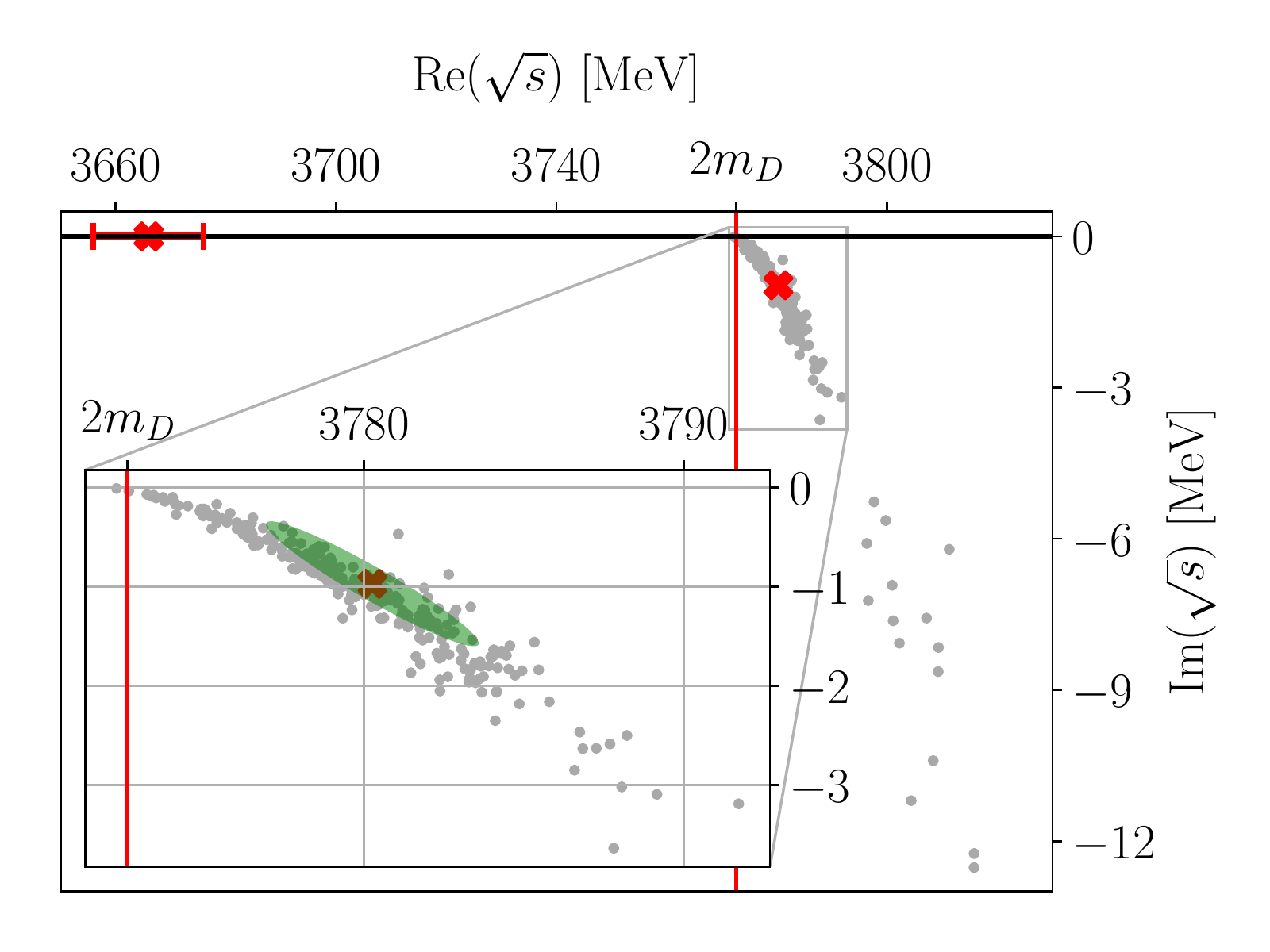}
  \caption{Position of the poles of the $\psi(3770)$ and $\psi(2S)$ in the complex energy plane for $\kappa_c=0.12522$. The red cross on the real axis with an error-bar is the pole of the $\psi(2S)$ appearing on the first Riemann sheet. The gray points represent the position of the $\psi(3770)$ pole on a given bootstrap sample on the second Riemann sheet. The green ellipse in the inset represents the error in the position of the $\psi(3770)$ pole, determined from the distribution of the points inside the inset, excluding the outliers. Each red cross corresponds to the position of a pole provided by the ensemble average, while the vertical red line marks the position of the scattering threshold. The horizontal axis is shifted and converted in physical units according to (\ref{mav}) as Re$(\sqrt{s}) = E^{\textrm{lat}} - M_{\textrm{av}}^{\textrm{lat}} + M_{\textrm{av}}^{\textrm{exp}}$.}\label{fig:poles_locations_2D}
\end{figure}

These parameters are verified by also considering the location of the resonance pole in the amplitude of Eq.~\ref{T1MM_3MM_double_pole_kc12522}, located at $m_2^{pole}+\tfrac{1}{2} \Gamma^{\textrm{pole}}=3780(4) - 0.97(63) i $ MeV. The pole is plotted in Fig.~\ref{fig:poles_locations_2D}. The value of $\Gamma^{pole}$ results from the formula on the right in Eq.~(\ref{bw}) to a coupling $g=17(9)$\footnote{The phase space  $p^2/s$ in (\ref{bw}) is extracted from  $m^{\textrm{pole}}\simeq3780$ MeV and threshold position 3772 MeV in Fig.~\ref{fig:poles_locations_2D} (phase space in simulation gives also compatible result).}, consistent with the value quoted in the previous paragraph.

The mass of the $\psi(2S)$ bound state is obtained from $p^3 \cot(\delta)/\sqrt{s}$ of Eq.~(\ref{T1MM_3MM_double_pole_kc12522}) by finding a pole for real energies below threshold according to Eq.~(\ref{bs1}): it corresponds to the crossing of the red and blue curves in Fig.~\ref{vector_channel_crossing_poles_kc12522}. The resulting mass in Table \ref{tab:states} is about $20~$MeV smaller than the experimental mass.

\item $\bold{ \kappa_c=0.12315}$ ($m_D = 1927(2)$ MeV): $\psi(3770)$ is a shallow bound state  for our heavier charm quark mass, in contrast to experiment where it is a strongly decaying resonance. It corresponds to a near-threshold pole in the first sheet of Fig.~\ref{complex_plane_vector_channel_kc12315_I}, arising from the fact that the crossing of our fits with the real axis occurs on the left of the axis origin (below threshold) in Fig.~\ref{double_pole_fit_12315_vector_channel}. The mass of the bound state $m=3776(7)$~MeV in Table \ref{tab:states} is determined according to Eq.~(\ref{bs1}) and corresponds to the near-threshold crossing of the red and blue lines in Fig.~\ref{double_pole_fit_12315_vector_channel}. The $\psi(3770)$ cannot strongly decay into $\bar DD$, but we still extract the coupling  $g=18.9({}^{+0.8}_{-0.7})$, which parametrizes the slope of $p^3 \cot\delta_1/\sqrt{s}$ near the real axes according to Eq.~(\ref{bw}).   
  
The $\psi(2S)$ corresponds to the lighter of the two bound states. Its mass $m\simeq 3687~$MeV,  determined from the bound-state condition of Eq.~(\ref{bs1}), has a large error due to the huge uncertainty of the coupling $G_2$ in the double-pole parametrization in Eq.~(\ref{T1MM_3MM_double_pole_kc12315}). This mass is consistent within errors with the value $m=3665(9)~$MeV obtained directly from the first-excited energy-level at  $\vec P=0$ and $L=32$. This agreement is expected as the influence of the threshold on a state which lies about 150 MeV below is very small. We quote the latter value as the final result for the $\psi(2S)$ mass at this $\kappa_c$. 
\end{itemize}

Our results for the masses of the $\psi(3770)$ and $\psi(2S)$  states are summarized  and compared to experiment in Fig.~\ref{fig:states} and Table \ref{tab:states}. The mass $m$ and the coupling $g$ for the $\psi(3770)$ agree with the experimental values  within errors, which applies to both charm-quark masses considered. The mass of the $\psi(2S)$ is  about $2\sigma$  below the experimental value, which is not unreasonable given the lack of continuum and quark-mass extrapolations to the physical values.

The above results~(that are based on a combined fit to two different ensembles) are subject to a technical subtlety. Both ensembles have been generated with the same bare lattice parameters, but with different lattice sizes. Hence sub-leading exponentially suppressed finite volume effects, which are generally neglected in the L\"uscher finite volume formalism, can sometimes lead to different physical situations for different box sizes. Such a delicate situation can happen when a pole singularity in the complex energy plane is expected to be very close to the threshold: a small volume may exhibit different physics from that of a larger volume. For example, in our case for the lighter $D$-meson mass, we see that our $N_L=24$ data have a tendency to prefer a bound state (unlike the $N_L=32$ data), although the data are compatible with a resonance within the large errors. In order to confirm that our combined fit does not lead to conclusions that are different from the physical situation in the larger volume~(where the neglected exponential corrections are small) we performed an analysis of the $N_L=32$ data alone. We find the mass and the coupling from such an analysis to be $m_2=3782(7)$ MeV and $g=15({}^{+2}_{-1})$, which are in agreement with the results from our combined fits.

 \begin{table*}
\begin{tabular}{cc| c||c|c || c}
 & $J^{PC}$ & lat (present work)& lat  (present work)  & exp  & lat \cite{Lang:2015sba} \\
  &                & $\kappa_c=0.12315$ &  $\kappa_c=0.12522$ & $\bar D^0D^0/D^+D^-$ \\
 & & & &   & \\
  $m_D$ [MeV] &                &$1927(2)$ &  $1762(2)~$  &   $\bar m_D\simeq 1867$~MeV  & $1763(22)(18)^*$\\
 $m_{D_s}$ [MeV]  &                & $1981(1)~$ &  $1818(1)~$ & $1968.34(7)$ & \\
    $M_{\textrm{av}}$ [MeV] &                & $3103(3)~$  & $2820(3)~$ & $3068.6(2)~$  & $3119(9)(33)^*$\\
    $m_\pi$ [MeV] &  & 280 & 280 &  $\bar m_\pi \simeq 137~$MeV & 266 \\
    \hline
  \hline
  $ \boldsymbol{\psi(3770)}$ & $1^{--}$  & bound st.& resonance & resonance \cite{Tanabashi:2018oca}  & resonance \\
  {\bf g} &   & $\mathbf{18.9({}^{+0.8}_{-0.7})}$ & $\mathbf{16.0({}^{+2.1}_{-0.2})}$ & $\mathbf{18.7(9)}$ & 13.2(1.2)  \\
  $m-M_{\textrm{av}}~$[MeV]  &    & 707(7) & 711(7)& 704.25(35) & 715(7) \\
    $m-2m_{D}~$[MeV]  &  & -43(8) & 9(7)& 38.52(35) & \\
  {\bf $\boldsymbol{m}~$[MeV]} &  & {\bf 3776(7)} & {\bf 3780(7)} & {\bf 3773.13(35)}\footnote{We consider PDG fit estimate as the experimental value for the mass of the $\psi(3770)$ resonance throughout this article.}  & 3784(7)\\
  \hline
    $ \boldsymbol{\psi(2S)}$ & $1^{--}$ & bound st. & bound st. & bound st.  \cite{Tanabashi:2018oca} & bound st. \\ 
      $m-M_{\textrm{av}}~$[MeV]  &  & 596(9) & 597(10)& 617.347(25)  & 605(6)\\
    $m-2m_{D}~$[MeV]  &   & -154(10)& -105(11)& -48.383(25) & \\
  {\bf $\boldsymbol{m}~$[MeV]}&  & {\bf 3665(9)} & {\bf 3666(10)}& {\bf 3686.097(25)} & 3674(6)\\
  \hline
   $ \boldsymbol{X(3842)} $ & $3^{--}$ & resonance & resonance & resonance \cite{Aaij:2019evc} \\ 
      $m-M_{\textrm{av}}~$[MeV]  &  &  754(${}^{+4}_{-7}$)& 762(${}^{+10}_{-16}$)  & 773.9(2) \\
    $m-2m_{D}~$[MeV]  &  & 4(${}^{+9}_{-3}$)& 59(${}^{+11}_{-16}$)& 108.2(2) \\   
   {\bf $\boldsymbol{m}~$[MeV]} &  & {\bf 3822}$\mathbf{({}^{+4}_{-7})}$ & {\bf 3831}$\mathbf{({}^{+10}_{-16})}$ & {\bf 3842.7(2)}   \\
    \hline  
\end{tabular} 
\caption{Summary of masses for resonances and bound states with $J^{PC}=1^{--}$ and $3^{--}$ studied in our lattice simulation  and compared to experiment \cite{Tanabashi:2018oca}. The dimensionless coupling $g$ parametrizes the width $\Gamma=g^2 p^3/(6\pi s)$ of the resonance decay $\psi(3770)\to \bar DD$ from Eq.~(\ref{bw}). Lattice results are given for two charm quark masses, where the corresponding values of  $m_D$   are given at the top of the table. The splittings $m-M_{\textrm{av}}~$ and $m-2m_D$ are obtained taking solely lattice or solely experimental values of masses. The final mass $m$ resulting from our simulation is quoted according to Eq.~(\ref{m}). The last column contains the lattice results from the  two-flavor  simulation of Ref.~\cite{Lang:2015sba} based on "fit (i)" described in Section \ref{sec:oldcomparison}. For the masses marked with (*) we quote the kinetic masses ($M_2$ in Eq.~(3.1) of Ref. \cite{Lang:2015sba}) which are the appropriate masses for the Fermilab approach \cite{ElKhadra:1996mp,Oktay:2008ex} pursued in Ref. \cite{Lang:2015sba}. These differ from the rest masses by discretization effects.}\label{tab:states}
\end{table*}

\begin{figure}
    \includegraphics[width=.49\textwidth]{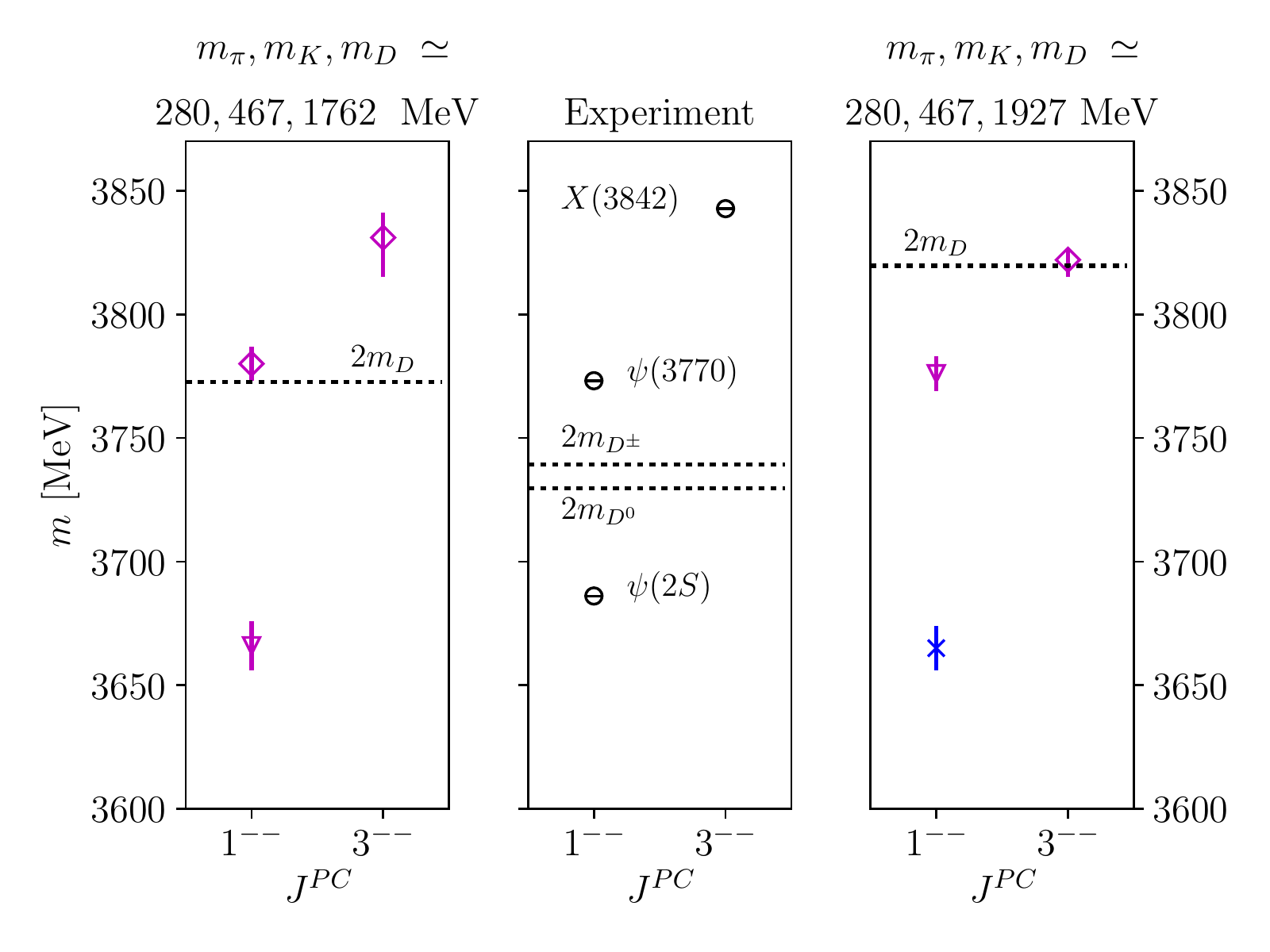}
    \caption{The masses~($m$) of $J^{PC}=1^{--}$ and $3^{--}$ states obtained for the two charm quark masses in our simulation, compared to experiment. The masses are determined  according to Eq.~(\ref{m}) and the location of the $\bar DD$ threshold, denoted by dotted lines, is presented in a similar way as $(2m_D-M_{\textrm{av}})^{\textrm{lat}}+(M_{\textrm{av}})^{\textrm{exp}}$. The magenta diamonds denote resonances extracted from $\bar DD$ scattering in partial waves $l=1$ or $l=3$. The magenta triangles indicate bound states extracted from $\bar DD$ scattering. The blue crosses denote states extracted as energy levels in the finite box.  All  masses  are also summarized in Table \ref{tab:states}.
%The resulting masses of $J^{PC}=1^{--}$ and $3^{--}$ states for two charm quark masses in our simulation, compared to the experimental masses. Masses from lattice simulation are quoted according to Eq.~\ref{m}.
}\label{fig:states}
\end{figure}

\subsection{$J^{PC}=3^{--}$ resonance}

The LHCb collaboration recently reported the discovery of the $X(3842)$ charmonium state in the $\bar DD$ invariant mass \cite{Aaij:2019evc} with a mass $m^{\textrm{exp}}\simeq  3842~$MeV, a very narrow width $\Gamma^{\textrm{exp}}\simeq 2.8~$MeV and likely quantum numbers $J^{PC}=3^{--}$. The presence of such a $3^{--}$ resonance was explored by fitting the parametrization of the $\tilde{K}$ matrix in Eq.~(\ref{eq:det}) to our energy levels including the Breit-Wigner form in Eq.~(\ref{del3BW}) for $l=3$.

The resulting mass for $\kappa_c=0.12522$~(see Eq.~(\ref{T1MM_3MM_double_pole_kc12522})) is 
\begin{equation}
  m_{3^{--}}=3831({}^{+10}_{-16})~\mathrm{MeV}\,.
\end{equation}
The mass obtained from our lattice simulation is consistent with the mass of the discovered $X(3842)$ state. This adds confidence to the interpretation of the $X(3842)$ as a $3^{--}$ charmonium resonance. The results for both values of the charm quark mass are compared to experiment in Table \ref{tab:states}. Our simulation is not sensitive to the width of this very narrow resonance since there are no $\bar DD$ eigenstates in the very narrow energy region $m\pm\Gamma$.\footnote{This applies for the width $\Gamma=\Gamma^{\textrm{exp}}$, while the width $\Gamma$ is even narrower on our lattice due to smaller phase space.}  In particular, the finite volume lattice energy by itself would lead to compatible results for the mass of the $X(3842)$.  For this reason we can not compare the width or the $X(3842)\to \bar DD$ coupling with
experiment.

\subsection{Comparison to previous results}\label{sec:oldcomparison}

Elastic $\bar{D}D$ scattering and its relation to the $\psi(2S)$ and $\psi(3770)$ resonances have previously been investigated in a lattice study in Ref. \cite{Lang:2015sba} for $m_\pi\simeq 156~$MeV and $266~$MeV. In this subsection we compare our current study to these older results, focusing on  those from the $m_\pi\simeq 266~$MeV ensemble, since the pion mass is similar and the errors are smaller. The most significant improvement over the calculation in Ref. \cite{Lang:2015sba} lies in the use of two lattice volumes at the same lattice spacing and light-quark masses, combined with calculations in moving frames. The latter is facilitated by a new approach to dealing with discretization effects described in Section \ref{approach} along with the spin identification techniques pioneered in Ref. \cite{Dudek:2009qf} and employed on our ensemble(s) in Ref. \cite{Padmanath:2018tuc}. The use of two volumes and several moving frames results in a significant increase in the number of energy levels obtained, which renders more information on the dependence of the $l=1,3$ scattering amplitudes on $E_{cm}$. 

One important consequence is that the $J^{PC}=3^{--}$ resonance, which contributes to the same irreducible representations on a hyper-cubic lattice, can now be included in our fits. In Ref. \cite{Lang:2015sba} it had to be assumed that after omission of an energy level related to the presence of this resonance, its effect on the remaining finite volume energies was negligible. For our current ensembles this assumption can instead be tested, and our results agree with the mass of the recently observed $X(3842)$ \cite{Aaij:2019evc}.

The $l=1$ scattering amplitude in Ref.~\cite{Lang:2015sba} was constrained only at three values of $E_{cm}$ and the results for $\psi(2S)$ and $\psi(3770)$ were summarized in Table V of Ref. \cite{Lang:2015sba} for two types of fits. Fit (ii) in  Eq.~(5.6)  was similar to the "quadratic fit"  used in the present work and does not satisfy the consistency check of Eq.~(\ref{sanity1}) for the $\psi(2S)$ bound state~(this constraint on the physical parameterizations was not checked in Ref. \cite{Lang:2015sba}). Fit (i) was the Breit-Wigner fit  of the energy region above threshold and rendered the $\psi(3770)$ resonance parameters summarized in the right column of Table \ref{tab:states}; the  mass is consistent with  our present result while the coupling is somewhat smaller. The $\psi(2S)$ mass $m=3674(6)~$MeV obtained from the naive energy level (second level in Table III of Ref. \cite{Lang:2015sba}) agrees within errors with our present result.  The two-pole fit with four parameters could not be preformed using just three energy levels in Ref. \cite{Lang:2015sba}.  

Another advantage of the present study is the use of the CLS gauge ensemble library, which in principle enables us to extend the current study to further ensembles with different lattice spacings and light quark masses, while maintaining our current setup.

\section{Conclusions}\label{sec:concl}

We determined the elastic $\bar{D}D$ scattering amplitude in the energy region of the $\psi(2S)$ bound state and 
the $\psi(3770)$ resonance from lattice QCD simulations using the finite volume L\"uscher method. The results we 
obtain are from simulations with 2+1 flavors of dynamical light quarks at unphysical masses and at a single 
lattice spacing. We investigate different parametrizations of the $p$-wave scattering amplitude and impose a 
consistency check on the extracted bound state poles to ensure the physical bound state constraint in 
Eq.~(\ref{sanity2}). Using a ``double pole'' parametrization defined in Eq.~(\ref{eq:doublefitform}), the bound 
state mass of the $\psi(2S)$ and the resonance parameters of the $\psi(3770)$ have been extracted from a 
simultaneous fit to the energy spectrum from three different moving frames and two lattice volumes. 

The lattice study of a near-threshold charmonium state away from the physical point requires a careful choice of 
the charm quark mass for the strong decay to $\bar{D}D$ to be allowed. To this end, on each gauge field ensemble, 
we employ two charm quark mass values leading to $D$-meson masses 100 MeV below and 60 MeV above the physical 
value. The masses of the $\psi(2S)$ and $\psi(3770)$, measured as splittings from the spin-averaged ground-state 
mass, are in good agreement with experiment for these values of the heavy quark mass. The result for the coupling 
$g$ of the $\psi(3770)$ resonance with the $\bar DD$ scattering channel is compatible with the experimental value, 
even for the charm quark mass where the $\psi(3770)$ is a bound state. The results from both heavy quark masses
employed in this work, the results from previous existing work and the corresponding experimental values are 
summarized in Table \ref{tab:states}. 

We also  investigate the $\bar DD$ scattering in partial wave $l=3$ that contains $3^{--}$ resonances. One of the 
aims is to investigate how this partial wave contributes to the finite volume spectra. Including the lattice 
energy levels related to the lowest $3^{--}$ state in the amplitude analysis and parametrizing the corresponding 
$\bar{D}D$ scattering amplitude with a Breit-Wigner form (Eq.~\ref{del3BW}), we are able to extract the mass of 
this state. Nice agreement is found with the mass of the recently discovered $X(3842)$, which is interpreted as 
a $3^{--}$ charmonium resonance. However, we are unable to reliably determine its coupling with the $\bar DD$ 
scattering channel as the finite volume spectrum of $\bar DD$ scattering channels are not sufficiently dense to 
be sensitive to its decay width. We also find our results for the vector channel remain unchanged with the 
exclusion of the $l=3$ partial wave from the amplitude analysis and hence conclude the influence of the low 
lying $3^{--}$ resonance on the lattice estimates for the $\psi(3770)$ resonance parameters is negligible in our calculation. 

The main challenge for future determinations of scattering amplitudes in the charmonium spectrum will be the presence 
of multiple coupled channels for the study of conventional and exotic resonances away from the $\bar{D}D$-threshold. 
A first preliminary analysis of the scalar channel has been presented in Ref.~\cite{Collins:2018myr}, and we plan to complete 
the study in a forthcoming publication. In addition it would be desirable to obtain a denser set of points by using more
and/or larger volumes, to reduce the discretization effects by calculations on CLS ensembles with a smaller lattice 
spacing, and to investigate the approach to the physical point by utilizing ensembles with lighter pion masses. In particular, discretization errors are an important source of systematic uncertainty of our current work. Using the CLS trajectory with physical strange-quark mass \cite{Bali:2016umi} would be particularly attractive, as the splitting between the $D$ and $D_s$ mesons on these ensembles is closer to the physical splitting and therefore results in a larger energy window for elastic $\bar{D}D$-scattering.

\section{Acknowledgements}
We thank our colleagues A. Sch\"afer and S. Weish\"aupl for help and support during the execution of the project.
We thank Ben H\"orz for his help and support with the package TwoHadronsInBox and C. B. Lang for his variational 
analysis codes. We are also thankful to Sinya Aoki, Zohreh Davoudi, Jozef Dudek, Andrew Hanlon and 
Akaki Rusetksy for valuable discussions. We are grateful to the Mainz Institute for Theoretical Physics (MITP) for 
its hospitality and its partial support during the  course of this work. We use the multigrid solver of 
Refs.~\cite{Heybrock:2014iga,Heybrock:2015kpy,Richtmann:2016kcq,Georg:2017diz} for the inversion of the Dirac operator. Our code implementing distillation is 
written within the framework of the Chroma software package~\cite{Edwards:2004sx}. The simulations were performed 
on the Regensburg iDataCool cluster, and the SFB/TRR 55 QPACE~2~\cite{Arts:2015jia} and QPACE~3 machines. We 
thank our colleagues in CLS for the joint effort in the generation of the gauge field ensembles  which form a 
basis for the here described computation. The Regensburg group was supported by the Deutsche Forschungs-gemeinschaft 
Grant No. SFB/TRR 55 and the European Union Innovative Training Network Grant No. 813942 (EuroPLEx). M. P. acknowledges support from the EU under grant no. MSCA-IF-EF-ST-744659 (XQCDBaryons). 
S. Prelovsek was supported by Slovenian Research Agency ARRS (research core funding No. P1-0035 and No. J1-8137).

\appendix

\section{Error determination\label{app1}}

\begin{figure}
 \includegraphics[width=.47\textwidth]{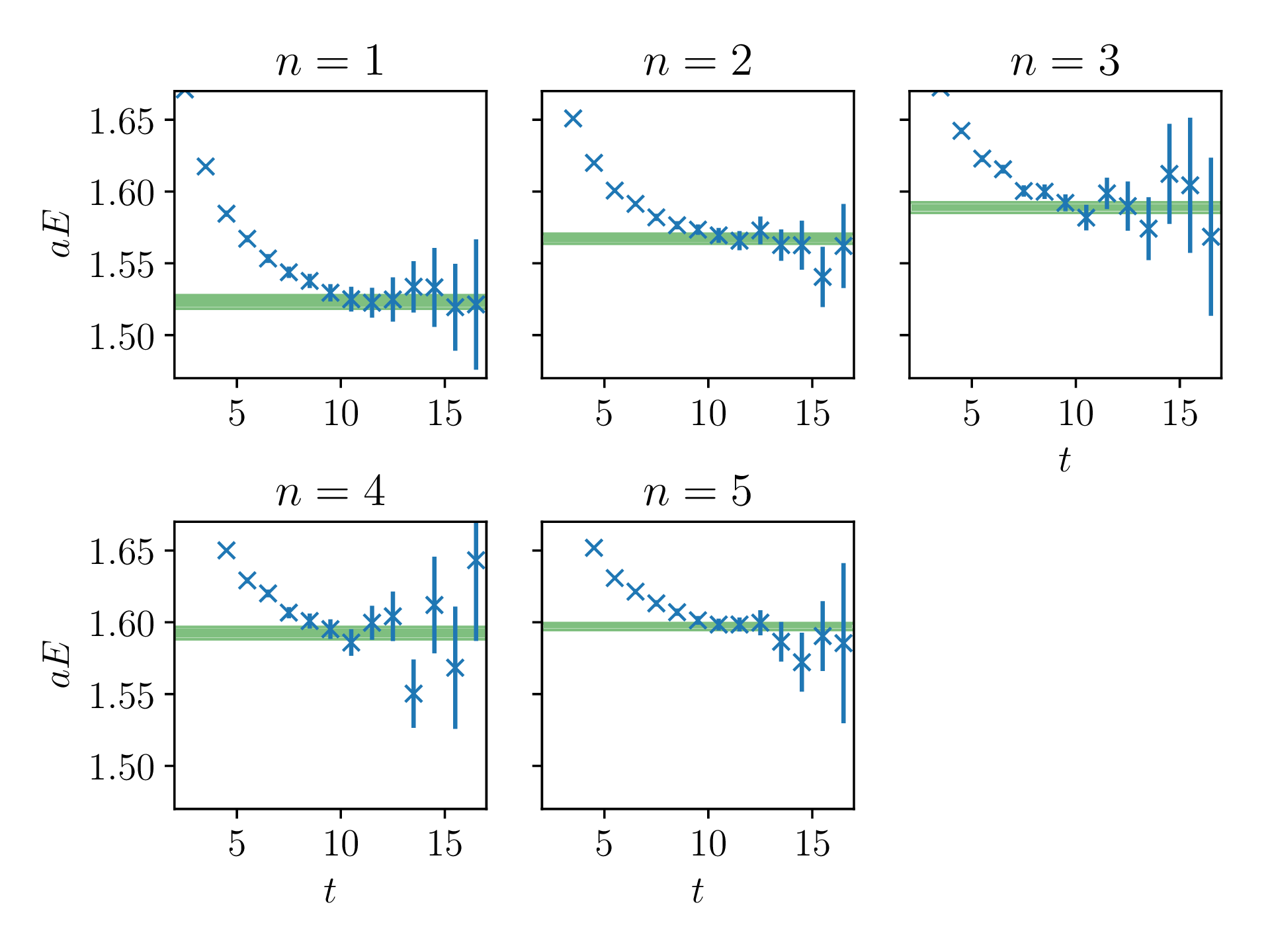}
  \caption{Effective masses of the eigenvalues of the GEVP for the first five excited states for ensemble H105 and $\kappa_c=0.12522$ in the irrep $A_1^-$ with total momentum $|\vec{P}|^2 = 2$. The masses obtained from fitting the eigenvalues with a single exponential are shown by the green band. The width of the band indicates the error on the mass.}\label{effmass_H105_A1_P110_Cm_kappa12522_nbin2}
\end{figure}

A combined fit of the phase shift from the energy spectrum in various lattice irreducible representations and ensembles requires a careful consideration of all the sources of correlations and systematic uncertainties. The fitting interval used to extract the energy levels from Eq.~(\ref{exp_lambda_decay}) is a well-known source of systematic error. An example of our fits is presented in Fig.~\ref{effmass_H105_A1_P110_Cm_kappa12522_nbin2}. We have performed fits for several different intervals and compared the stability of the raw energy shifts as defined in Eq.~(\ref{delE}). We observed that the dependency of the energy shifts on the fitting interval of the scattering $D$-mesons is relatively small. There is a dependence on the starting point of the fit of the excited charmonium state, that stabilizes when the excited state contamination is under control. The raw correlators are binned to take into account the autocorrelations in our data and we have chosen a bin-size equal to two, after observing that the error is stable if larger bin-sizes are used.

\begin{figure}
  \includegraphics[width=.47\textwidth]{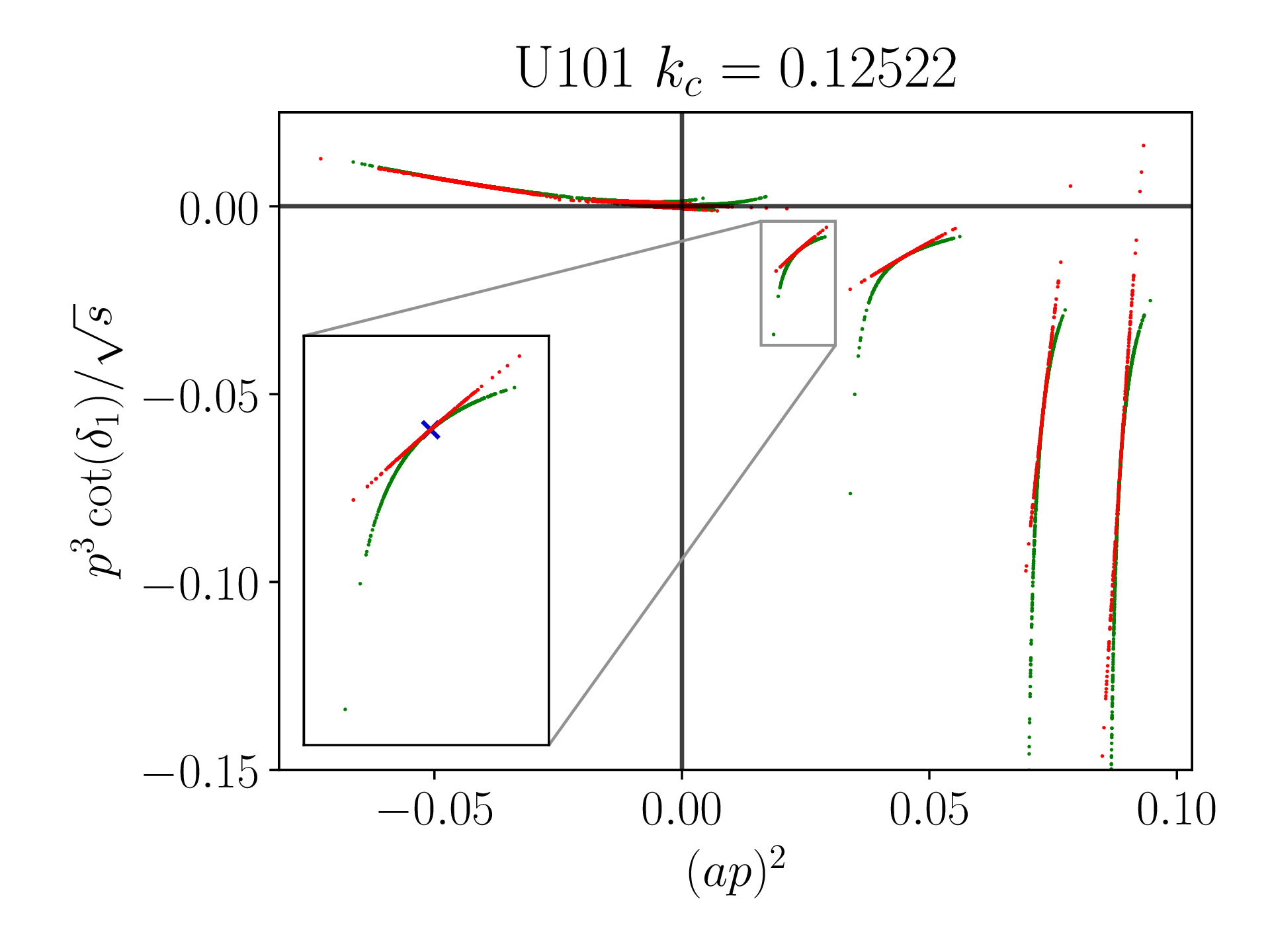}
  \caption{Comparison of the error on the phase shift in the vector channel as determined from bootstrapping and jackknife for the U101 ensemble. Every green~(red) point represents the phase shift determined on a given bootstrap~(jackknife) sample. The blue cross represents the ensemble average. For jackknife, we rescale the resampling points by a factor $\sqrt{N_{\textrm{conf}}}$.}\label{jackknife_vs_bootstrapping}
\end{figure}

\begin{figure}
  \includegraphics[width=.23\textwidth]{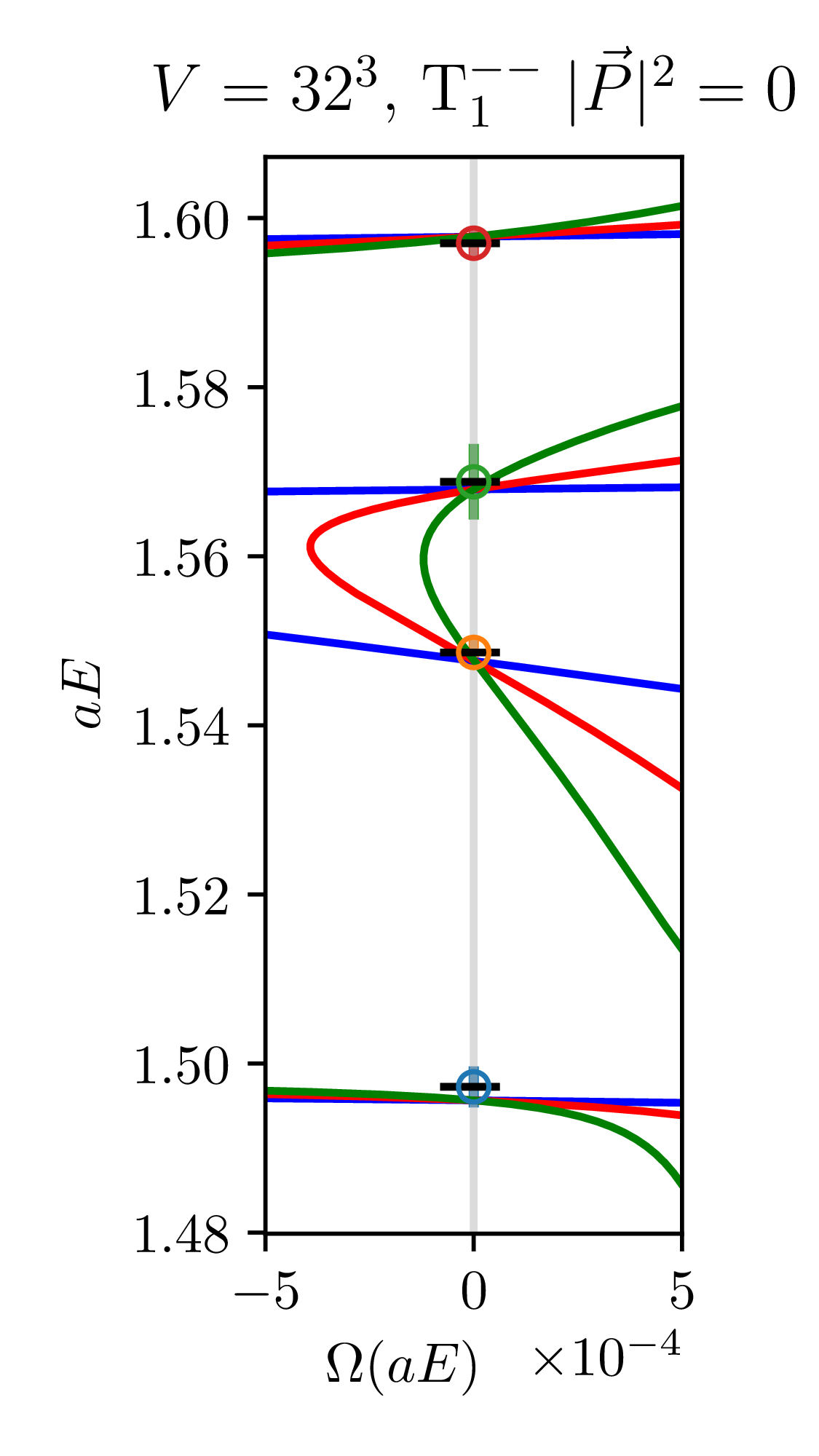}
  \caption{Comparison of the $\Omega$-function for the ``double pole'' fit-form for $\kappa_c=0.12522$ for different values of the regularization parameter $\mu$ (green for $\mu=16$, blue for $\mu=2$ and red for $\mu=8$). The crossings with the origin on the vertical axis are not affected by the choice of $\mu$.}\label{multiple_mu_comparison}
\end{figure}

Jackknife and bootstrapping are methods widely employed in lattice QCD to determine the errors on secondary quantities extracted or fitted from the raw data. Each bootstrapping resample consists of 255 and 492 randomly chosen configurations from the U101 and H105, respectively. The first step in the analysis is the determination of the covariance matrix for the correlated $\chi^2$ minimization of our $K$-matrix parametrization. This is estimated simply as the standard correlation between two variables as given by bootstrapping. The minimum of the correlated $\chi^2$, defined from the $\Omega$-function of Eq.~(\ref{omega_function}) as given in Eqs. (38) and (44) of Ref.~\cite{Morningstar:2017spu}, is recomputed for each bootstrapping resample. We extract from here the distribution of secondary derived quantities such as the parameters of the $K$-matrix or the positions of the poles in the complex plane. The $\Omega$-function depends on an additional regularization parameter $\mu$, and we have verified that varying $\mu$ in the range between 2 and 16 does not change the central value nor the error of the fitted parameters significantly. The crossing with the vertical axis, determining the expected spectrum, does not change significantly if $\mu$ is varied in the above interval, see Fig.~\ref{multiple_mu_comparison}. Note however that the value of $\mu$ determines the slope and the curvature of the $\Omega$-function at the crossing.

Given that our fits combine two different ensembles with a different number of configurations, single elimination jackknife is not an optimal choice. However, it is worth to compare bootstrapping and jackknife if we restrict the fits to a single ensemble, say U101. Looking to Fig.~\ref{jackknife_vs_bootstrapping}, we can see that jackknife~(for which the spread of the resamples is smaller by a factor $\sqrt{N_{\textrm{conf}}}$ compared to bootstrapping) determines the error as a linear approximation of the L\"uscher $\mathcal{Z}$-functions around the central value of the phase shift. It might happen however that a given energy level lies very close to a turning point of the $\mathcal{Z}$-function, as in the case of the four rightmost points of Fig.~\ref{jackknife_vs_bootstrapping}. In this situation, the strong non-linear nature of the phase-shift analysis could result in a systematic underestimation of the error when using jackknife. For the same reason, the bootstrap distribution of the phase shift is not symmetric and has long tails. We therefore conclude that bootstrapping provides a more reliable determination of the uncertainty compared to jackknife for phase-shift studies. 

The error on the fitted parameters, as well as on the couplings and masses of the charmonium states, is determined as the standard deviation of the bootstrap samples, except for the case when the distribution is asymmetrical or has long tails. In this situation, we quote asymmetric uncertainties as determined by the interval that fits 68~\% of the bootstrap samples, {\it i.e.} by an interval that excludes 16~\% of the bootstrap samples from the left and right tail.

\section{Correlation matrices of the fitted parameters}

In this appendix we include the correlation matrix $C$ between the parameters $\kappa_i$ of the ``double pole'' fits described in Sec.~\ref{sec:fits} which fulfill the consistency checks of Sec.~\ref{sec:consistency_check}. The correlation matrix is defined as
\begin{equation}
 C_{ij}= \frac{\langle (\kappa_i - \bar{\kappa}_i) (\kappa_j - \bar{\kappa}_j) \rangle}{\sqrt{\langle (\kappa_i - \bar{\kappa}_i)^2 \rangle\langle (\kappa_j - \bar{\kappa}_j)^2 \rangle}}
\end{equation}
where the expectation values appearing in the numerator and the denominator are computed directly from the bootstrapping distributions and $\bar{\kappa}_i = \langle \kappa_i \rangle$.

For the ``double pole'' fit form, the six parameters $\kappa_i$ are given in the order $m_2$, $m_1$, $m_3$, $G_2$, $G_1$ and $g_3$. The correlation matrices are
\begin{equation*}
 C = \left( \begin{matrix}
    1 &   -0.07   & -0.33 & -0.12 & -0.11 & 0.06 \\
    -0.07 & 1     & -0.04 & 0.    & -0.61 & -0.02 \\
    -0.33 & -0.04 & 1     & -0.14 & -0.14 & -0.23 \\
    -0.12 & 0     & -0.14 & 1     & 0.06  & -0.19 \\
    -0.11 & -0.61 & 0.14  & 0.06  & 1     & -0.06 \\
    0.06  & -0.02 & -0.23 & -0.19 & -0.06 & 1
 \end{matrix} \right)\,,
\end{equation*}
for $\kappa_c = 0.12522$, and
\begin{equation*}
 C = \left( \begin{matrix}
 1     & -0.05 & -0.68 & -0.26 & 0.12 & -0.01 \\
 -0.05 & 1     & -0.21 & 0.04  & 0.60 &  -0.01\\
 -0.68 & -0.21 &  1    & 0.02  &-0.26 & -0.04 \\
 -0.26 & 0.04  &  0.02 &  1    & 0.07 &  0.06 \\
  0.12 & 0.60  & -0.26 &  0.07 &  1   & -0.13 \\
 -0.01 & -0.01 & -0.04 &  0.06 & -0.13&  1
 \end{matrix} \right)\,,
\end{equation*}
for $\kappa_c = 0.12315$.

\section{Alternative parametrization aiming to describe the full vector charmonium spectrum\label{app3}}

\begin{figure}[b!]
  \subfigure[$\kappa_c =0.12315$]{\includegraphics[width=0.175\textwidth]{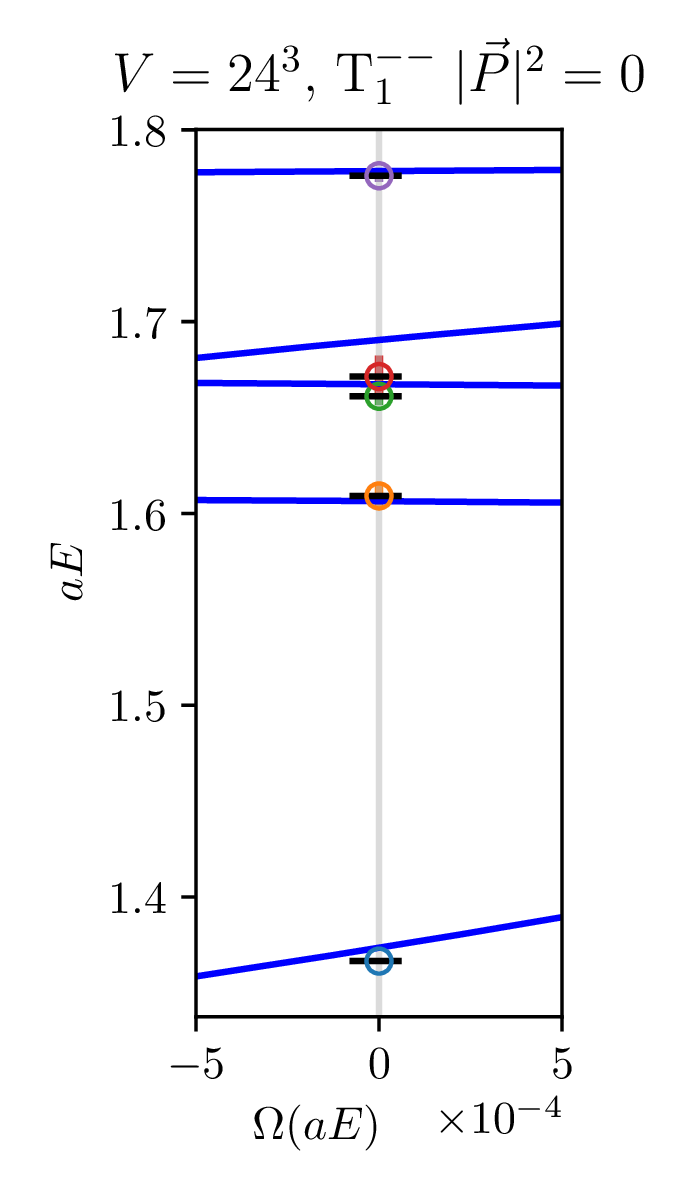}}
  \subfigure[$\kappa_c =0.12315$]{\includegraphics[width=0.175\textwidth]{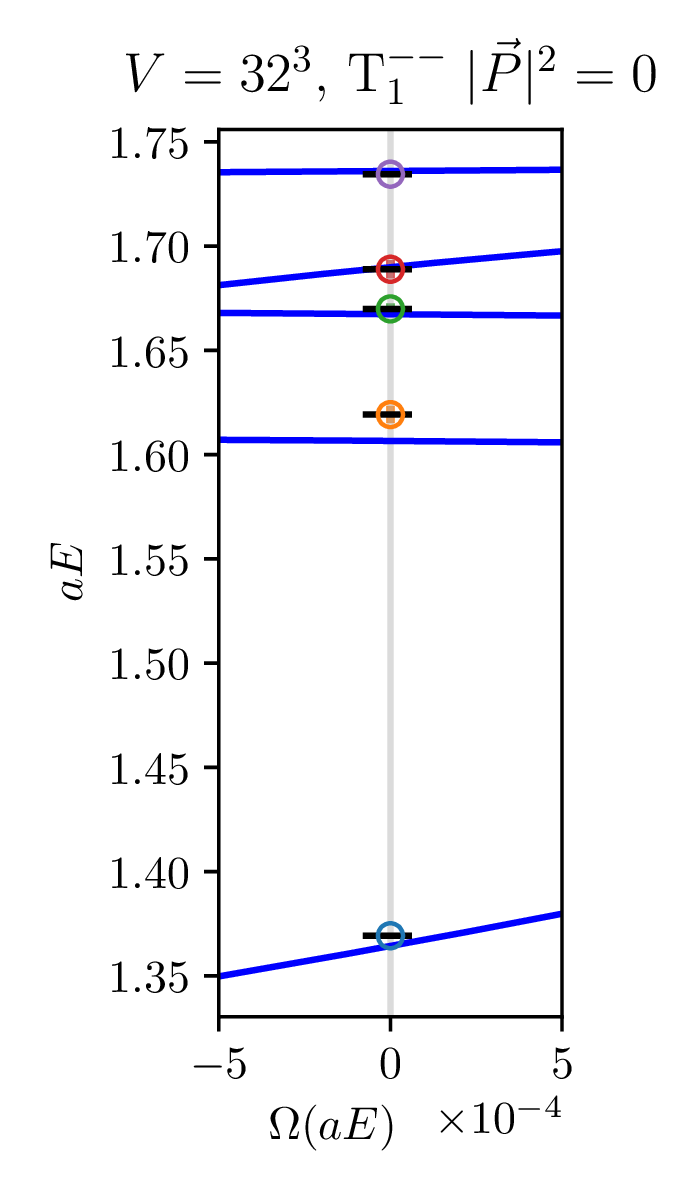}}
  \subfigure[$\kappa_c =0.12522$]{\includegraphics[width=0.175\textwidth]{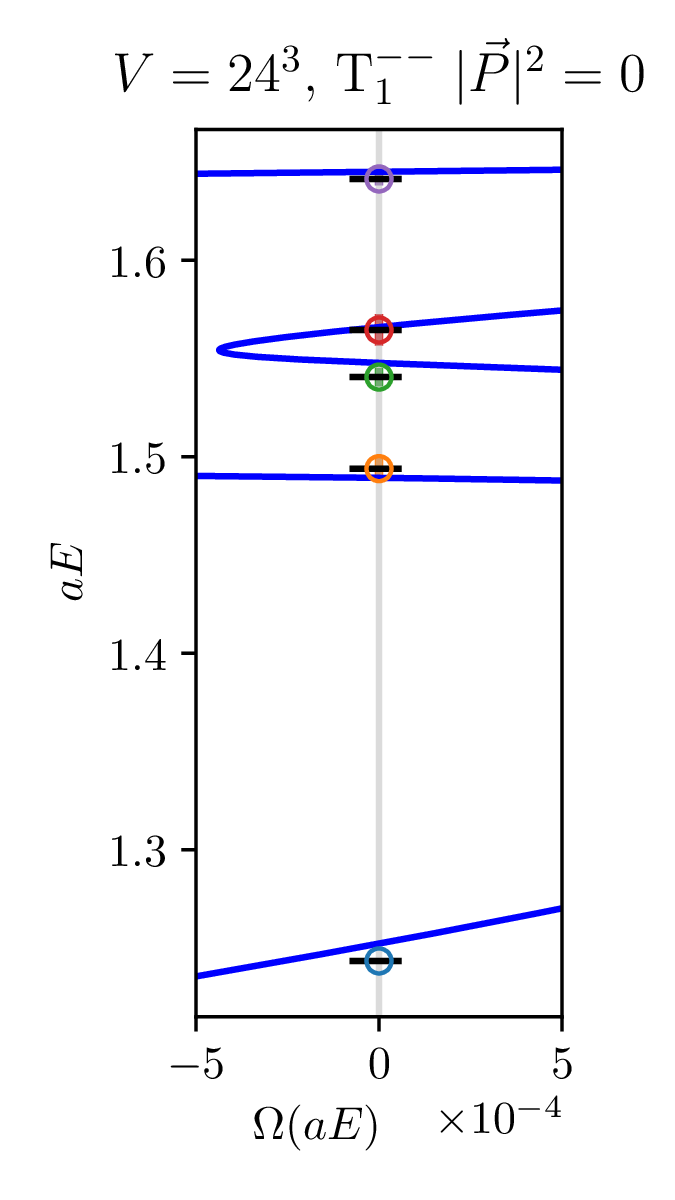}}
  \subfigure[$\kappa_c =0.12522$]{\includegraphics[width=0.175\textwidth]{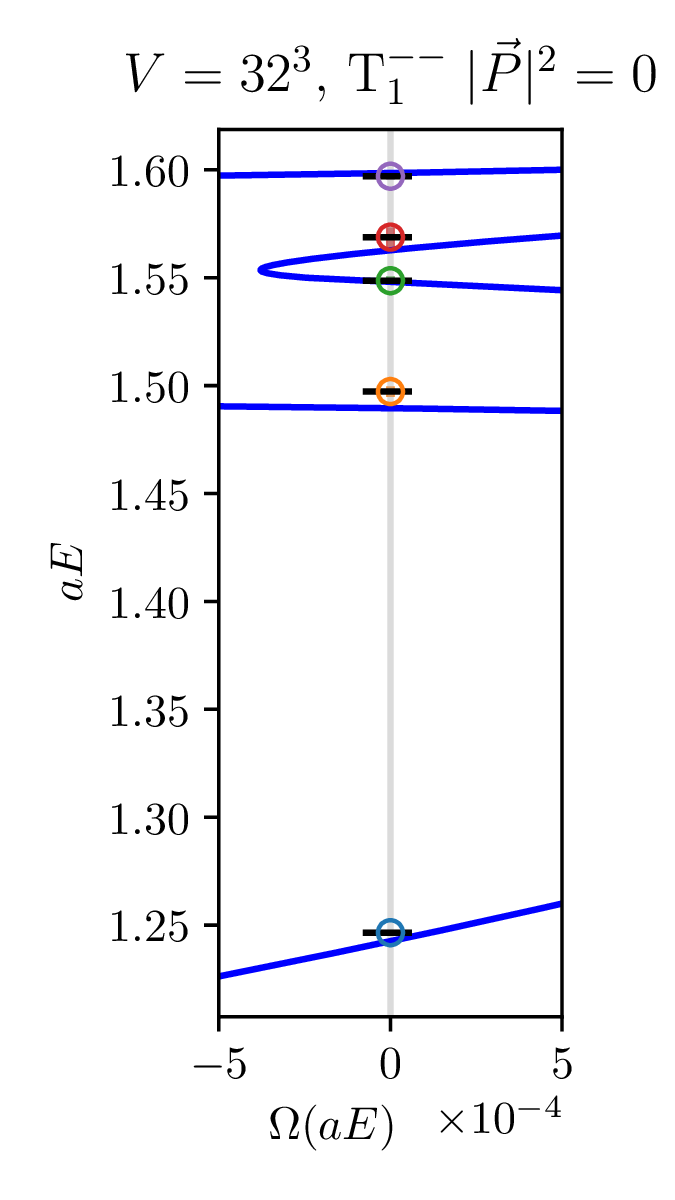}}
  \caption{$\Omega$-function in the $T_1^{--}$ lattice irreducible representation in rest frames for the vector channel resulting from the fit of all energy levels including the $J/\psi$ using the modified ``double pole'' of Eq.~(\ref{double_pole_mod}) fit ansatz.}\label{alternative_param}
\end{figure}

The ``double pole'' fit form is able to reproduce the charmonium spectrum including the $\psi(2S)$, as discussed in Sec.~\ref{sec:fits}. The presence of an additional crossing of the $\Omega$-function below the $\psi(2S)$ is an interesting feature that suggests the possibility of describing the full charmonium spectrum including even the $J/\psi$. For example, by using a slightly modified ``double pole'' fit form with an additional linear parameter $\rho$
\begin{equation}\label{double_pole_mod}
 \frac{p^3 \cot(\delta_1)}{\sqrt{s}} = \left(\frac{G_1^2}{m_1^2 - s} + \frac{G_2^2}{m_2^2 - s}+ \rho\right)^{-1}\,.
\end{equation}
The corresponding $\Omega$-function determined from the fits are presented in Fig.~\ref{alternative_param}. While the couplings and the masses $G_1$, $G_2$, $m_1$ and $m_2$ are consistent within error with Eq.~(\ref{T1MM_3MM_double_pole_kc12522}) and (\ref{T1MM_3MM_double_pole_kc12315}) after the inclusion of the $J/\psi$, the parameter $\rho$ fluctuates strongly across the bootstrapping resamples. Given that the spectrum below the $\psi(2S)$ is well separated from the two resonances we are investigating, we can safely exclude the $J/\psi$ from our main analysis. We emphasize again that our focus is the scattering amplitude in the energy region above the $\psi(2S)$ (the consistency check of Sec.~\ref{sec:consistency_check} would not be satisfied for the $J/\psi$ with the parameterization \ref{double_pole_mod}).

\bibliography{VectorCharmoniumResonances}

\end{document}